\documentclass[pdflatex,sn-basic,iicol]{sn-jnl}

\usepackage{graphicx}%
\usepackage{multirow}%
\usepackage{amsmath,amssymb,amsfonts}%
\usepackage{amsthm}%
\usepackage{mathrsfs}%
\usepackage[title]{appendix}%
\usepackage{xcolor}%
\usepackage{textcomp}%
\usepackage{manyfoot}%
\usepackage{booktabs}%
\usepackage{algorithm}%
\usepackage{algorithmicx}%
\usepackage{algpseudocode}%
\usepackage{listings}%
\usepackage{enumitem}%

\begin{document}

\title[White dwarf spectral evolution]{The spectral evolution of white dwarfs: where do we stand?}

\author*{\fnm{Antoine} \sur{B\'edard}}\email{antoine.bedard@warwick.ac.uk}

\affil{\orgdiv{Department of Physics}, \orgname{University of Warwick}, \orgaddress{\city{Coventry}, \postcode{CV4 7AL}, \country{UK}}}

\abstract{White dwarfs are the dense, burnt-out remnants of the vast majority of stars, condemned to cool over billions of years as they steadily radiate away their residual thermal energy. To first order, their atmosphere is expected to be made purely of hydrogen due to the efficient gravitational settling of heavier elements. However, observations reveal a much more complex situation, as the surface of a white dwarf (1) can be dominated by helium rather than hydrogen, (2) can be polluted by trace chemical species, and (3) can undergo significant composition changes with time. This indicates that various mechanisms of element transport effectively compete against gravitational settling in the stellar envelope. This phenomenon is known as the spectral evolution of white dwarfs and has important implications for Galactic, stellar, and planetary astrophysics. This invited review provides a comprehensive picture of our current understanding of white dwarf spectral evolution. We first describe the latest observational constraints on the variations in atmospheric composition along the cooling sequence, covering both the dominant and trace constituents. We then summarise the predictions of state-of-the-art models of element transport in white dwarfs and assess their ability to explain the observed spectral evolution. Finally, we highlight remaining open questions and suggest avenues for future work.}


\maketitle

\section{Introduction}
\label{sec:introduction}

White dwarfs are the end products of the life cycle of the wide majority of stars and, as such, contain a wealth of information on the history of the Milky Way. The archetypical white dwarf has a mass of $\simeq 0.60\,M_{\odot}$ within a radius of only $\simeq 0.013\,R_{\odot}$, resulting in an extremely high density. The stellar interior is thus strongly electron-degenerate, giving rise to a peculiar but well-defined (and hence very useful) mass--radius relation. Furthermore, the intense gravitational field results in a highly stratified chemical structure: a large carbon--oxygen core, a thin helium mantle, and an even thinner hydrogen layer. The helium and hydrogen shells making up the stellar envelope are generally thought to account for $\simeq 10^{-2}$ and $\simeq 10^{-4}$ of the total mass. The tenuous observable atmosphere, comprising a mere $\simeq 10^{-14}$ of the total mass, is usually composed of hydrogen to a high degree of purity due to the efficient gravitational settling of heavier elements. Because nuclear burning no longer occurs in their interior, white dwarfs continuously cool and fade over time as they slowly radiate away their residual thermal energy. The effective temperature decreases from $\simeq 100,000$\,K to $\simeq 3000$\,K in about 10\,Gyr, thereby serving as a direct proxy for the cooling age. Thanks to this unique property, white dwarfs act as reliable cosmic clocks, routinely used to measure the age of stellar populations and the stellar formation history of the Milky Way \citep{winget1987,oswalt1996,leggett1998,fontaine2001,hansen2007,garcia-berro2010,kalirai2012,tremblay2014,kilic2017,fantin2019,isern2019,cukanovaite2023}.

Given these basic characteristics, white dwarfs are often viewed as very simple astrophysical objects. However, upon closer scrutiny, one is faced with the paradox that the applications arising from their relative simplicity require a detailed understanding of their structure and evolution, which in turn involves many complex considerations. Building an accurate model of a white dwarf requires a multitude of microphysical ingredients (equation of state, radiative and conductive opacity, diffusion coefficients, etc.) over an exceptionally wide range of conditions, including some highly challenging physical regimes \citep{saumon2022}. Moreover, the cooling process is not as uneventful as one may naively believe: it is markedly impacted by phenomena as diverse as residual nuclear burning, neutrino emission, core crystallisation, and convective coupling \citep{fontaine2001,althaus2010,chen2021}. The cooling rate also depends on the masses of the hydrogen and helium layers as well as on the relative amounts of carbon and oxygen in the core, which are dictated by complex (and still poorly understood) mechanisms taking place in previous evolutionary stages \citep{fontaine2001,althaus2010}. In hindsight, exploiting the potential of stellar remnants as cosmic clocks proves to be a highly non-trivial endeavour.

One particularly intricate and surprising property of white dwarfs as a group is the diverse and changing nature of their surface composition. Although most objects possess a standard hydrogen atmosphere, many instead exhibit a helium atmosphere, indicating that they have lost the bulk of their outer hydrogen layer in an earlier evolutionary phase. Furthermore, in both cases, it is not uncommon for elements other than the main constituent to be present in small amounts, despite the expected efficiency of gravitational settling. Even more puzzling is the fact that the surface composition of white dwarfs can change with time as they cool, a phenomenon referred to as spectral evolution. These variations involve not only the abundances of the trace contaminants, but also the nature of the dominant constituent, with atmospheres transitioning from helium-rich to hydrogen-rich and vice versa. This suggests that the transport of elements in the envelope of white dwarfs is in fact much more complicated than a simple sedimentation process. Indeed, additional transport mechanisms such as convection, winds, and accretion may effectively compete against gravitational settling and thus alter the chemical makeup of the observable layers \citep{fontaine1987}.

The study of spectral evolution can reveal valuable information on the envelope of white dwarfs, which in turn has crucial implications in various areas of astrophysics. As mentioned above, the envelope composition influences the cooling rate and must therefore be well constrained to ensure the reliability of stellar age-dating applications \citep{fontaine2001}. White dwarf envelopes also bear the chemical imprint of complex physical processes occurring in their progenitors and can thus help improve our knowledge of previous phases of stellar evolution \citep{werner2006}. Moreover, as we will see later, spectral evolution finds a remarkable application in the field of exoplanets as it can provide unique information on the internal structure of these objects \citep{jura2014}. Finally, the changing surface composition of a white dwarf represents a direct window on the physics of element transport in stars \citep{salaris2017}.

The spectral evolution of white dwarfs has been a topic of active research for several decades, with efforts directed along two main lines of investigation. The first approach aims to provide an empirical characterisation of the phenomenon, that is, to constrain the observed variations in atmospheric composition along the cooling sequence. This undertaking rests mostly on the analysis of observational data from large astronomical surveys; in particular, key advances have been made possible by the Sloan Digital Sky Survey \citep{york2000} and, more recently, the Gaia mission \citep{gaia2016}. The second approach is theoretical in nature and seeks to identify and understand the physical mechanisms giving rise to spectral evolution. This involves the calculation of white dwarf models considering the effects of various element transport processes, the predictions of which can then be compared to empirical constraints. These two fields are complementary and have often driven one another, with new observational findings motivating further modelling efforts and vice versa. The ultimate goal is to develop a complete theory of spectral evolution, offering a consistent explanation for all the observed features \citep{fontaine1987}.

The purpose of this review is to provide an exhaustive picture of our current understanding of white dwarf spectral evolution, highlighting recent advances and remaining challenges. The paper is structured in two main parts relating to the two approaches outlined above: Section~\ref{sec:observations} focuses on the empirical evidence, while Section~\ref{sec:theory} describes the theoretical interpretation. Finally, concluding remarks are given in Section~\ref{sec:conclusion}.

\section{Empirical evidence}
\label{sec:observations}

Except in a few rare cases mentioned below, the spectral evolution of white dwarfs is usually not witnessed in real time. Rather, the existence of this phenomenon is deduced from the fact that large samples of white dwarfs show statistical variations of surface composition as a function of effective temperature (and thus as a function of cooling age). By studying these variations, one can infer how the atmospheric composition of individual objects changes with time along the cooling sequence. This is the empirical approach to spectral evolution, where one attempts to constrain the net effect of the various element transport mechanisms on the observable outer layers. As such an endeavour rests heavily on the study of white dwarf spectral types and atmospheric properties, we first review some basic notions of spectral classification and analysis. We then provide a comprehensive overview of the latest findings regarding the surface composition of white dwarfs, in terms of both the dominant constituent and the trace contaminants.

\subsection{Preliminaries: spectral classification and atmospheric characterisation}
\label{sec:observations_prelim}

\begin{figure*}[t]
\centering
\includegraphics[width=2.0\columnwidth,clip=true,trim=3.00cm 15.00cm 2.75cm 4.25cm]{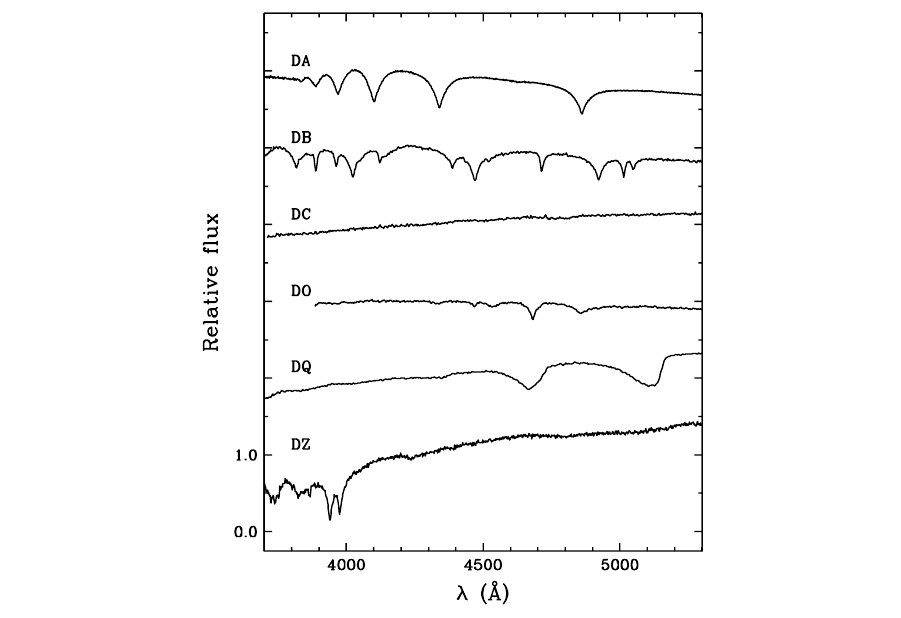}
\caption{Examples of optical spectra of DA, DB, DC, DO, DQ, and DZ white dwarfs. The spectra are normalised at $\lambda = 4200$\,\AA\ and shifted vertically from each other by an arbitrary amount. The data are taken from \citet{bergeron1997}, \citet{bergeron2011}, \citet{giammichele2012}, and \citet{subasavage2017}.}
\label{fig:spectra}
\end{figure*}

The standard spectral classification system for white dwarfs is described in detail in \citet{sion1983} and \citet{wesemael1993}. Spectral types always begin with the letter D (to indicate the degenerate nature of the star) and include a second letter identifying the most prominent lines in the optical region of the spectrum. The main spectral classes and their associated features are as follows:
\begin{itemize}
\item DA: hydrogen features;
\item DB: neutral helium features;
\item DC: no features (continuous spectrum);
\item DO: ionised helium features;
\item DQ: carbon features\footnote{In their original classification system, \citet{sion1983} included an exception by which the DQ spectral type could be assigned based on the presence of carbon features not only in the optical, but in any part of the electromagnetic spectrum. We discourage the use of this exception, as it would create a two-tier classification system where the distinction between the DC and DQ spectral types would largely depend on the availability of an ultraviolet spectrum. As discussed later in this review, most optically featureless white dwarfs are expected to exhibit carbon features in the ultraviolet, however ultraviolet spectra are currently (and will likely remain) unavailable in the overwhelming majority of cases. We believe that these objects should be referred to as DC white dwarfs regardless of their ultraviolet features, so that the main spectral designation remains independent of sparsely available ultraviolet spectroscopy.};
\item DZ: other metal features.
\end{itemize}
Figure~\ref{fig:spectra} shows examples of white dwarf optical spectra for each of these classes.

The spectral type is a rough indicator of the surface composition but does not always reflect the nature of the main chemical constituent. For instance, a white dwarf with a pure-hydrogen atmosphere is indeed of the DA type over most of the sequence cooling, but is of the DC type at $T_{\mathrm{eff}} \lesssim 5000$\,K because hydrogen transitions are not excited at these low temperatures. Likewise, a white dwarf with a pure-helium atmosphere is classified as DO at $T_{\mathrm{eff}} \gtrsim 45,000$\,K, DB at $45,000\,\mathrm{K} \gtrsim T_{\mathrm{eff}} \gtrsim 11,000$\,K, and DC at $T_{\mathrm{eff}} \lesssim 11,000$\,K. The shift from DO to DB is due to the recombination of the helium ions, while the shift from DB to DC results from the disappearance of helium transitions. Finally, the atmosphere of DQ- and DZ-type stars actually consists mostly of helium and contains only small amounts of carbon or other metals. Their spectral appearance is attributable to their low effective temperature ($T_{\mathrm{eff}} \lesssim 11,000$\,K): under these conditions, helium is invisible while heavier elements remain visible, even at low abundances\footnote{There are a few exceptions. First, a small fraction of DQ-type white dwarfs have a carbon-dominated atmosphere (see Section~\ref{sec:observations_trace2}). Second, at $T_{\mathrm{eff}} \lesssim 5000$\,K, the DZ class includes both hydrogen-rich and helium-rich atmospheres (since hydrogen lines in turn disappear at these temperatures).}.

Additional letters can be appended to indicate the presence of secondary spectral features. For instance, a white dwarf spectrum showing both strong hydrogen lines and weak ionised helium lines is classified as DAO. Among the many possible hybrid types, some of the most common (and thus most relevant for the present review) are DAO, DBA, DAZ, and DBZ. Occasionally, spectral signatures from more than two elements can be detected, giving rise to even more specific classes such as DBAZ\footnote{In addition, other letters can be used to indicate the presence of emission features (``e''), a magnetic field (``H''), or photometric variability (``V'').}.

Although the spectral type is a useful classification label, it is not sufficient in itself to characterise the atmosphere of a white dwarf. Extracting the physical information encoded in spectral observations generally involves two analysis steps: the calculation of atmosphere models from first principles, and a quantitative comparison between the models and data. This allows the determination of the atmospheric parameters of a given star, namely, its effective temperature, surface gravity, and atmospheric composition. This procedure comes in two main variants, the so-called spectroscopic and photometric techniques, differing in the kind of observational data being analysed. The spectroscopic technique relies on the detailed shape of the spectral lines and thus requires a relatively high signal-to-noise spectrum \citep{bergeron1992,finley1997,liebert2005,bergeron2011}, while the photometric technique uses the overall spectral energy distribution built from apparent magnitudes together with a trigonometric parallax measurement \citep{bergeron1997,bergeron2001,bergeron2019,gentile-fusillo2019}. The photometric method has the advantage of being more widely applicable (as it does not require spectroscopic observations) but the inconvenience of being much less sensitive to the surface composition (which is thus often assumed rather than inferred). The measured atmospheric parameters can be combined to theoretical mass-radius relations and cooling calculations to derive other stellar properties of interest, such as the mass, radius, luminosity, and cooling age \citep{renedo2010,camisassa2016,bedard2020,salaris2022,bauer2023}.

\subsection{The main constituent: hydrogen or helium?} 
\label{sec:observations_main}

Despite their rich diversity of spectral appearances, nearly all white dwarfs fall into one of two main categories: those with a hydrogen-dominated atmosphere, and those with a helium-dominated atmosphere. Therefore, the most basic questions to be answered are: what are the respective proportions of these two groups among the white dwarf population, and how do these proportions change along the cooling sequence? In practice, a common strategy consists in determining the atmospheric parameters of a large number of white dwarfs and calculating the fraction of helium-rich objects as a function of effective temperature. This simplified approach, where one focuses solely on the dominant chemical constituent and ignores the possible presence of trace elements, provides a first-order picture of spectral evolution. The assessment of the helium-atmosphere fraction has been an active research topic for several decades \citep{sion1984,fleming1986,greenstein1986,fontaine1987,dreizler1996,bergeron1997,napiwotzki1999,bergeron2001,eisenstein2006a,tremblay2008,krzesinski2009,bergeron2011,giammichele2012,reindl2014b,limoges2015}.

In recent years, the number, scope, and quality of such studies have significantly increased thanks to the advent of the Gaia mission \citep{gaia2016}. In 2018, the second data release of Gaia \citep{gaia2018a} provided, for the first time, exquisite astrometry and photometry for more than 250,000 high-confidence white dwarf candidates \citep{gentile-fusillo2019}. This represented an almost tenfold increase in the number of known white dwarfs \citep{kepler2019} and a thousandfold increase in the number of available parallax measurements \citep{bedard2017}. The use of Gaia parallaxes in conjunction with optical photometry, either from Gaia or other large surveys such as the Sloan Digital Sky Survey (SDSS; \citealt{york2000}) and the Panoramic Survey Telescope and Rapid Response System (Pan-STARRS; \citealt{chambers2016}), enabled the characterisation of white dwarf atmospheres on an unprecedented scale \citep{jimenez-esteban2018,bergeron2019,genest-beaulieu2019a,gentile-fusillo2019,ourique2019,tremblay2019b}. Nevertheless, as mentioned above, these photometric analyses are usually insensitive to the surface composition, which is consequently assumed rather than derived. For spectral evolution studies, additional information is needed to unambiguously determine the nature of the dominant atmospheric constituent, a requirement that necessarily limits the size of the usable sample.

The most direct and reliable option is, of course, spectroscopy. \citet{genest-beaulieu2019b}, \citet{ourique2019}, and \citet{bedard2020} made use of SDSS spectroscopy, which remains at the time of writing the largest available source of medium-resolution white dwarf spectra (nearly 40,000; \citealt{kepler2019,kepler2021}). \citet{blouin2019b} and \cite{caron2023} focused on cool white dwarf samples ($T_{\mathrm{eff}} \lesssim 10,000$\,K) built from a mix of SDSS and archival spectroscopic data from various sources. \citet{mccleery2020} and \citet{obrien2024} relied on the nearly complete spectroscopic follow-up of white dwarfs in the local 40 pc volume. At intermediate temperatures ($25,000\,\mathrm{K} \gtrsim T_{\mathrm{eff}} \gtrsim 10,000$\,K), an alternative possibility is the use of ultraviolet or narrow-band blue photometry, which is sensitive to the presence or absence of the Balmer jump and thus allows to discriminate between hydrogen-rich and helium-rich atmospheres. \citet{cunningham2020} and \citet{lopez-sanjuan2022} exploited this method using data from the Galaxy Evolution Explorer (GALEX; \citealt{martin2005}) and the Javalambre Photometric Local Universe Survey (J-PLUS; \citealt{cenarro2019}), respectively. Among all these studies, those that rely on SDSS observations tend to have a larger sample size and/or a broader $T_{\mathrm{eff}}$ coverage, but suffer from complex selection biases. The other works typically trade off lower-number statistics for a better completeness.

More recently, the third data release of Gaia \citep{gaia2021,gaia2023a} increased the number of high-confidence white dwarf candidates to more than 350,000 \citep{gentile-fusillo2021}. Most importantly for the purpose of this review, it also provided low-resolution spectra for about 100,000 of these objects \citep{gaia2023b}, thereby producing the largest sample so far for which a rough estimate of the surface composition can be made. \citet{jimenez-esteban2023}, \citet{torres2023}, and \citet{vincent2024} leveraged this new data set to measure the helium-atmosphere fraction across a broad $T_{\mathrm{eff}}$ range with improved precision and/or completeness (depending on the adopted sample). However, the accuracy of these results may still suffer from systematic classification errors due to the very low resolution of the Gaia spectra \citep{garcia-zamora2023,vincent2024}.

\begin{figure*}[t]
\centering
\includegraphics[width=2.0\columnwidth,clip=true,trim=3.00cm 14.00cm 2.75cm 4.25cm]{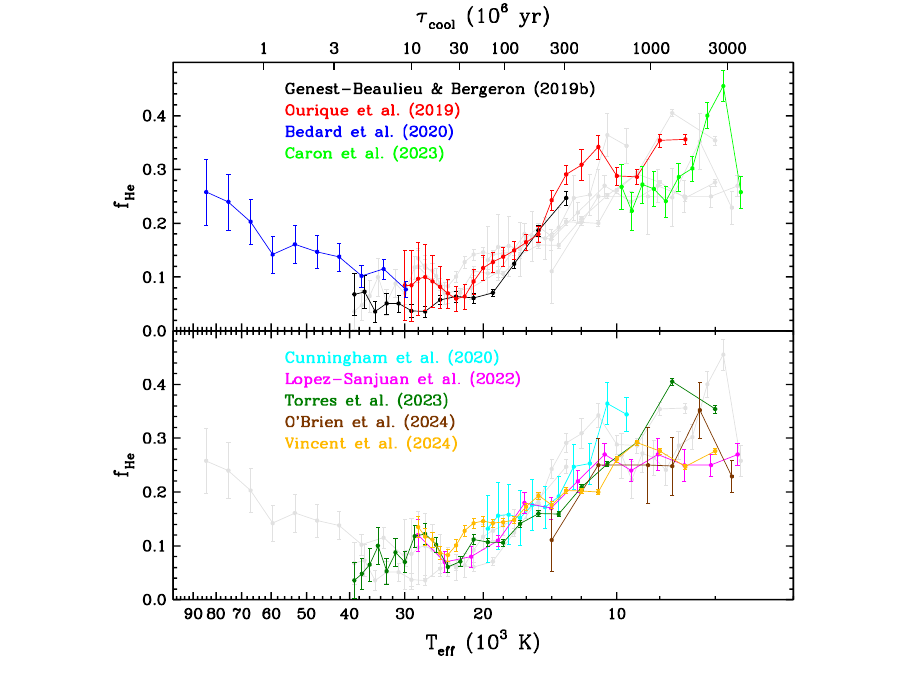}
\caption{Fraction of helium-atmosphere white dwarfs as a function of effective temperature in various post-Gaia studies. For clarity, the studies are arbitrarily divided into those that rely on SDSS spectroscopy (top panel) and those that do not (bottom panel). The curves emphasised in one panel are displayed in light grey in the other panel to allow comparison. Values at $T_{\mathrm{eff}} > 90,000$\,K and $T_{\mathrm{eff}} < 5000$\,K have been purposefully excluded because they are considered unreliable. The results of \citet{blouin2019b}, \citet{mccleery2020}, and \citet{jimenez-esteban2023} are not shown as they have been superseded more recently by those of \citet{caron2023}, \citet{obrien2024}, and \citet{torres2023}, respectively. For reference, the top axis gives the cooling age of a standard 0.60\,$M_{\odot}$ hydrogen-rich white dwarf according to the theoretical evolutionary calculations of \citet{bedard2020}.}
\label{fig:fraction}
\end{figure*}

Figure~\ref{fig:fraction} shows the fraction of helium-dominated white dwarfs as a function of effective temperature as determined in these works. Let us first focus on the global behaviour of the function at $T_{\mathrm{eff}} \gtrsim 10,000$\,K, where the agreement between different papers is most satisfactory. At the beginning of the cooling sequence, 20--30\% of white dwarfs have a helium-rich atmosphere\footnote{At $T_{\mathrm{eff}} \gtrsim 100,000$\,K, the helium-rich fraction is actually much higher (80--90\%; \citealt{bedard2020}), but this is likely not representative of the true incidence of hydrogen-rich and helium-rich atmospheres. This is because helium-dominated white dwarfs cool much more slowly than their hydrogen-dominated counterparts in this early phase \citep{werner2019}, and thus the effective temperature is not a good proxy for the cooling age.}. This proportion then gradually decreases with decreasing temperature and reaches a minimum value of 5--15\%, which subsequently remains roughly constant within the range $40,000\,\mathrm{K} \gtrsim T_{\mathrm{eff}} \gtrsim 20,000$\,K. This part of the cooling sequence is historically known as the DB gap, in reference to the scarcity of helium-atmosphere DB white dwarfs relative to their hydrogen-atmosphere DA counterparts \citep{fontaine1987}. At lower temperatures, the helium-rich fraction gradually re-increases and reaches 20--35\% at $T_{\mathrm{eff}} \simeq 10,000$\,K, approximately the same value as for very hot white dwarfs.

This V-shaped function tells a compelling story: the surface composition of some objects must evolve from helium to hydrogen and then back to helium as they cool. More specifically, Figure~\ref{fig:fraction} suggests that there exist three distinct evolutionary channels among the white dwarf population at $T_{\mathrm{eff}} \gtrsim 10,000$\,K:
\begin{enumerate}[label=(\arabic*)]
\item stars that have and retain a hydrogen atmosphere (DA type);
\item stars that have and retain a helium atmosphere (DO then DB type);
\item stars that initially have a helium atmosphere (DO type) but then experience a helium-to-hydrogen (DO-to-DA) transition at high temperature and a hydrogen-to-helium (DA-to-DB) transition at low temperature.
\end{enumerate}
From the extrema of the helium-dominated fraction quoted above, we can deduce that these three channels account for about 75, 10, and 15\% of the white dwarf population, respectively. These approximate numbers arise from the fact that the fraction of hydrogen-rich objects is always at least about 75\% while the fraction of helium-rich objects is always at least about 10\%.

Although this interpretation successfully explains the overall variation of the helium-rich fraction, some details remain unclear. The slope of the decrease at high temperatures is uncertain, as there is only one modern study in this domain and it has large error bars. The same can be said of the increase at lower temperatures but for a different reason: here multiple results are available but they exhibit significant scatter. In particular, note that the curves of \citet{torres2023} and \citet{vincent2024} based on Gaia spectrophotometry achieve a very high precision but disagree with each other at the $\simeq 2\sigma$ level over most of this range. Also of interest is the possible presence of substructures in Figure~\ref{fig:fraction}, most notably a small bump in the helium-rich fraction within the DB gap. All in all, more work is needed to clarify the fine points of the spectral evolution function.

At $T_{\mathrm{eff}} \lesssim 10,000$\,K, the current situation is much less satisfactory, with different papers reporting wildly inconsistent results. While some find that the helium-atmosphere fraction remains roughly constant at 20--30\%, others instead find that it sharply increases and then decreases with cooling. Among the latter, the location and amplitude of the spike varies from one study to another; the most dramatic variation is seen in \citet{caron2023}, where the helium-rich fraction reaches almost 50\% at $T_{\mathrm{eff}} \simeq 6000$\,K. If real, such a feature would imply that some white dwarfs that had previously maintained a hydrogen atmosphere (channel 1 above) develop a helium atmosphere at very low temperature and thereby undergo a DA-to-DC transition. Better constraints on the incidence of this spectral transformation would be highly desirable. However, we point out that the volume-complete sample of \citet{obrien2024}, which is the least affected by selection biases, shows little evidence of spectral evolution at $T_{\mathrm{eff}} \lesssim 10,000$\,K.

Finally, note that we chose to limit the results displayed in Figure~\ref{fig:fraction} to $T_{\mathrm{eff}} > 5000$\,K. Although some attempts to probe spectral evolution at lower temperatures have been made \citep{blouin2019b,elms2022,caron2023,vincent2024}, these results should be taken with caution. In this regime, distinguishing hydrogen-rich and helium-rich atmospheres becomes very challenging, as both hydrogen and helium lines disappear and one must instead rely on subtle differences in the spectral energy distribution. Furthermore, severe problems are known to affect current atmospheric models of very cool white dwarfs and thus to jeopardise the accuracy of the inferred parameters, including the effective temperature \citep{hollands2018b,bergeron2022,caron2023,obrien2024}. For these reasons, we refrain from speculating on spectral evolution at $T_{\mathrm{eff}} < 5000$\,K.

\subsection{Trace elements: hydrogen and helium} 
\label{sec:observations_trace1}

Although white dwarf atmospheres appear highly pure compared to other types of stars, they are not perfectly pure: elements other than the main constituent are often present in small amounts. The nature and abundances of these contaminants vary significantly along the cooling sequence, thereby adding a layer of complexity to the problem of spectral evolution. These composition variations, although more subtle than those discussed above, still provide a wealth of information on the transport processes at work in the stellar envelope. The study of trace elements at the surface of white dwarfs requires detailed spectroscopic analyses, which have so far been feasible only for the SDSS sample and a few smaller targeted surveys. We first focus on the objects exhibiting either a hydrogen-dominated, helium-polluted atmosphere or a helium-dominated, hydrogen-polluted atmosphere, which we henceforth refer to as hybrid white dwarfs. Observationally, these compositions can translate into various spectral types, namely, DAO, DOA, DAB, DBA, DA, and DC, depending on the effective temperature and relative elemental abundances.

\begin{figure*}[t]
\centering
\includegraphics[width=2.0\columnwidth,clip=true,trim=3.00cm 17.75cm 2.75cm 4.25cm]{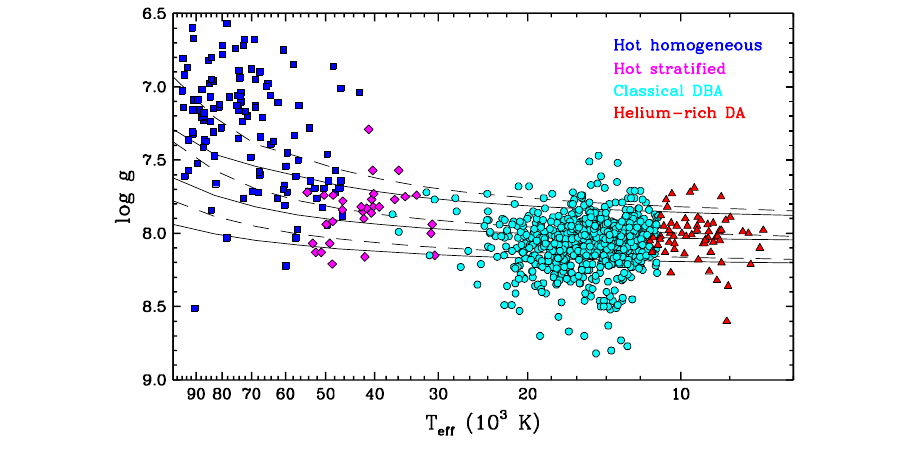}
\caption{Surface gravity as a function of effective temperature for the four main groups of hybrid-atmosphere white dwarfs: the hot homogeneous stars (blue squares, taken from \citealt{gianninas2010} and \citealt{bedard2020}), the hot stratified stars (magenta diamonds, taken from \citealt{bedard2020}), the classical DBA stars (cyan circles, taken from \citealt{rolland2018} and \citealt{genest-beaulieu2019b}), and the helium-rich DA stars (red triangles, taken from \citealt{rolland2018} and \citealt{coutu2019}). In all cases, objects suspected to be unresolved double white dwarf systems have been excluded. For the stars from \citet{genest-beaulieu2019b} and \citet{coutu2019}, the photometric parameters are used and only the objects with a parallax error smaller than 20\% are displayed. A small number of objects interpreted as DBA white dwarfs with $T_{\mathrm{eff}} > 30,000$\,K in \citet{genest-beaulieu2019b} do not appear here as they have been reanalysed in \citet{bedard2020} and found to have either pure-helium or stratified atmospheres. The cool DBA stars from \citet{rolland2018} that suffer from the spectroscopic high-$\log g$ problem have been excluded. The helium-rich DA stars from \citet{rolland2018} were assumed to have $\log g = 8.0$ in that paper, hence the $\log g$ values shown here are taken from other sources \citep{dufour2017,gentile-fusillo2021,caron2023}. Also displayed for visual guidance are theoretical evolutionary sequences representative of hydrogen-rich (dashed curves) and helium-rich (solid curves) white dwarfs with stellar masses of 0.50, 0.60, and 0.70\,$M_{\odot}$ (from top to bottom), taken from \citet{bedard2020}.}
\label{fig:hybrid1}
\end{figure*}

In terms of physical properties, hybrid white dwarfs can be divided into four main groups occupying different parts of the cooling sequence, with very little overlap between them. These are:
\begin{itemize}
\item the hot homogeneous white dwarfs with $T_{\mathrm{eff}} \gtrsim 55,000$\,K;
\item the hot stratified white dwarfs with $55,000\,\mathrm{K} \gtrsim T_{\mathrm{eff}} \gtrsim 30,000$\,K;
\item the classical DBA white dwarfs with $30,000\,\mathrm{K} \gtrsim T_{\mathrm{eff}} \gtrsim 11,000$\,K;
\item the helium-rich DA white dwarfs with $T_{\mathrm{eff}} \lesssim 11,000$\,K.
\end{itemize}
Figure~\ref{fig:hybrid1} shows the location of these objects in the temperature--gravity plane. 

In order to discuss the two groups at $T_{\mathrm{eff}} \gtrsim 30,000$\,K, we must first make a brief digression. Although hybrid white dwarfs obviously contain both hydrogen and helium, the distribution of these elements with depth in the atmosphere is not known a priori. On one hand, the simplest possible configuration is a chemically homogeneous atmosphere, where hydrogen and helium are uniformly mixed. In this case, the surface composition can be parameterised by a conventional elemental abundance, typically the hydrogen-to-helium number ratio, $N_{\mathrm{H}}/N_{\mathrm{He}}$, when hydrogen is trace, and the inverse ratio when helium is trace. On the other hand, given the expected efficiency of gravitational settling in white dwarfs, another plausible configuration is a chemically stratified atmosphere, where a very thin (and thus partially transparent) hydrogen layer floats on top of a helium layer. Because the composition changes with depth, a more natural quantity to describe such an atmosphere is the mass of the hydrogen layer, or rather the ratio of this mass to the total stellar mass, $M_{\mathrm{H}}/M$. As it turns out, the shape of the hydrogen and helium lines is sensitive to the chemical stratification, so a detailed spectroscopic analysis can reveal whether a given star has a homogeneous or stratified atmosphere \citep{jordan1986,vennes1992b,bergeron1994,barstow1998a,manseau2016,bedard2020}. 

Among hot white dwarfs with $T_{\mathrm{eff}} \gtrsim 30,000$\,K, both the homogeneous and stratified configurations are observed, but in nearly separate regions of the temperature--gravity plane, as shown in Figure~\ref{fig:hybrid1}. At $T_{\mathrm{eff}} \gtrsim 55,000$\,K, all known hybrid white dwarfs possess a chemically homogeneous atmosphere. Almost all of them belong to the DAO spectral class, meaning that their surface is dominated by hydrogen. These homogeneous DAO stars are relatively common: overall, they represent about 25--30\% of the hot hydrogen-rich white dwarf population \citep{gianninas2010,bedard2020}. This fraction is however strongly temperature-dependent, going from about 50\% at $T_{\mathrm{eff}} \simeq 80,000$\,K to less than 10\% at $T_{\mathrm{eff}} \simeq 55,000$\,K \citep{bedard2020}. The measured helium abundances span the range $-3.5 \lesssim \log N_{\mathrm{He}}/N_{\mathrm{H}} \lesssim -0.5$, where the lower bound corresponds to the optical detection threshold \citep{napiwotzki1999,good2004,gianninas2010,tremblay2011,bedard2020,reindl2023}. This parameter is also strongly correlated with the surface temperature and luminosity, with hotter and brighter objects showing higher levels of helium pollution \citep{napiwotzki1999,gianninas2010}. Finally, hot DAO white dwarfs tend to have lower-than-average stellar masses, as is apparent in Figure~\ref{fig:hybrid1}, although this may be partially due to systematic errors in spectroscopic analyses \citep{gianninas2010,bedard2020,reindl2023}. All in all, these properties indicate that the physical mechanism maintaining helium homogeneously in the outer layers is particularly efficient in hot, low-mass white dwarfs and gradually weakens with cooling.

While hot DAO stars are fairly common, very few of their DOA counterparts are observed, and one may wonder whether helium-dominated, hydrogen-polluted atmospheres are intrinsically rare at the beginning of the cooling sequence. This is likely not the case: the DAO/DOA dichotomy is actually the result of a spectroscopic detection bias. Because ionised helium is a hydrogenic ion, all hydrogen lines are blended with ionised helium lines, which makes it very challenging to detect hydrogen in a hot helium-dominated atmosphere, with a typical visibility limit of $\log N_{\mathrm{H}}/N_{\mathrm{He}} \gtrsim -1.0$ \citep{werner1996b,bedard2020}. Therefore, the hydrogen content of the hot helium-rich white dwarf population is largely unconstrained by spectroscopic observations. Nevertheless, we will see later that this hydrogen content must be nonzero given our current interpretation of spectral evolution.

At $55,000\,\mathrm{K} \gtrsim T_{\mathrm{eff}} \gtrsim 30,000$\,K, most hybrid white dwarfs possess a chemically stratified atmosphere, indicative of efficient gravitational settling. Estimates of the fractional mass of the surface hydrogen layer fall within the range $-18 \lesssim \log M_{\mathrm{H}}/M \lesssim -15$ \citep{manseau2016,bedard2020}. This is simply the range over which both hydrogen and helium lines are detectable; objects with thinner and thicker hydrogen layers appear as pure-helium and pure-hydrogen atmosphere white dwarfs, respectively. As a result of their broad $T_{\mathrm{eff}}$ and $M_{\mathrm{H}}/M$ ranges, stratified white dwarfs exhibit a diverse collection of spectral types: DAO, DOA, DAB, and DBA. Unlike the hotter homogeneous DAO stars, they have average stellar masses, which suggests that these two groups do not share an evolutionary link, as white dwarfs cool at constant mass \citep{manseau2016,bedard2020}.

At $T_{\mathrm{eff}} \lesssim 30,000$\,K, the hydrogen-rich and helium-rich white dwarf populations are once again characterised by a sharp dichotomy, but a real one this time. While traces of helium in hydrogen-dominated atmospheres are quite rare \citep{gianninas2011,tremblay2011,kepler2019}, traces of hydrogen in helium-dominated atmospheres are the rule rather than the exception. In the overwhelming majority of cases, the outer layers are found to be chemically homogeneous. These objects can be further divided into two spectral groups: the classical DBA white dwarfs with $30,000\,\mathrm{K} \gtrsim T_{\mathrm{eff}} \gtrsim 11,000$\,K, and the helium-rich DA white dwarfs with $T_{\mathrm{eff}} \lesssim 11,000$\,K.

\begin{figure*}[t]
\centering
\includegraphics[width=2.0\columnwidth,clip=true,trim=3.00cm 17.75cm 2.75cm 4.25cm]{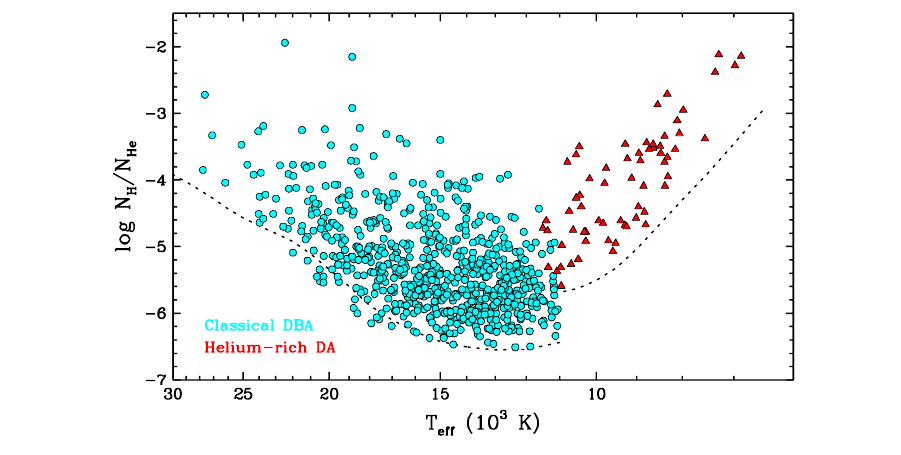}
\caption{Atmospheric hydrogen abundance (by number relative to helium) as a function of effective temperature for the classical DBA stars and helium-rich DA stars shown in Figure~\ref{fig:hybrid1}. The dotted lines correspond to the optical spectroscopic detection limit for signal-to-noise ratios of $\simeq 100$ ($T_{\mathrm{eff}} > 11,000$\,K) and $\simeq 20$ ($T_{\mathrm{eff}} < 11,000$\,K), which are representative of the best-observed objects in the DBA and helium-rich DA samples, respectively.}
\label{fig:hybrid2}
\end{figure*}

The classical DBA stars (so called to distinguish them from the few hotter stratified DBA stars mentioned above) are by far the most numerous and best studied type of hybrid white dwarfs. It is estimated that they represent as much as 60--75\% of the helium-dominated population in this part of the cooling sequence \citep{koester2015b,rolland2018,genest-beaulieu2019b}. Spectroscopic analyses reveal hydrogen abundances in the range $-6.5 \lesssim \log N_{\mathrm{H}}/N_{\mathrm{He}} \lesssim -3.0$, although the lower bound depends on the visibility of the hydrogen lines, which changes with effective temperature \citep{voss2007,bergeron2011,koester2015b,rolland2018,genest-beaulieu2019b,cukanovaite2021}. Figure~\ref{fig:hybrid2} shows the hydrogen abundance as a function of effective temperature for a sample of DBA stars. Most objects are found at $T_{\mathrm{eff}} \lesssim 20,000$\,K and $\log N_{\mathrm{H}}/N_{\mathrm{He}} \lesssim -4.0$. Aside from the hydrogen contamination, the other properties of DBA white dwarfs, such as their mass distribution, are similar to those of their pure-helium DB counterparts \citep{voss2007,bergeron2011,genest-beaulieu2019b}.

At $T_{\mathrm{eff}} \lesssim 11,000$\,K, helium lines disappear and thus hybrid white dwarfs display a DA-type spectrum, even if hydrogen is not the dominant species. However, the presence of helium can still be inferred from its effect on the H$\alpha$ line, which appears unusually broad and shallow as a consequence of van der Waals interactions \citep{bergeron2001,rolland2018,kilic2020,caron2023}. Therefore, these so-called helium-rich DA stars can be easily distinguished from standard pure-hydrogen DA stars, and their surface composition can be measured from the shape of the hydrogen lines. The difference between these objects and their hotter DBA siblings is not purely one of spectral appearance: they also tend to have higher hydrogen abundances, as shown in Figure~\ref{fig:hybrid2} \citep{rolland2018,coutu2019,kilic2020}. This is partially a visibility effect: the hydrogen lines become increasingly difficult to detect with decreasing effective temperature, so cool white dwarfs with a low hydrogen content appear as featureless DC stars. Nevertheless, there is a clear temperature gap between the most hydrogen-polluted members of the two groups, indicating that these objects do not undergo a constant-composition evolution from DBA to helium-rich DA.

Although nearly all hybrid white dwarfs can be assigned to one of the four categories outlined above, a few peculiar objects do not fit any of these typical descriptions. Three well-known examples are PG 1210+533 \citep{bergeron1994,gianninas2010}, HS 0209+0832 \citep{heber1997,wolff2000}, and GD 323 \citep{koester1994,pereira2005}. These stars share a few common characteristics: they are relatively hot ($T_{\mathrm{eff}} \simeq 46,000$, 36,000, and 29,000\,K, respectively) and their atmosphere appears to be neither perfectly homogeneous nor perfectly stratified. Even more interestingly, the strength of their hydrogen and helium lines is observed to vary on time scales of hours (GD 323) to years (PG 1210+533 and HS 0209+0832). More recently, similar but more extreme variations were discovered in two new peculiar hybrid white dwarfs, one of which successively shows a pure DA and pure DB spectrum over its rotation period of 15 minutes \citep{caiazzo2023,moss2024}. Such horizontal surface inhomogeneities may arise from an atmospheric transformation taking place in the presence of a weak asymmetric magnetic field, and thus these stars may be real-time manifestations of spectral evolution. Besides, we note that a few tens of objects exhibiting a DAB/DBA-type spectrum were actually shown to be unresolved binary systems containing a DA and a DB white dwarf \citep{wesemael1994,bergeron2002,limoges2009,limoges2010,bergeron2011,tremblay2011,genest-beaulieu2019b}.

\subsection{Trace elements: carbon and other metals} 
\label{sec:observations_trace2}

Elements heavier than hydrogen and helium are also observed to pollute the surface of many white dwarfs. Among these, carbon has a special status, as it is often detected independently of other metals. We will consequently address carbon-polluted and metal-polluted white dwarfs in turn.

\begin{figure*}[t]
\centering
\includegraphics[width=2.0\columnwidth,clip=true,trim=3.00cm 17.75cm 2.75cm 4.25cm]{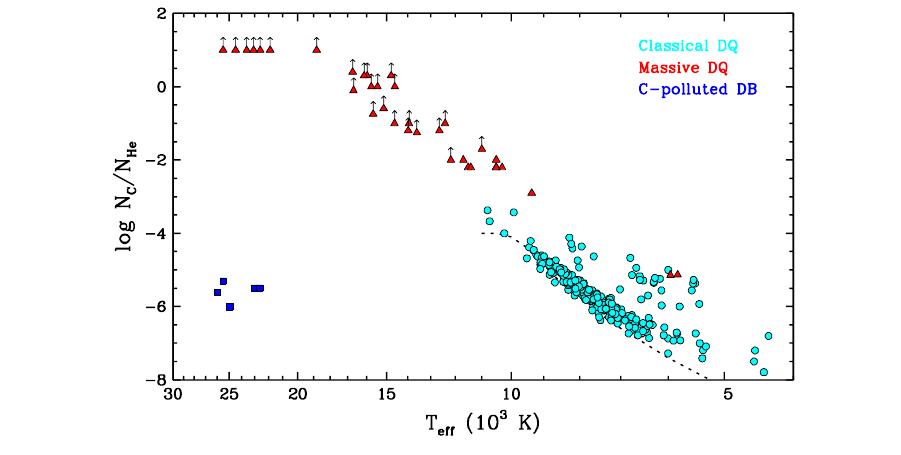}
\caption{Atmospheric carbon abundance (by number relative to helium) as a function of effective temperature for the three main groups of carbon-bearing white dwarfs: the classical DQ stars (cyan circles, taken from \citealt{coutu2019} and \citealt{blouin2019c}), the massive DQ stars (red triangles, taken from \citealt{koester2019}, \citealt{koester2020}, and \citealt{blouin2019c}), and the carbon-polluted DB stars (blue squares, taken from \citealt{petitclerc2005} and \citealt{koester2014b}). An upward-pointing arrow indicates that the displayed carbon abundance is a lower limit rather than a true determination. The classical and massive DQ white dwarfs are arbitrarily defined as having stellar masses lower and higher than 0.80\,$M_{\odot}$, respectively. In all cases, objects suspected to be unresolved double white dwarf systems have been excluded. Among the stars in \citet{coutu2019} and \citet{blouin2019c}, only those with a parallax error smaller than 20\% are displayed. Among the stars in \citet{petitclerc2005} and \citet{koester2014b}, those that also exhibit traces of other metals have been excluded. The dotted line corresponds to the optical spectroscopic detection limit for a signal-to-noise ratio of $\simeq 20$ ($T_{\mathrm{eff}} < 11,000$\,K), which is the average value for the cool DQ sample.}
\label{fig:carbon}
\end{figure*}

Most carbon-bearing white dwarfs have a cool ($T_{\mathrm{eff}} \lesssim 10,000$\,K), helium-dominated atmosphere. Although carbon is merely a trace constituent, the invisibility of helium at low temperatures means that their optical spectrum only exhibits atomic lines and/or molecular bands of carbon, defining the DQ spectral class. These objects are often referred to as the cool DQ or classical DQ stars, and they represent about 20\% of the helium-rich population at $10,000\,\mathrm{K} \gtrsim T_{\mathrm{eff}} \gtrsim 5000$\,K \citep{bergeron2019,mccleery2020,caron2023,obrien2024}. A remarkable feature of classical DQ white dwarfs is that their carbon contamination is far from random: the surface composition is tightly correlated with the effective temperature, such that cooler objects contain less carbon \citep{dufour2005,koester2006b,blouin2019c,coutu2019,koester2019,caron2023}. This is illustrated in Figure~\ref{fig:carbon}, where we reproduce the temperature--abundance diagram of a large DQ sample. The typical carbon abundance sharply decreases from $\log N_{\mathrm{C}}/N_{\mathrm{He}} \simeq -4.0$ at $T_{\mathrm{eff}} \simeq 10,000$\,K to $\log N_{\mathrm{C}}/N_{\mathrm{He}} \simeq -7.5$ at $T_{\mathrm{eff}} \simeq 5000$\,K. The narrowness of the observed sequence is actually a visibility effect: as shown in Figure~\ref{fig:carbon}, the bottom of the sequence perfectly coincides with the optical detection threshold of carbon. This suggests that cool helium-rich atmospheres with lower amounts of carbon do exist but give rise to featureless DC-type optical spectra; and indeed, some DC stars do show carbon lines in the ultraviolet, where the detection limit is lower \citep{weidemann1995,dufour2011}. Still, the correlation seen in Figure~\ref{fig:carbon} necessarily implies that the physical mechanism responsible for carbon pollution becomes less effective with cooling. Another interesting property of cool DQ white dwarfs is that they tend to have slightly lower-than-average stellar masses, with a distribution peaking at $\simeq 0.55\,M_{\odot}$ rather than the canonical $\simeq 0.60\,M_{\odot}$ value \citep{blouin2019c,coutu2019,koester2019,caron2023}.

There also exists a second, more exotic kind of DQ-type white dwarfs, which differ from their classical counterparts in several ways. For reasons that will become obvious, these objects are often called the massive DQ, warm DQ, or hot DQ stars\footnote{Strictly speaking, ``warm DQ" and ``hot DQ" usually denote objects cooler and hotter than $T_{\mathrm{eff}} \simeq 18,000$\,K, respectively, but the distinction is mostly historical.}. From a purely atmospheric point of view, they are both hotter ($25,000\,\mathrm{K} \gtrsim T_{\mathrm{eff}} \gtrsim 10,000$\,K) and more carbon-rich ($\log N_{\mathrm{C}}/N_{\mathrm{He}} \gtrsim -3.0$). This can be seen in Figure~\ref{fig:carbon}, where they form a second sequence, clearly separate from that of the classical DQ stars \citep{dufour2007b,dufour2008,coutu2019,koester2019}. In fact, the surface of most objects either is or could be dominated by carbon rather than helium (and interestingly, traces of hydrogen are often present as well; \citealt{koester2020,kilic2024}). In addition, they are much more massive ($0.90\,M_{\odot} \lesssim M \lesssim 1.20\,M_{\odot}$; \citealt{coutu2019,koester2019}) and tend to have unusually high space velocities \citep{cheng2019,coutu2019,kawka2023}, rotation rates \citep{williams2016,macfarlane2017}, and magnetic fields \citep{dufour2013,williams2013}. Taken together, these characteristics suggest that massive DQ white dwarfs are the products of white dwarf mergers rather than standard single-star evolution \citep{dunlap2015,kawka2023}.

Finally, a third category of carbon-polluted white dwarfs, although more marginal, deserves to be mentioned. A few relatively hot DB stars showing a seemingly pure-helium atmosphere in the optical reveal traces of carbon (and nothing else) in the ultraviolet. These objects cluster around $T_{\mathrm{eff}} \simeq 25,000$\,K and $\log N_{\mathrm{C}}/N_{\mathrm{He}} \simeq -5.5$ in Figure~\ref{fig:carbon} \citep{provencal2000,petitclerc2005,dufour2010a,koester2014b}. Only a handful of them are currently known due to the limited availability of ultraviolet spectra. Unlike the hot DQ white dwarfs, they appear to have typical stellar properties, including masses near $\simeq 0.60\,M_{\odot}$ \citep{koester2014b}. As an aside, note that all three classes of carbon-polluted white dwarfs belong to the wider family of hydrogen-deficient white dwarfs. In contrast, carbon contamination of hydrogen-dominated atmospheres is an extremely rare phenomenon \citep{liebert1983,hollands2020,kilic2024}.

Let us now move on to the case of other heavy elements, which is completely different from that of carbon. Metal pollution affects both hydrogen-rich and helium-rich white dwarfs over the entire cooling sequence, giving rise to an array of spectral types including DAZ, DBZ, DOZ, and DZ. Although this phenomenon is frequently detected in the optical through weak lines of one or two elements, it produces a much stronger signature in the ultraviolet, where many additional species can be identified and measured. Consequently, most of what we know about the metal abundance pattern of white dwarfs comes from detailed analyses of ultraviolet spectra of a relatively small number of objects. For reasons that will appear in due course, we will discuss white dwarfs hotter and cooler than $T_{\mathrm{eff}} \simeq 30,000$\,K separately.

At $T_{\mathrm{eff}} \gtrsim 30,000$\,K, the detectability of heavy elements in the optical is particularly poor, hence metal lines are usually seen exclusively in the ultraviolet. Despite this observational challenge, an astonishing number of elements have been identified at the surface of hot white dwarfs. The most common species are carbon, nitrogen, oxygen, silicon, phosphorus, sulfur, iron, and nickel \citep{holberg1993,werner1994,dreizler1996,dreizler1999b,barstow2003a,good2005,vennes2006,preval2013,barstow2014,werner2017,preval2019}, but the full list of detected elements also includes aluminium, argon, chromium, manganese, cobalt, copper, zinc, gallium, germanium, arsenic, selenium, bromine, krypton, strontium, zirconium, molybdenum, indium, tin, antimony, tellurium, iodine, xenon, cesium, and barium \citep{chayer2005,vennes2005,werner2007,werner2012,rauch2013,rauch2014a,rauch2014b,rauch2015b,rauch2015a,rauch2016a,rauch2016b,hoyer2017,rauch2017a,rauch2017b,hoyer2018,werner2018b,werner2018a,lobling2020,rauch2020,chayer2023}. The measured abundances span several orders of magnitude and can vary significantly from one object to another, even when the atmospheric parameters are otherwise similar. Despite this large dispersion, a few global trends stand out. First, the incidence of heavy-element pollution declines along the cooling sequence, from 60--70\% at $T_{\mathrm{eff}} \gtrsim 60,000$\,K to 20--30\% at $T_{\mathrm{eff}} \simeq 30,000$\,K \citep{barstow2014}. The level of contamination seems to follow an analogous but weaker correlation, with cooler objects generally having slightly lower abundances. Furthermore, at a given temperature, helium-dominated white dwarfs tend to be more polluted than their hydrogen-dominated counterparts \citep{dreizler1996,dreizler1999b,barstow2003a,good2005,vennes2006,barstow2014,werner2018a,rauch2020}.

Although metals usually produce spectral features only in the ultraviolet, they can manifest themselves indirectly in the optical through their influence on the hydrogen and helium lines. In particular, several hot DA white dwarfs are afflicted by the so-called Balmer-line problem, wherein the core of the H$\alpha$ and H$\beta$ lines is poorly reproduced by atmosphere models assuming a pure-hydrogen composition \citep{napiwotzki1992,bergeron1994,napiwotzki1994,napiwotzki1999,gianninas2010}. A similar issue affects the ionised helium features of hot DO white dwarfs \citep{dreizler1996,reindl2014b,werner2014a,bedard2020,reindl2023}. The Balmer-line problem was shown to vanish when heavy elements are properly included in the atmosphere models, as they alter the temperature stratification of the outer layers, which in turn impacts the shape of the spectral features \citep{bergeron1993,werner1996a,gianninas2010}. In accordance with ultraviolet studies of metal pollution, the incidence of the Balmer-line problem quickly decreases along the cooling sequence, such that most affected objects have $T_{\mathrm{eff}} \gtrsim 60,000$\,K \citep{gianninas2010,bedard2020}. Interestingly, the phenomenon is more common and more severe in (homogeneous) DAO stars than in DA stars, indicating that the helium and metal contents are correlated and thus due to the same physical mechanism \citep{bergeron1994,napiwotzki1999,good2005,gianninas2010,bedard2020,reindl2023}.

At $T_{\mathrm{eff}} \lesssim 30,000$\,K, the fraction of white dwarfs exhibiting traces of heavy elements remains roughly constant at 20--30\% along the rest of the cooling sequence \citep{zuckerman1998,zuckerman2003,zuckerman2010,koester2014a,manser2024,obrien2024}. Because the visibility of metals increases as the temperature decreases, optical lines become more common in this range, especially those of calcium. It is therefore possible to estimate surface calcium abundances for large samples of DAZ, DBZ, and DZ stars. The measured abundances roughly span $-10.0 \lesssim \log N_{\mathrm{Ca}}/N_{\mathrm{H}} \lesssim -6.0$ for hydrogen-atmosphere white dwarfs and $-12.0 \lesssim \log N_{\mathrm{Ca}}/N_{\mathrm{He}} \lesssim -6.0$ for helium-atmosphere white dwarfs \citep{zuckerman1998,zuckerman2003,koester2005,dufour2007a,zuckerman2010,koester2011,koester2015b,hollands2017,coutu2019,blouin2022a,badenas-agusti2024}. The difference is simply due to the higher transparency of a helium plasma compared to a hydrogen plasma at the same temperature, which enables the detection of smaller amounts of contaminants. Moreover, in both cases, the lower limits are attainable only at $T_{\mathrm{eff}} \lesssim 10,000$\,K, where the transparency of the atmosphere is highest. Worthy of mention is the fact that DBZ and DZ white dwarfs are more common than their DAZ counterparts, a disparity that can be explained by the above-mentioned visibility effect \citep{koester2005,mccleery2020,blouin2022a,caron2023,obrien2024}. Another interesting trend specifically concerns the DB population, among which the presence of calcium is correlated with the presence of hydrogen; in other words, several (classical) DBA stars are in fact DBAZ stars \citep{koester2005,koester2015b,gentile-fusillo2017}. This suggests that the trace hydrogen and metals observed in helium-atmosphere white dwarfs may have a common origin.

As in hotter white dwarfs, blue or ultraviolet spectroscopy allows the detection of many additional species, most notably oxygen, sodium, magnesium, aluminium, silicon, titanium, chromium, manganese, iron, and nickel. This list includes all the major rock-forming elements. In fact, and quite interestingly, a plethora of studies have revealed that the metal abundance pattern of cool white dwarfs typically roughly resembles that of the Earth's interior \citep{zuckerman2007,dufour2010b,klein2010,farihi2011,klein2011,melis2011,zuckerman2011,dufour2012,gansicke2012,jura2012,farihi2013,xu2013,jura2014,koester2014a,xu2014,raddi2015,wilson2015,farihi2016a,farihi2016b,gentile-fusillo2017,xu2017,hollands2018a,blouin2019a,swan2019,xu2019,fortin-archambault2020,hoskin2020,hollands2021,izquierdo2021,kaiser2021,klein2021,elms2022,hollands2022,johnson2022,doyle2023,swan2023,rogers2024,vennes2024}. Besides, note that calcium and magnesium are frequently observed in cool white dwarfs but never in their hotter siblings discussed above, which indicates a fundamental difference in the source of heavy-element pollution.

\subsection{The bifurcation in the Gaia colour--magnitude diagram} 
\label{sec:observations_GaiaHRD}

\begin{figure*}[t]
\centering
\includegraphics[width=2.0\columnwidth,clip=true,trim=3.00cm 17.75cm 2.75cm 4.25cm]{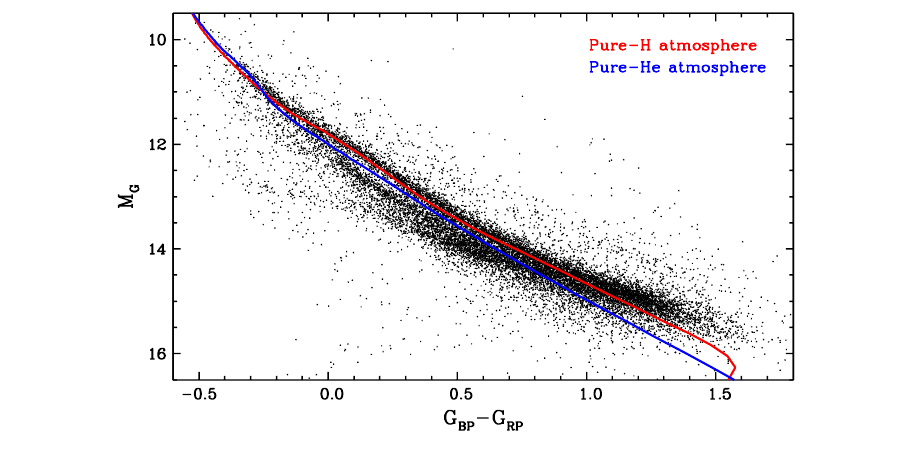}
\caption{Gaia colour--magnitude diagram of white dwarfs located within 100 pc of the Sun (black dots, taken from the Montreal White Dwarf Database; \citealt{dufour2017}). Only the objects with a parallax error smaller than 10\% are displayed. Also shown are theoretical evolutionary sequences for pure-hydrogen atmosphere (red curve) and pure-helium atmosphere (blue curve) white dwarfs with a stellar mass of 0.60\,$M_{\odot}$, calculated using the atmosphere models of \citet{bergeron2011}, \citet{tremblay2011}, and \citet{blouin2019b}.}
\label{fig:bifurcation1}
\end{figure*}

Spectroscopy is essential to unravel the nature and abundances of the trace elements present at the surface of individual white dwarfs. However, as it turns out, photometry can also provide insights on the overall atmospheric contamination of large samples of objects. A now famous example of this is the bifurcation of the white dwarf sequence into two distinct branches, denoted A and B, in the Gaia colour--magnitude diagram, reproduced here in Figure~\ref{fig:bifurcation1} \citep{gaia2018b}. The A and B branches were immediately interpreted as the cooling tracks of hydrogen-dominated and helium-dominated white dwarfs, respectively, offset in colour and magnitude due to the different atmospheric opacities. One one hand, pure-hydrogen atmosphere models indeed predict that a typical 0.60\,$M_{\odot}$ white dwarf should evolve along the A branch. On the other hand, it was found that the analogous pure-helium atmosphere model sequence does not coincide with the B branch and instead falls in between the two branches, where very few stars are observed \citep{jimenez-esteban2018,kilic2018,bergeron2019,gentile-fusillo2019}.

The bifurcation appears at relatively low effective temperatures, $11,000\,\mathrm{K} \gtrsim T_{\mathrm{eff}} \gtrsim 7000$\,K, and the subsample of white dwarfs with SDSS spectroscopy confirms that the lower branch consists mostly of helium-rich objects with spectral types DC, DQ, and DZ \citep{bergeron2019,gentile-fusillo2019,gentile-fusillo2020}. The location of these stars below the theoretical 0.60\,$M_{\odot}$ pure-helium cooling track suggests that they have either higher masses (making them smaller and thus less luminous) or atmospheric contaminants providing an additional source of opacity \citep{jimenez-esteban2018,kilic2018,bergeron2019,gentile-fusillo2019,ourique2019,ourique2020}. The first hypothesis can be rejected on the grounds that the direct precursors of B-branch white dwarfs, the slightly hotter DB(A)-type stars, almost all have typical masses near 0.60\,$M_{\odot}$. On the other hand, the second explanation is supported by the observation that part of the B-branch population is made up of DQ and DZ stars, which obviously do not have a pure-helium surface. And indeed, it is well-known that carbon and other metals not only produce spectral features, but also affect the overall energy distribution and thus the colours of cool helium-rich white dwarfs \citep{dufour2005,dufour2007a,caron2023}. However, this does not completely solve the problem, as DQ and DZ stars collectively account for only 40--50\% of the cool helium-dominated population, while the bulk of the remainder consists of featureless DC white dwarfs seemingly devoid of contaminants \citep{bergeron2019,mccleery2020,caron2023,obrien2024}.

The key realisation was that some elements may be present in amounts too low to produce optical lines but still high enough to alter the energy distribution. This is the case of hydrogen, which becomes very difficult to detect spectroscopically at $T_{\mathrm{eff}} \lesssim 10,000$\,K (as shown in Figure~\ref{fig:hybrid2}) but still influences the continuum opacity of the atmosphere. It was demonstrated that a surface abundance of $\log N_{\mathrm{H}}/N_{\mathrm{He}} \simeq -5.0$ is sufficient to bring the helium-atmosphere model sequence into agreement with the observed B branch \citep{bergeron2019,gentile-fusillo2020,kilic2020,gentile-fusillo2021}. Moreover, carbon was shown to have such an effect as well, with surface abundances $\simeq$ 1--2\,dex lower than measured in classical DQ stars being sufficient in this case \citep{blouin2023a,camisassa2023}. Therefore, the bifurcation hints that the majority of cool DC white dwarfs have undetected traces of hydrogen and/or carbon in their atmosphere\footnote{On the other hand, metals other than carbon cannot be invoked to explain the location of the B branch, because the surface abundances required to alter the Gaia magnitudes would produce clearly visible calcium features \citep{blouin2023a}.}. However, the true amounts of hydrogen and/or carbon cannot be determined using Gaia data alone, as the effects of these elements are degenerate. Fortunately, we will see later that theoretical modelling of the transport of hydrogen and carbon in helium-rich white dwarfs sheds considerable light on this question.

\section{Theoretical interpretation}
\label{sec:theory}

The first part of this review was devoted to the description of the numerous pieces of empirical evidence for spectral evolution. This multifaceted picture is the net observable result of an array of physical mechanisms that modify the distribution of chemical elements in the envelope of cooling white dwarfs. In this second part, we approach the subject from a theoretical angle: given a plausible initial composition, what do current models of element transport predict, and how well do these predictions match the empirical evidence? We ultimately seek a unified theory of spectral evolution, constraining the importance of the relevant physical processes and explaining all the observed trends at once. This field of research relies on detailed numerical simulations where both white dwarf cooling and element transport are considered simultaneously and self-consistently, which can be particularly challenging given the extremely wide range of characteristic time and space scales involved. We will first review the fundamentals of chemical transport in stellar envelopes, and then describe in detail how the chemical structure of white dwarfs is expected to change along the cooling sequence according to state-of-the-art simulations. As the chain of events depends on the initial composition, our discussion will be divided into three sections corresponding to three different types of white dwarf progenitors.

\subsection{Preliminaries: element transport mechanisms} 
\label{sec:theory_prelim}

The transport of chemical elements in stars is a rich and complex topic on which there are many comprehensive reviews, such as those of \citet{michaud2015} and \citet{salaris2017}. Here, we only give a brief overview of the transport mechanisms thought to be important in white dwarf envelopes: atomic diffusion, convection, radiative levitation, radiative winds, and external accretion. These transport mechanisms can be divided into two categories: diffusion and levitation are microscopic processes, acting on individual chemical species differentially, while convection, winds, and accretion are macroscopic processes, impacting the stellar plasma as a whole. 

Generally speaking, \textbf{diffusion} denotes the net movement of the chemical constituents of a fluid arising from interparticle collisions in the presence of a spatial gradient in some physical quantity. In stars, the three main types of diffusive transport are chemical diffusion, gravitational settling, and thermal diffusion, which are induced by composition, pressure, and temperature gradients, respectively. On one hand, gravitational settling and thermal diffusion cause heavier particles to move towards regions of higher pressure and temperature, that is, towards the centre of the star. These mechanisms thus tend to sort the elements vertically according their atomic weight, with lighter ones floating above heavier ones. On the other hand, chemical diffusion tends to oppose this separation, as it is directed against the composition gradient and thereby mixes the different species. In the specific case of white dwarfs, gravitational settling is by far the dominant type of diffusive transport. At the surface, the time scale for this process is of the order of a year, much shorter than the time scale for the cooling of the star \citep{paquette1986b,dupuis1992,koester2006a,koester2009,bauer2019,heinonen2020}. This means that complete element separation is achieved practically instantaneously in the absence of competing mechanisms, as we have already mentioned a few times. However, it is important to note that the diffusion time scale increases with depth and reaches hundreds of millions of years at the bottom of the envelope, such that the composition there remains unaltered for a significant portion of the cooling process \citep{fontaine1979,dehner1995}. Thermal diffusion and chemical diffusion also occur in white dwarfs but have minor effects. Thermal diffusion slightly accelerates element separation in the envelope of hot white dwarfs but becomes negligible as the temperature decreases \citep{paquette1986b,althaus2004}. Chemical diffusion induces mixing at the boundary between regions of different compositions, resulting in narrow but continuous transition zones \citep{arcoragi1980,vennes1988}.

\textbf{Radiative levitation} is the name given to selective chemical transport arising from interactions between matter and radiation. As they propagate throughout the envelope, photons can transfer part of their net outward momentum to highly absorbing atoms, which are thereby supported against gravitational settling. The efficiency of this support is determined by the radiative force, which itself depends on the flux of radiation and the opacity of a given element. Consequently, this transport mechanism mainly affects metals (which are very opaque in the ultraviolet) in luminous stars (which have strong radiation fields). This is also true for white dwarfs: theoretical calculations show that radiative levitation can indeed maintain small amounts of heavy elements at the surface of hot, luminous white dwarfs \citep{chayer1995a,chayer1995b,dreizler1999a,dreizler1999b,schuh2002,rauch2013,chayer2014,koester2014a}. Although the magnitude and behaviour of the contamination are specific to each chemical species, the predicted abundances are typically highest at the beginning of the cooling sequence and decrease with decreasing effective temperature. Furthermore, the amount of levitating elements is expected to be higher in helium-rich white dwarfs than in their hydrogen-rich counterparts, mainly because of the higher metal-line opacity in a helium-dominated plasma. The cooling of the star eventually makes radiative levitation completely inefficient; this cutoff occurs at $T_{\mathrm{eff}} \simeq 20,000$--40,000\,K depending on the element \citep{chayer1995a,chayer1995b}. We note that existing calculations assume a strict equilibrium between gravitational and radiative forces, which results in a unique abundance pattern for a given set of atmospheric parameters.

\begin{figure*}[t]
\centering
\includegraphics[width=2.0\columnwidth,clip=true,trim=3.00cm 15.00cm 2.75cm 4.25cm]{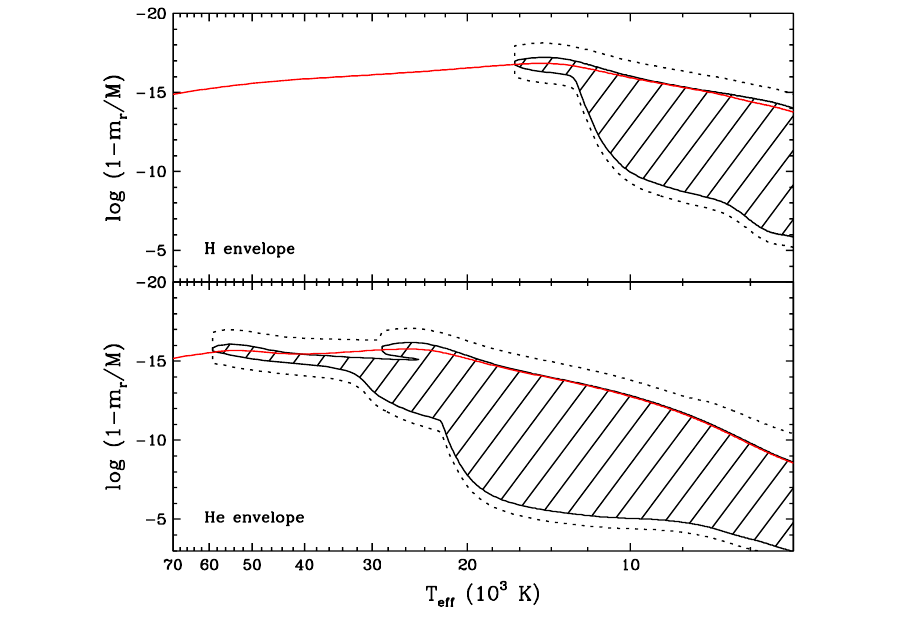}
\caption{Vertical extent of the convection zone (hatched area) as a function of effective temperature in hydrogen-rich (top panel) and helium-rich (bottom panel) white dwarf envelope models. The position in the star is measured in terms of the fraction of the total mass located outside a given radius, such that the surface is towards the top and the core is towards the bottom. These models were computed with the STELUM code \citep{bedard2022a} assuming a stellar mass of 0.60\,$M_{\odot}$ and the ML2 version of the mixing-length theory \citep{tassoul1990}. In each panel, the red line shows the position of the photosphere. As a very rough indication of the expected extent of convective overshoot, the dotted lines denote two pressure scale heights above and below the formally convective region.}
\label{fig:convzone}
\end{figure*}

Let us now turn to macroscopic transport, starting with \textbf{convection}, arguably one of the most important physical mechanisms in stellar astrophysics. In stellar envelopes, convection can be described as a large-scale, cyclic flow of matter whereby hot fluid ascends while cool fluid descends, hence causing very efficient heat transport towards the surface. This process arises in regions where heat transport through radiation is ineffective, for instance as a result of a large overall opacity. The rapid fluid motions also generate element transport, more specifically by thoroughly mixing the various chemical species present within the convective region. In other words, convection completely wipes out the segregation effect of diffusion and produces a perfectly flat composition profile. In a white dwarf, the envelope develops a convection zone as the temperature decreases and the main chemical constituent recombines and thus becomes very opaque. For most applications, convection is modelled using the so-called Schwarzschild criterion and mixing-length theory, a one-dimensional approximation of the inherently three-dimensional process \citep{fontaine1976,tassoul1990,macdonald1991,benvenuto1997,althaus1998,koester2009,rolland2018,bauer2019,bedard2022a}. Figure~\ref{fig:convzone} shows the predicted vertical extent of the convective region in hydrogen-rich (top panel) and helium-rich (bottom panel) white dwarf envelope models as a function of effective temperature. The position in the star is measured in terms of the mass $m_{r}$ located inside radius $r$, such that $1-m_{r}/M$ represents the fraction of the total mass located outside radius $r$. In a hydrogen-rich white dwarf, the convection zone appears at $T_{\mathrm{eff}} \simeq 16,000$\,K but is initially limited to the atmospheric layers. It then starts to expand significantly at $T_{\mathrm{eff}} \simeq 12,000$\,K and ends up spanning a large portion of the envelope. An analogous behaviour is seen in a helium-rich white dwarf but at a higher temperature: the convection zone appears at $T_{\mathrm{eff}} \simeq 60,000$\,K and expands downward at $T_{\mathrm{eff}} \simeq 20,000$\,K.

The mixing-length theory provides a very coarse representation of convection, where the fluid motions abruptly stop at the top and bottom boundaries of the convective region. In reality, the convective flows are expected to overshoot beyond the boundaries, thereby inducing chemical mixing over a larger region. In one-dimensional white dwarf models such as those displayed in Figure~\ref{fig:convzone}, the extent of convective overshoot is essentially treated as a free parameter and thus introduces significant uncertainty \citep{paxton2011,rolland2018,koester2020,bedard2022a}. In the last decade or so, a more accurate description of convection in white dwarfs has become possible thanks to elaborate three-dimensional radiation-hydrodynamics simulations \citep{tremblay2013,cukanovaite2018,kupka2018,cunningham2019}. These are too demanding to be coupled to white dwarf cooling calculations, but can be used to calibrate the mixing-length theory \citep{tremblay2015,cukanovaite2019} and to estimate the extent of convective overshoot \citep{kupka2018,cunningham2019}. The latter works reveal that overshoot can generate chemical mixing up to a distance of a few pressure scale heights beyond the convection zone. For illustrative purposes, the dotted lines in Figure~\ref{fig:convzone} indicates the depths corresponding to two pressure scale heights above and below the formally convective region. However, such hydrodynamics-based constraints have so far been obtained only for relatively hot white dwarfs with shallow convection zones ($\log (1-m_{r}/M) \lesssim -12.0$), and therefore our understanding of convective overshoot remains largely incomplete. Besides, another convection-like process that can generate chemical mixing in formally non-convective regions is the so-called thermohaline or fingering instability. This instability arises in the presence of an inverted composition gradient, where heavier material sits on top of a lighter fluid, and induces a slow homogenisation of the different chemical layers. One-dimensional models predict that thermohaline mixing should occur in the envelope of accreting (see below) hydrogen-rich white dwarfs \citep{deal2013,wachlin2017,bauer2018,bauer2019,wachlin2022,dwomoh2023}, although these results have been disputed \citep{koester2015a}.

\textbf{Radiative winds} constitute another transport mechanism believed to be important in the context of spectral evolution, especially at the beginning of the cooling sequence. These winds are essentially an extreme, macroscopic version of radiative levitation: the outward radiative force due to metal absorption exceeds that of gravity, such that hydrostatic equilibrium becomes impossible. Consequently, the outer layers of the star experience an outward radial motion and are successively peeled off, resulting in continuous mass loss. This outflow imparts the same velocity to all chemical species present in the envelope, thereby preventing element separation through diffusion, similarly to convection. The quantity typically employed to measure the strength of a stellar wind is the rate of mass loss. For radiation-driven winds, the mass-loss rate is determined primarily by the stellar luminosity, and secondarily by the heavy-element content; the wind is stronger in a more luminous and/or more metal-rich star \citep{kudritzki2000}. The question of whether hot white dwarfs can power such winds in spite of their high gravity has received very little theoretical attention, but the few available calculations indicate they indeed can at the very beginning of the cooling sequence \citep{unglaub2000,unglaub2008}. The predicted mass-loss rates, $\simeq 10^{-12}$--$10^{-11}\,M_{\odot}\,\mathrm{yr}^{-1}$, are too low to produce a spectroscopic signature but sufficiently high to considerably slow down gravitational settling in the outer layers. Due to the general lack of wind calculations, models of chemical transport in hot white dwarfs must rely on simplistic analytical expressions for the mass-loss rate, usually considering only the luminosity dependence and calibrated on previous phases of stellar evolution \citep{unglaub1998,unglaub2000,quirion2012,bedard2022a}.

Finally, the external \textbf{accretion} of matter can alter the surface composition of white dwarfs. This process can not only enhance the abundance of elements that are already there, but also, unlike the other transport mechanisms, add new species to the chemical reservoir of the star. Possible sources of accretion material are the interstellar medium \citep{alcock1980,macdonald1991,dupuis1993a} and the immediate circumstellar environment, including a stellar or substellar companion \citep{sion1999,debes2002,veras2021}. In white dwarfs, it is latter source that turns out to be prevalent, as we will see below.

\subsection{Hydrogen-rich progenitor}
\label{sec:theory_hydrogen}

The chemical evolution of a given white dwarf is largely dictated by the initial condition, that is, by the envelope composition of its immediate precursor. This initial reservoir not only determines the nature of the elements that will potentially appear at the surface, but also impacts the efficiency of transport mechanisms, most notably convection, along the cooling sequence. We will begin by considering the most common and straightforward scenario, arising from a standard hydrogen-rich progenitor. According to stellar evolution models, the majority of stars should retain a relatively thick hydrogen layer of $\simeq 10^{-4}\,M$ as they become white dwarfs \citep{iben1984,renedo2010,romero2012}. This necessarily implies that the surface should remain dominated by hydrogen along the entire cooling sequence. Indeed, among the transport mechanisms discussed in Section~\ref{sec:theory_prelim}, only convection is powerful enough to change the bulk atmospheric composition altogether. As shown in Figure~\ref{fig:convzone}, the convection zone present in the hydrogen layer at $T_{\mathrm{eff}} \lesssim 16,000$\,K is always limited to the outer $\simeq 10^{-5}\,M$ and thus never reaches the underlying helium mantle. Therefore, the hydrogen and helium layers never mix, implying that the hydrogen-rich nature of the atmosphere cannot be altered. This canonical scenario naturally explains the first of the three main evolutionary channels identified in Section~\ref{sec:observations_main}, namely, that involving enduring DA stars and representing about 75\% of the white dwarf population \citep{tremblay2008,rolland2018,bedard2020,cunningham2020}.

Of course, transport processes other than convection can still affect the surface chemistry by acting as sources of minor contaminants. At the hot end of the cooling sequence, the presence of helium and heavier elements in most hydrogen-dominated white dwarfs indicates that the primordial solar mixture has not yet been fully purified by gravitational settling. At first glance, two transport mechanisms appear as plausible candidates to explain this observation: radiative levitation and radiative winds. In both scenarios, it is the strong outward radiation field that prevents the sedimentation of trace elements, and the question essentially boils down to whether it does so in or out of hydrostatic equilibrium. Given the high surface gravity of white dwarfs, the occurrence of a full-blown stellar wind historically appeared unlikely, and thus radiative levitation was initially considered as the most sensible hypothesis. However, theoretical calculations demonstrated that radiative levitation alone is glaringly insufficient to account for the amount of helium observed in hot DAO stars \citep{vennes1988,macdonald1991}. Furthermore, the assumption of a strict equilibrium between gravitational and radiative forces leads to the prediction of a unique surface composition for a given set of atmospheric parameters, which is at odds with the large spread in measured heavy-element abundances \citep{chayer1995a,chayer1995b,barstow2003a,barstow2014}.

This leaves a radiative wind as the only viable explanation for the helium contamination of hot DAO white dwarfs. Although the theoretical basis is admittedly uncertain, the wind scenario is fully consistent not only with the inferred helium abundances \citep{unglaub1998,unglaub2000}, but also with several other properties of the hot DAO population. First, these objects are observed to have a chemically homogeneous atmosphere, as expected in the presence of a wind. Second, the strength of a radiative wind is mainly determined by the surface luminosity, with more luminous stars generating stronger winds. This naturally explains why helium contamination becomes both less frequent (as measured by the overall fraction of DAO stars) and less significant (as measured by abundances of individual objects) along the cooling sequence. This may also explain why hot DAO white dwarfs tend to have low masses (corresponding to large radii and thus high luminosities). Third, as the wind is driven by metals, we necessarily expect a correlation between the presence of helium and heavier elements, which is indeed observed through severe forms of the Balmer-line problem in many hot DAO stars \citep{bergeron1994,napiwotzki1999,gianninas2010,bedard2020}.

The exact effective temperature at which radiative winds vanish is unknown and likely varies from one object to another depending on the stellar mass and the metal content. However, we can conservatively assert that such winds cannot exist at $T_{\mathrm{eff}} \lesssim 60,000$\,K \citep{unglaub2000,unglaub2008}. This crude theoretical limit is in line with the disappearance of hot homogeneous DAO stars, which then turn into normal DA stars. Nevertheless, a significant fraction of these cooler hydrogen-rich white dwarfs still show traces of heavy elements, indicating that other transport mechanisms take over in supporting or replenishing the atmospheric metal content. Radiative levitation may play a role, as the observed decrease in pollution incidence along the cooling sequence is roughly consistent with the expected decline in radiative support \citep{chayer1995a,chayer1995b,barstow2003a,barstow2014}. However, this process alone still fails to account for the large object-to-object variations in measured abundances and for all instances of metal pollution below $T_{\mathrm{eff}} \simeq 20,000$\,K (at best). Therefore, the only possible interpretation is that heavy elements are currently being accreted (or have very recently been accreted) at the surface of these white dwarfs. Historically, the source of this external material was believed to be the interstellar medium \citep{dupuis1993a,dupuis1993b,zuckerman2003,koester2005,koester2006a}. It is now well established that white dwarfs actually accrete the remains of their planetary systems: the orbiting planets and asteroids are disrupted by the strong tidal forces and subsequently fall onto the stellar surface \citep{zuckerman2007,farihi2010,barstow2014,jura2014,koester2014a,farihi2016a,schreiber2019,veras2021}. This naturally explains why the metal abundance pattern of many cool white dwarfs is consistent to first order with the bulk composition of the Earth. The accretion scenario also provides the element of randomness that characterises the occurrence and magnitude of metal pollution.

The accretion of planetary debris onto white dwarfs provides a unique way to probe the internal structure of exoplanets. Nevertheless, the abundances measured in the atmosphere of the white dwarf do not strictly reflect the original composition of the accreted body, because the material deposited at the surface is affected by the transport mechanisms at work in the envelope. On one hand, the accreted matter sinks into the star due to gravitational settling, and this process unfolds at different rates for different elements. On the other hand, the sinking material can be slowed down by radiative levitation (in sufficiently hot white dwarfs), by mixing within a convection zone, and possibly by mixing arising from the thermohaline instability. By modelling this interplay between accretion, diffusion, and mixing \citep{dupuis1993a,koester2009,bauer2019,wachlin2022}, it is possible to trace back the mass and bulk composition of the accreted asteroid or planet (see the references given in the last paragraph of Section~\ref{sec:observations_trace2}). In some cases, it may even be possible to constrain the original core, mantle, and crust stratification \citep{jura2014,harrison2018,bonsor2020,harrison2021,buchan2022,swan2023}.

\subsection{Helium-rich progenitor: the effect of residual hydrogen}
\label{sec:theory_helium1}

As we have seen in Section~\ref{sec:observations_main}, about 25\% of all stars enter the cooling sequence with a strongly hydrogen-deficient atmosphere. These objects have apparently lost the greater part of their residual hydrogen layer, implying a non-standard evolutionary history. The main such evolutionary pathway is thought to be the so-called born-again scenario, where a soon-to-be white dwarf experiences a late helium-shell flash and thus rapidly grows back to giant dimensions. This violent event triggers extensive convection in the envelope, such that the hydrogen is deeply engulfed into the star and thereby almost completely burned, while some carbon and oxygen are dredged up to the surface. When nuclear burning ceases and the star contracts again to become a white dwarf (for good this time), the outcome is an envelope made of helium, carbon, and oxygen in similar proportions \citep{iben1983,herwig1999,blocker2001,althaus2005a,lawlor2006,miller-bertolami2006a,miller-bertolami2006b,werner2006}. This scenario is strongly supported by observations, as most hydrogen-deficient pre-white dwarfs indeed exhibit a hot helium--carbon--oxygen atmosphere, corresponding to the PG 1159 spectral class\footnote{A minority of helium-rich pre-white dwarfs show unexpectedly low carbon and oxygen abundances, corresponding to the O(He) spectral class; the origin of these objects is still debated \citep{rauch1998,reindl2014a}.} \citep{werner1991,dreizler1998,hugelmeyer2005,hugelmeyer2006,werner2006,werner2014a,werner2014b,reindl2023}. A few objects known as hybrid PG 1159 stars also show traces of hydrogen, indicating that they did not get rid of their whole hydrogen content \citep{werner2006,werner2014a}. The measured abundances are relatively large ($\log N_{\mathrm{H}}/N_{\mathrm{He}} \simeq 0.0$), but given the difficulty of detecting hydrogen in a hot helium-dominated atmosphere, lower amounts of hydrogen may be present in most born-again stars \citep{althaus2005a,lawlor2006,miller-bertolami2006a,miller-bertolami2017}. In short, the appropriate initial condition to model this spectral evolution channel is a helium-rich envelope containing trace hydrogen as well as significant carbon and oxygen. The initial abundances of these elements are additional parameters that can be varied over plausible ranges, thereby making the predicted behaviours much more complex and diverse than in the case of a standard hydrogen-rich progenitor. This was first demonstrated by \citet{althaus2005a} thanks to full evolutionary calculations linking the born-again episode to the white dwarf phase. In this section, we will first focus on the effect of residual hydrogen.

Before diving into the topic, we note that the considerations of the previous section regarding the origins of heavy elements in hydrogen-dominated white dwarfs also apply to their helium-dominated counterparts (with the obvious exceptions of carbon and oxygen). The primordial metals are initially supported in the atmosphere by the radiative wind, and then by radiative levitation once the wind has faded. These mechanisms are expected to be more efficient in helium-rich white dwarfs, in line with the observed trend for these stars to have higher metal abundances \citep{chayer1995a,unglaub2000}. At lower effective temperature, the detected heavy elements originate from the accretion of tidally disrupted asteroids and planets. A key difference for the interpretation of the observed surface abundances is that the convection zone of helium-dominated objects is much deeper at a given temperature (see the bottom panel of Figure~\ref{fig:convzone}). The accreted material is thus mixed within a larger region, but also remains there longer as the settling time scales of metals below the deep convection zone reach millions of years \citep{paquette1986b,dupuis1993a,koester2009}. The extended convective mixing also greatly diminishes the role of thermohaline mixing compared to hydrogen-rich white dwarfs \citep{deal2013,bauer2019}.

\begin{figure*}[t]
\centering
\includegraphics[width=2.0\columnwidth,clip=true,trim=3.00cm 16.75cm 2.75cm 4.25cm]{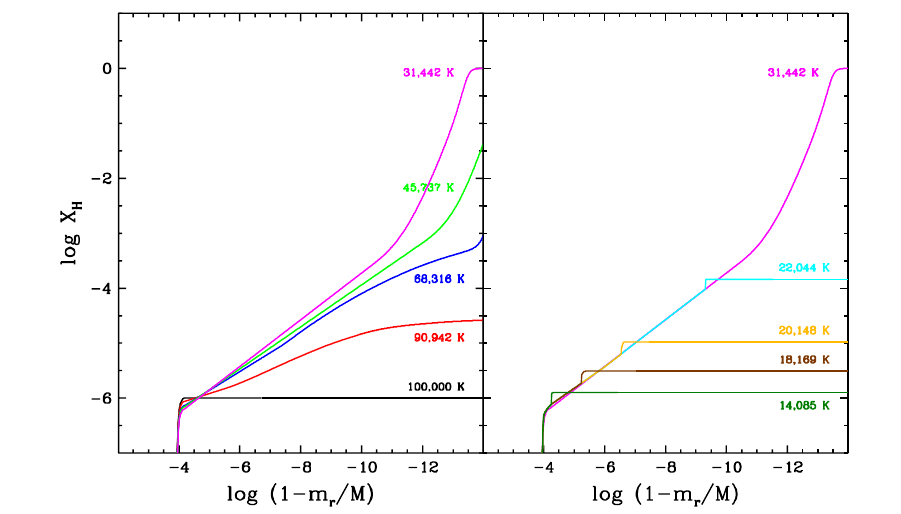}
\caption{Hydrogen mass fraction profile at various effective temperatures in a typical simulation of element transport in helium-rich white dwarfs. The position in the star is measured in terms of the fraction of the total mass located outside a given radius, such that the surface is towards the right and the core is towards the left. This particular simulation is taken from \citet{bedard2023}, and assumes a stellar mass of 0.60\,$M_{\odot}$ and an initial hydrogen mass fraction $X_{\mathrm{H}} = 10^{-6}$. The effective temperature decreases monotonically with time; the left panel illustrates the float-up process at high temperature, while the right panel illustrates the convective dilution process at low temperature.}
\label{fig:dilution1}
\end{figure*}

Let us now come back to the main question of this section: what is the expected chemical evolution of a helium-dominated progenitor harbouring residual hydrogen? To aid the discussion, Figure~\ref{fig:dilution1} displays the results of a theoretical simulation of element transport in such an object; more specifically, the predicted hydrogen abundance profile is shown at various stages of the cooling process \citep{bedard2023}. The position in the star is measured in terms of the quantity $1-m_{r}/M$ introduced earlier in Section~\ref{sec:theory_prelim}. This particular model assumes a relatively low initial hydrogen content: a uniform mass fraction $X_{\mathrm{H}} = 10^{-6}$ in the outer 10$^{-4}\,M$ of the star (corresponding to a total hydrogen mass of 10$^{-10}\,M$). As shown in the left panel of Figure~\ref{fig:dilution1}, the hydrogen diluted within the envelope floats towards the surface due to gravitational settling. However, this so-called float-up process is initially slowed down by the radiative wind, as the latter tends to prevent any chemical differentiation in the outer part of the envelope. As the white dwarf cools and the wind accordingly fades, diffusion becomes increasingly efficient and eventually leads to the formation of a thin pure-hydrogen layer at the surface \citep{unglaub2000,althaus2005a,althaus2005b,althaus2020a,bedard2022a,bedard2023}. Note, however, that this layer comprises only a small fraction of the total hydrogen content of the white dwarf, the rest of which is still located at great depths where diffusion time scales are much longer \citep{rolland2020,bedard2023}. From an observational point of view, the helium-atmosphere DO star has transformed into a hydrogen-atmosphere DA star. In the simulation shown in Figure~\ref{fig:dilution1}, this DO-to-DA transition occurs at a relatively advanced stage, $T_{\mathrm{eff}} \simeq 35,000$\,K, due to the low initial hydrogen abundance assumed. A progenitor with a higher amount of residual hydrogen develops a hydrogen-rich atmosphere earlier. For instance, the initial condition $X_{\mathrm{H}} = 10^{-3}$ gives rise to a DO-to-DA transition at $T_{\mathrm{eff}} \simeq 70,000$\,K. Nevertheless, the speed of the float-up process also depends on the strength of the radiative wind, which is poorly constrained and thus introduces significant uncertainty in the predicted transition temperature \citep{unglaub1998,unglaub2000,bedard2023}.

The newly-formed thin hydrogen layer is short-lived, however, as it is condemned to be wiped out by convection. At this point, two distinct scenarios may take place depending on the exact thickness of the hydrogen shell built by diffusion. On one hand, if the latter extends over less than the outer $\simeq 10^{-14}\,M$ of the star, the underlying helium-rich envelope becomes convective (see the bottom panel of Figure~\ref{fig:convzone}). Overshoot above the convection zone then erodes the hydrogen layer from below and dilutes it within the helium-rich envelope. Observationally, the result is a helium-dominated, hydrogen-contaminated atmosphere, namely, a DBA white dwarf. As the star cools and the convection zone grows deeper, the hydrogen is gradually diluted within an increasingly large region, such that surface hydrogen abundance decreases with time. This process, illustrated in the right panel of Figure~\ref{fig:dilution1}, is generally referred to as convective dilution\footnote{For numerical convenience, the transport simulation shown in Figure~\ref{fig:dilution1} does not include the outer 10$^{-14}\,M$ of the star. As a consequence, the thickness of the surface hydrogen layer at $T_{\mathrm{eff}} \simeq 31,400$\,K is overestimated and the onset of the convective dilution process cannot be modelled per se. This limitation is circumvented by inducing convective dilution artificially, which still provides a reliable prediction of the ensuing chemical evolution (see \citealt{bedard2023} for details).} \citep{macdonald1991,rolland2018,rolland2020,bedard2023}. Depending on the thickness of the hydrogen layer, the DA-to-DBA transition is predicted to occur in the range $30,000\,\mathrm{K} \gtrsim T_{\mathrm{eff}} \gtrsim 14,000$\,K, with larger layers giving rise to lower transition temperatures. That said, the exact relation between these two quantities remains significantly uncertain due to our poor knowledge of convective overshoot \citep{cunningham2020,rolland2020}.

On the other hand, a hydrogen shell more massive than $\simeq 10^{-14}\,M$ inhibits the onset of convection in the helium-dominated envelope and thus remains intact a little longer. Nevertheless, the hydrogen layer itself inevitably becomes convective at lower effective temperature (see the top panel of Figure~\ref{fig:convzone}). When the convection zone reaches the underlying helium-rich region, helium is efficiently carried into the outer hydrogen-rich layer; the convection zone accordingly grows deeper, such that more helium is dredged up to the surface, and so on. This runaway process results in a thorough, almost instantaneous mixing of the hydrogen shell within the helium-rich envelope. This phenomenon is known as convective mixing and takes place in the range $14,000\,\mathrm{K} \gtrsim T_{\mathrm{eff}} \gtrsim 6000$\,K \citep{baglin1973,koester1976,dantona1989,macdonald1991,althaus1998,tremblay2008,chen2011,rolland2018,cunningham2020,bedard2022a,bergeron2022}. Similarly to convective dilution, the observational outcome is a white dwarf with a helium-dominated, hydrogen-contaminated atmosphere. However, given the lower effective temperature and the larger hydrogen content, such an object may appear as a helium-rich DA star rather than a DBA star \citep{chen2011,rolland2018,bedard2022a}.

The theoretical expectations described above can be linked to several observational features detailed in Section~\ref{sec:observations}. First, let us recall that while about 25\% of all white dwarfs initially have a helium-dominated atmosphere, only about 10\% retain this surface composition along the entire cooling sequence. The other 15\% experience a helium-to-hydrogen-to-helium transition, as revealed by the V-shaped variation of the helium-rich fraction in the range $75,000\,\mathrm{K} \gtrsim T_{\mathrm{eff}} \gtrsim 10,000$\,K (Figure~\ref{fig:fraction}). This evolutionary channel is naturally explained by the float-up process at high temperature and the convective dilution and mixing mechanisms at low temperature \citep{fontaine1987,macdonald1991,dreizler1996,bergeron2011,rolland2018,genest-beaulieu2019b,ourique2019,bedard2020,cunningham2020,rolland2020,lopez-sanjuan2022,bedard2023,vincent2024}. An interesting corollary is that about 60\% of helium-rich progenitors contain sufficient residual hydrogen to undergo spectral evolution, while the rest of them do not. Furthermore, the fact that the variation of the helium-rich fraction is gradual implies that different objects experience their atmospheric transformations at different temperatures and thus possess different hydrogen contents. In principle, the range of initial hydrogen abundances can be estimated by comparing the transition temperatures predicted from transport simulations (Figure~\ref{fig:dilution1}) to those ``observed'' in the white dwarf population (Figure~\ref{fig:fraction}). Such an exercise indicates that the observed spectral evolution arises from helium-rich progenitors with mass fractions in the range $10^{-7} \lesssim X_{\mathrm{H}} \lesssim 10^{-3}$. This rather broad range partially reflects the large uncertainties in the current modelling of radiative winds and convective overshoot, and thus the true range is likely narrower \citep{bedard2023}.

Furthermore, the transport of hydrogen in helium-rich white dwarfs readily explains the various groups of hybrid-atmosphere objects (with the exception of the very hot homogeneous DAO stars, which we discussed in the last section). In the course of the float-up process, there is inevitably a brief phase where the newly-formed surface hydrogen layer is thin enough that the underlying helium is still visible. Hot white dwarfs exhibiting such a chemically stratified atmosphere can therefore be interpreted as transitional objects in the midst of a DO-to-DA transformation. And indeed, the existence of stratified white dwarfs and the decrease in the helium-rich fraction coincide in terms of temperature range (Figures \ref{fig:fraction} and \ref{fig:hybrid1}), strengthening the idea that these are two manifestations of the same phenomenon \citep{manseau2016,bedard2020,bedard2022a,bedard2023}.

\begin{figure*}[t]
\centering
\includegraphics[width=2.0\columnwidth,clip=true,trim=3.00cm 17.75cm 2.75cm 4.25cm]{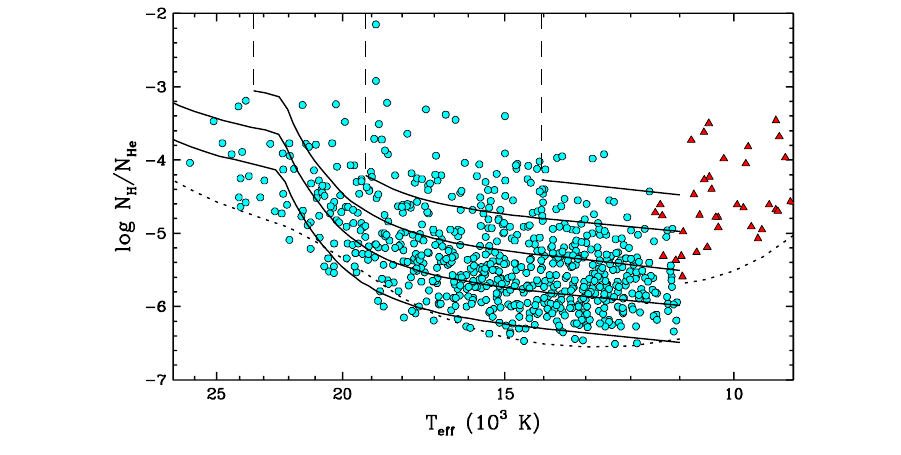}
\caption{Zoomed-in version of Figure~\ref{fig:hybrid2} including theoretical predictions for the convective dilution of hydrogen. These simulations assume a stellar mass of 0.60\,$M_{\odot}$ and initial hydrogen mass fractions $X_{\mathrm{H}} = 10^{-7.0}$, $10^{-6.5}$, $10^{-6.0}$, $10^{-5.5}$, and $10^{-5.0}$ (from bottom to top). For each simulation, the transition temperature (dashed part of the curve) is based on the approximate method of \citet{rolland2020}, while the resulting hydrogen abundance (solid part of the curve) is taken from \citet{bedard2023}.}
\label{fig:dilution2}
\end{figure*}

Further along the cooling sequence, the classical DBA stars can be interpreted as products of the convective dilution and mixing mechanisms, as their measured temperatures and compositions are well reproduced by current transport simulations \citep{rolland2020,bedard2023}. This is demonstrated quantitatively in Figure~\ref{fig:dilution2}, which is a zoomed-in version of the temperature--abundance diagram where theoretical predictions for various initial conditions were added (the middle curve corresponds to the simulation shown in Figure~\ref{fig:dilution1}). We stress once again that our poor knowledge of helium-rich convective overshoot results in significant uncertainties on both the predicted DA-to-DBA transition temperatures ($\pm$1000--5000\,K; \citealt{rolland2020}) and DBA abundances ($\pm$0.5\,dex; \citealt{bedard2023}). For this reason, it would be inadvisable to use the model curves to attempt to infer the total hydrogen content of individual objects. Nevertheless, the overall agreement seen in Figure~\ref{fig:dilution2} strongly supports the view that DBA white dwarfs represent an inevitable stage of spectral evolution. Moreover, we recall that DBA stars account for 60--75\% of the DB population, which is consistent with our previous inference that about 60\% of helium-rich progenitors go through the DO--DA--DBA channel. Although the theoretical predictions displayed in Figure~\ref{fig:dilution2} are limited to $T_{\mathrm{eff}} > 11,000$\,K, the atmospheric composition is expected to remain roughly constant at lower temperatures, because the base of the convection zone then barely moves (see the bottom panel of Figure~\ref{fig:convzone}). Therefore, we can deduce from Figure~\ref{fig:dilution2} that DBA white dwarfs are bound to become featureless DC white dwarfs, although those with relatively high hydrogen abundances will first temporarily appear as helium-rich DA stars \citep{rolland2018,rolland2020,bedard2023}.

For the past three decades or so, the origin of hydrogen in classical DBA white dwarfs has been a topic of considerable debate. Indeed, it has often been proposed that the source of hydrogen is external rather than internal \citep{macdonald1991,voss2007}, and more recently that it can be attributed to the accretion of water-rich planets, asteroids, or even comets \citep{jura2010,klein2010,farihi2011,farihi2013,veras2014,raddi2015,gentile-fusillo2017,xu2017,hoskin2020,izquierdo2021}. This scenario has been preferred by many for two main reasons. First, for a few individual objects, the observed hydrogen and metal abundance pattern indicates ongoing or recent accretion of rocky, water-bearing material \citep{farihi2013,raddi2015,gentile-fusillo2017,hoskin2020,izquierdo2021}. Second, until very recently, simulations of the convective dilution and mixing processes completely failed to reproduce the measured hydrogen abundances of DBA stars, leaving the accretion hypothesis as the only viable alternative \citep{macdonald1991,voss2007,bergeron2011,koester2015b,rolland2018,genest-beaulieu2019b,cunningham2020}. The extension of this scenario to the entire DBA population would lead to the drastic implication that 60--75\% of all white dwarfs must accrete water-rich bodies within a few 100 Myr on the cooling sequence. Nevertheless, the latest works on convective dilution have identified and corrected a significant shortcoming of previous generations of models (see \citealt{bedard2023} for details), thereby bringing the predictions in excellent agreement with the observations (Figure~\ref{fig:dilution2}). Given this improvement, the internal transport of primordial hydrogen now appears as a natural explanation for the existence of the bulk of DBA stars \citep{rolland2020,bedard2023}. However, water accretion must not be totally ruled out: the convective dilution and mixing scenarios predict baseline hydrogen abundances of the right order of magnitude, but it is still plausible that accretion contributes to the observed range. In particular, the model curves of Figure~\ref{fig:dilution2} do not pass through the most hydrogen-rich DBA white dwarfs, which thus require such an external contribution. Finally, we recall that there exists a correlation between hydrogen and metal pollution among the DB population. This suggests that some objects do acquire hydrogen alongside heavier elements through the accretion of planetary debris \citep{gentile-fusillo2017}.

At $T_{\mathrm{eff}} \lesssim 10,000$\,K, there is marginal empirical evidence for the occurrence of convective mixing. The possible further increase in the helium-atmosphere fraction (Figure~\ref{fig:fraction}) and the existence of cool helium-rich DA stars with high hydrogen abundances (Figure~\ref{fig:hybrid2}) suggest that some white dwarfs undergo a late hydrogen-to-helium atmospheric transformation \citep{fontaine1987,bergeron1997,bergeron2001,tremblay2008,chen2012,giammichele2012,limoges2015,rolland2018,blouin2019b,mccleery2020,caron2023,obrien2024}. The location of the helium-rich DA stars in the temperature--abundance diagram is well reproduced by simulations of convective mixing (not shown here) assuming relatively thick surface hydrogen layers between $\simeq 10^{-10}\,M$ and 10$^{-8}\,M$ \citep{chen2011,rolland2018,bedard2022a,bergeron2022}. In the bigger picture of spectral evolution, such objects may descend from the most hydrogen-contaminated of the helium-dominated progenitors, namely, the hybrid PG 1159 stars. As they cool further, helium-rich DA white dwarfs are expected to rapidly turn into DC white dwarfs due to the disappearance of hydrogen lines.

\begin{figure*}[t]
\centering
\includegraphics[width=2.0\columnwidth,clip=true,trim=3.00cm 16.75cm 2.75cm 4.25cm]{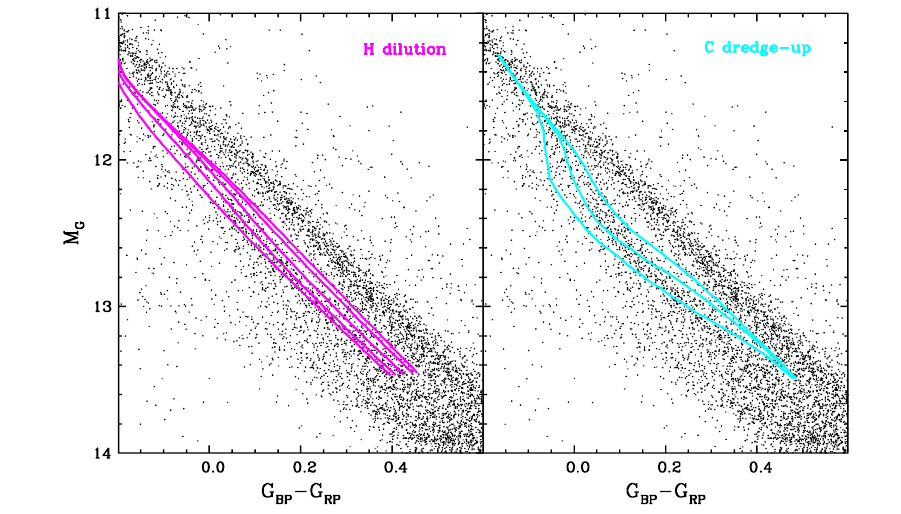}
\caption{Zoomed-in version of Figure~\ref{fig:bifurcation1} including theoretical evolutionary sequences for helium-rich white dwarfs contaminated by hydrogen due to convective dilution (left panel) and contaminated by carbon due to convective dredge-up (right panel). These sequences were calculated using the atmosphere models of \citet{blouin2023a}, assuming a stellar mass of 0.60\,$M_{\odot}$ and the hydrogen (left panel) and carbon (right panel) abundances predicted by the simulations shown in Figures \ref{fig:dilution2} and \ref{fig:dredgeup2}, respectively.}
\label{fig:bifurcation2}
\end{figure*}

A key implication of the chemical evolution model outlined in this section is that the majority of cool, helium-atmosphere DC stars should contain spectroscopically undetectable hydrogen. This is especially interesting in the context of the bifurcation in the Gaia colour--magnitude diagram, as the B branch can be explained if DC white dwarfs have $\log N_{\mathrm{H}}/N_{\mathrm{He}} \simeq -5.0$ \citep{bergeron2019,gentile-fusillo2020,kilic2020,gentile-fusillo2021}. However, this specific requirement proves to be inconsistent with our global understanding of spectral evolution. The direct precursors of most DC stars are DB and DBA stars, for which the surface hydrogen abundance is well constrained. Because the composition of a given DB or DBA white dwarf remains roughly constant as it cools, the DC population should be characterised by a similar range of abundances. The expected trajectory of these objects in the Gaia colour--magnitude diagram is illustrated in the left panel of Figure~\ref{fig:bifurcation2}; the five sequences displayed here assume hydrogen abundances taken from the five simulations of convective dilution shown in Figure~\ref{fig:dilution2}. At first glance, the overall agreement with the observed B branch appears satisfactory, but a closer inspection reveals that this is true only for the three lower curves, corresponding to the three most hydrogen-rich models. The two most hydrogen-poor sequences, which are representative of a significant portion of the DBA population (see Figure~\ref{fig:dilution2}), fall in between the A and B branches, similarly to the pure-helium atmosphere sequence previously displayed in Figure~\ref{fig:bifurcation1}. In other words, in the descendants of hydrogen-poor DB and DBA stars ($\log N_{\mathrm{H}}/N_{\mathrm{He}} \lesssim -6.0$), the amount of hydrogen is too low to affect the spectral energy distribution. This issue is confirmed by detailed population synthesis calculations assuming a realistic hydrogen abundance distribution, which predict a significant number of objects in between the A and B branches and thus fail to reproduce a clear separation \citep{blouin2023a}. The conclusion is that the Gaia bifurcation cannot be explained solely by the presence of trace hydrogen in helium-rich white dwarfs.

\subsection{Helium-rich progenitor: the effect of residual carbon} 
\label{sec:theory_helium2}

As mentioned in the previous section, most hydrogen-deficient PG 1159-type pre-white dwarfs have large amounts of carbon and oxygen in their envelope as a result of their born-again evolution. A range of surface abundances is observed among the PG 1159 population, that is, $0.20 \lesssim X_{\mathrm{C}} \lesssim 0.60$ and $0.02 \lesssim X_{\mathrm{O}} \lesssim 0.20$ in mass fractions \citep{werner2006,werner2014a}. The purpose of this section is to examine the impact of these elements on the spectral evolution of helium-rich white dwarfs. We will see that oxygen plays a very minor role, while carbon is a critical piece of the puzzle. To alleviate the discussion, we will first ignore the possible presence of trace hydrogen and concentrate on the transport of carbon. We will then come back to the general case where both of these elements are simultaneously present. It will become clear that the transport of hydrogen and the transport of carbon are largely independent from each other, and thus this segmental approach is justified.

\begin{figure*}[t]
\centering
\includegraphics[width=2.0\columnwidth,clip=true,trim=3.00cm 16.75cm 2.75cm 4.25cm]{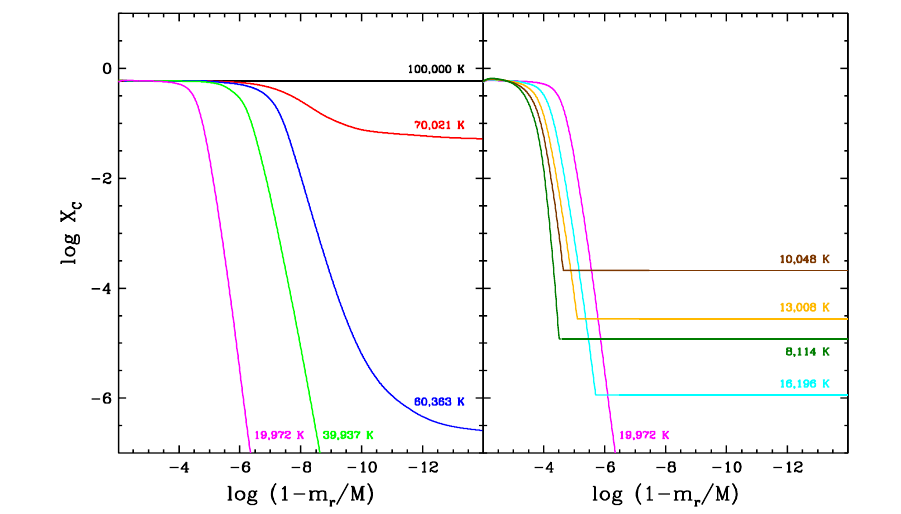}
\caption{Carbon mass fraction profile at various effective temperatures in a typical simulation of element transport in helium-rich white dwarfs. The position in the star is measured in terms of the fraction of the total mass located outside a given radius, such that the surface is towards the right and the core is towards the left. This particular simulation is taken from \citet{blouin2023a}, and assumes a stellar mass of 0.55\,$M_{\odot}$ and an initial carbon mass fraction $X_{\mathrm{C}} = 0.60$. The effective temperature decreases monotonically with time; the left panel illustrates the settling process at high temperature, while the right panel illustrates the convective dredge-up process at low temperature. In the latter case, notice the non-monotonic behaviour of the surface carbon abundance, which first increases (cyan, yellow, and brown curves) and then decreases (dark green curve).}
\label{fig:dredgeup1}
\end{figure*}

The chemical evolution of a \mbox{helium-,} \mbox{carbon-,} and oxygen-rich progenitor is governed by the same three main physical mechanisms invoked in the last section: radiative wind, gravitational settling, and convection. To illustrate this, Figure~\ref{fig:dredgeup1} shows the carbon abundance profile as a function of effective temperature in a simulation starting from a PG 1159-type composition \citep{bedard2022a,blouin2023a}. In the absence of processes opposing diffusion, the carbon and oxygen would be expected to sink out of sight within a few years. In reality, they are supported in the outer layers for much longer by the radiative wind. Indeed, the fact that PG 1159 stars are observed down to $T_{\mathrm{eff}} \simeq 75,000$\,K is considered as another conclusive empirical proof of the existence of winds at the beginning of the cooling sequence \citep{unglaub2000,werner2006,quirion2012,bedard2022a,bedard2022b}. The left panel of Figure~\ref{fig:dredgeup1} shows that as the wind fades, the carbon and oxygen rapidly sink into the star, thereby producing a growing pure-helium layer at the surface and accordingly changing the spectral character to DO and then DB \citep{dehner1995,fontaine2002,althaus2004,althaus2005a,althaus2009b,camisassa2017,bedard2022a,bedard2022b}. With further cooling, the convection zone appears and sharply expands downward at $T_{\mathrm{eff}} \simeq 20,000$\,K (see the bottom panel of Figure~\ref{fig:convzone}). The convective flows eventually catch up with the settling carbon and oxygen, which are thus efficiently dredged up to the surface. In practice, the dredged-up material is significantly richer in carbon than in oxygen, because the latter is less abundant in the progenitor to begin with and also sinks faster due do its higher atomic weight. For this reason, only carbon may become visible again, giving rise to a DQ white dwarf \citep{koester1982,fontaine1984,pelletier1986,macdonald1998,althaus2004,althaus2005a,dufour2005,scoccola2006,brassard2007,camisassa2017,koester2020,bedard2022a,bedard2022b,blouin2023a}. The right panel of Figure~\ref{fig:dredgeup1} shows that the atmospheric carbon abundance gradually increases as the convective region expands and reaches farther into the deep carbon reservoir\footnote{Comparing the right panels of Figures \ref{fig:dilution1} and \ref{fig:dredgeup1}, it can be seen that the extent of the convectively mixed region, corresponding to the flat part of the abundance profiles, is not exactly the same at a given effective temperature. This is due to the assumption of different stellar masses and the use of different prescriptions for the efficiency of convective overshoot \citep{bedard2022a,bedard2023}.}. Once the base of the convection zone stabilises at $T_{\mathrm{eff}} \simeq 10,000$\,K, carbon sinks out of the convection zone and thus the surface abundance starts decreasing again \citep{pelletier1986,dufour2005,brassard2007,bedard2022a,bedard2022b,blouin2023a}.

The quantitative predictions of this scenario can be compared to observations of carbon-bearing white dwarfs, starting with the classical DQ stars. For instance, we can now explain why carbon pollution is detected almost exclusively in helium-dominated objects: their hydrogen-dominated counterparts do not have carbon-enriched progenitors and have much shallower convection zones. More importantly, we will see below that the expected rate of carbon depletion at $T_{\mathrm{eff}} \lesssim 10,000$\,K is perfectly consistent with the diagonal sequence formed by the classical DQ stars in the temperature--abundance diagram (Figure~\ref{fig:carbon}). Therefore, the existence of these objects can be unambiguously attributed to the convective dredge-up of carbon \citep{pelletier1986,dufour2005,koester2006b,brassard2007,coutu2019,koester2019,koester2020,bedard2022a,bedard2022b,blouin2023a}. Nevertheless, as with the dilution of hydrogen, the predicted carbon abundance depends sensitively on the efficiency of convective overshoot, which is essentially a free parameter and thus limits the predictive power of the models \citep{koester2020,bedard2022b}.

In this context, a reasonable approach is to turn the problem around, that is, to rely on the observed narrow DQ sequence to calibrate the extent of overshoot. The main complication in this procedure is that the amount of dredged-up carbon is also influenced by other parameters, most notably the stellar mass and the composition of the progenitor. More specifically, carbon contamination is predicted to be more significant when the mass is lower and the initial carbon content is higher \citep{pelletier1986,bedard2022b,blouin2023a}. The question, then, is what values should be assumed for these parameters while adjusting overshoot to fit the theoretical and empirical carbon abundances. Fortunately, the masses of DQ white dwarfs and the carbon abundances of their PG 1159 progenitors are well constrained by independent observations. First, although the typical white dwarf mass is $\simeq 0.60\,M_{\odot}$, most cool DQ stars have inferred masses closer to $\simeq 0.55\,M_{\odot}$, hence this slightly lower value is likely more appropriate \citep{coutu2019,bedard2022b,caron2023}. Second, as mentioned above, the PG 1159 population is characterised by $0.20 \lesssim X_{\mathrm{C}} \lesssim 0.60$, and it can be argued that the progenitors of DQ white dwarfs probably lie at the upper end of this range. Indeed, we have already remarked earlier that the observed DQ sequence sits just above the optical detection limit of carbon (Figure~\ref{fig:carbon}), suggesting that DQ stars actually represent the most carbon-rich members of a larger carbon-polluted population \citep{weidemann1995,dufour2005,bedard2022b}.

\begin{figure*}[t]
\centering
\includegraphics[width=2.0\columnwidth,clip=true,trim=3.00cm 17.75cm 2.75cm 4.25cm]{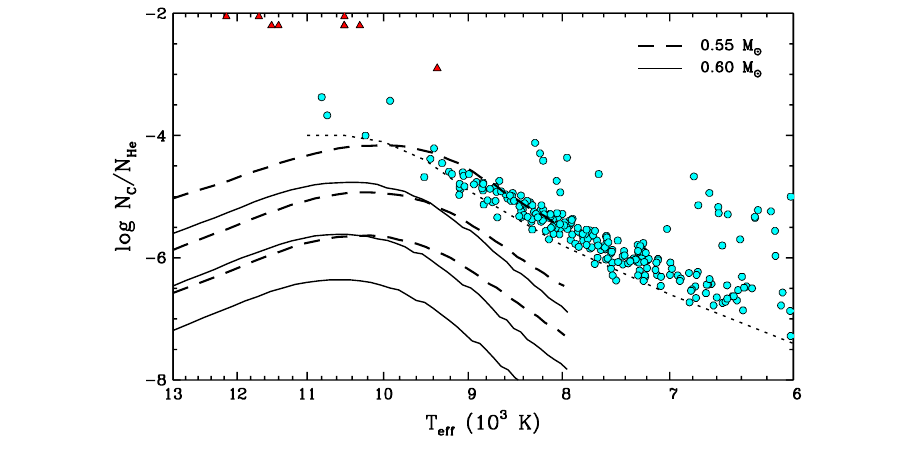}
\caption{Zoomed-in version of Figure~\ref{fig:carbon} including theoretical predictions for the convective dredge-up of carbon. These simulations are taken from \citet{blouin2023a}, and assume stellar masses of 0.55\,$M_{\odot}$ (dashed curves) and 0.60\,$M_{\odot}$ (solid curves) and initial carbon mass fractions $X_{\mathrm{C}} = 0.20$, 0.40, and 0.60 (from bottom to top).}
\label{fig:dredgeup2}
\end{figure*}

Figure~\ref{fig:dredgeup2} presents a zoomed-in version of the temperature--abundance diagram including predictions from various transport simulations\footnote{The model predictions shown in Figure~\ref{fig:dredgeup2} do not extend below $T_{\mathrm{eff}} \simeq 8000$\,K due to the lack of low-temperature carbon opacity data \citep{bedard2022b}.} \citep{blouin2023a}. The reference model assuming a mass of 0.55\,$M_{\odot}$ along with the initial condition $X_{\mathrm{C}} = 0.60$ corresponds to the uppermost dashed line (this is the simulation shown in detail in Figure~\ref{fig:dredgeup1}). Of course, the perfect match to the empirical DQ sequence is the outcome of the fine-tuning of convective overshoot. Still, overshoot mainly impacts the height of the theoretical curve in the diagram, but affects its overall shape to a much lesser extent \citep{bedard2022b}. Therefore, the fact that the slope of the predicted decrease at $T_{\mathrm{eff}} \lesssim 10,000$\,K aligns with the DQ sequence constitutes an independent validation of the simulation, at least in this regime. Given this anchor point, the stellar mass and initial carbon content can then be varied to quantitatively assess their influence on the dredge-up process. The two lower dashed lines correspond to models with $X_{\mathrm{C}} = 0.40$ and 0.20 but still assuming 0.55\,$M_{\odot}$. The set of solid curves uses the same three values for the initial carbon abundance while assuming a more standard mass of 0.60\,$M_{\odot}$. As stated above, a higher mass and a less carbon-rich progenitor both translate into lighter carbon pollution at low effective temperature.

The results shown in Figure~\ref{fig:dredgeup2} have three important implications. First, the fitting of the most carbon-polluted model on the observed DQ sequence provides a rough constraint on convective overshoot in cool ($T_{\mathrm{eff}} \lesssim 10,000$\,K) helium-rich white dwarfs. It is estimated that chemical mixing due to overshoot only extends over $\simeq 0.2$ pressure scale height, or $\simeq 0.1$\,dex in $\log (1-m_{r}/M)$, below the convection zone\footnote{The constraint quoted in \citet{bedard2022b}, $\simeq 0.8$ pressure scale height, is different from that given here, $\simeq 0.2$ pressure scale height, as the latter is based on the improved models presented in \citet{blouin2023a}.}. This distance is much smaller than that obtained from three-dimensional hydrodynamics calculations for hotter white dwarfs with shallower convection zones, indicating that overshoot becomes less significant as the convective region deepens. Second, the mass dependence of the dredge-up process naturally explains why classical DQ stars tend to have slightly lower-than-average masses. Carbon dredge-up likely occurs in most cool helium-dominated white dwarfs, but only those at the lower end of the mass distribution are polluted enough to show optical features earning them the DQ classification. Third, as DQ white dwarfs represent only the tip of the expected carbon abundance distribution at a given temperature, the vast majority of featureless DC white dwarfs should have unseen traces of carbon in their envelope \citep{bedard2022b}.

The latter inference naturally takes us back to the topic of the Gaia bifurcation. We have seen in the previous section that the presence of hydrogen in most cool helium-rich white dwarfs is insufficient to explain the clear gap between the A and B branches. Nevertheless, the atmosphere of DC stars is predicted to contain not only hydrogen, but also carbon, which can further alter the spectral energy distribution. The right panel of Figure~\ref{fig:bifurcation2} illustrates how the dredge-up of carbon impacts the course of helium-atmosphere white dwarfs in the Gaia colour--magnitude diagram; the three sequences displayed here assume evolving carbon abundances following the three 0.60\,$M_{\odot}$ simulations shown in Figure~\ref{fig:dredgeup2}. Unlike in the case of hydrogen, the successive increase and decrease of the atmospheric carbon content result in a significant deviation, such that the predicted curves satisfactorily coincide with the B branch. Accordingly, full population synthesis calculations taking carbon pollution into account successfully reproduce the separation between the A and B branches \citep{blouin2023a,camisassa2023}. Therefore, the Gaia bifurcation can be interpreted as the observational manifestation of the ubiquity of carbon dredge-up among the helium-rich white dwarf population. This remarkable conclusion has been independently corroborated by ultraviolet GALEX photometry, on which the opacity of carbon produces an even more distinctive signature \citep{blouin2023b}. However, the agreement between the observed and predicted B branch is not perfect, which has been attributed to uncertainties in current models of convective dredge-up. In particular, the theoretical curves shown in Figure~\ref{fig:dredgeup2} depend on the shape of the carbon diffusion tail at the bottom of the envelope, which is itself determined by the ionisation state of carbon at this depth \citep{pelletier1986,macdonald1998,blouin2023a}. The equation of state of partially ionised carbon is thus an important model ingredient, yet it is currently poorly known, and this is especially problematic for the ascending part of the theoretical curves \citep{blouin2023a}. Ultraviolet abundance measurements for cool DB white dwarfs with $15,000\,\mathrm{K} \gtrsim T_{\mathrm{eff}} \gtrsim 11,000$\,K could provide useful empirical constraints on the rate of carbon enrichment.

Recently, \citet{farihi2022} presented an alternative view of the origin of classical DQ white dwarfs, where these objects are products of binary stellar evolution. Among other arguments, they allege that the relatively low helium contents (or high carbon contents) and low stellar masses of DQ stars constitute evidence of a binary origin. This interpretation is at odds with the considerations of the present section. While it is true that DQ white dwarfs descend from carbon-enriched (hence moderately helium-deficient) progenitors, this is by no means a conundrum: such a PG 1159-type composition is the natural outcome of a well-known variant of single-star evolution, namely, the born-again scenario. This connection was first established in a seminal paper by \citet{althaus2005a}, who modelled the entire evolutionary channel from the zero-age main sequence, through the late helium-shell flash, and to the end of the cooling sequence. More recently, and as emphasised here, it has become clear that cool DQ stars are not fundamentally different from their DC and DZ counterparts, as both the DQ abundance distribution (Figure~\ref{fig:dredgeup2}) and the Gaia bifurcation (Figure~\ref{fig:bifurcation2}) indicate that most helium-atmosphere white dwarfs experience carbon dredge-up. Furthermore, we have already pointed out that the relatively low masses of DQ white dwarfs can be explained by the mass dependence of the convective dredge-up process\footnote{\citet{farihi2022} further argue that most DQ stars have such low masses that their inferred total ages are larger than the age of the Galactic thin disk. However, this inference is highly uncertain as the main-sequence lifetimes of low-mass white dwarfs are notoriously poorly constrained \citep{heintz2022}. For instance, had they used the initial-to-final mass relation of \citet{elbadry2018} rather than that of \citet{cummings2018}, they would have found the vast majority of DQ white dwarfs to be consistent with single-star evolution. Besides, it is also possible that the absolute mass scale of DQ stars is currently slightly underestimated due to inaccuracies in atmosphere models \citep{coutu2019}.} (Figure~\ref{fig:dredgeup2}). It is beyond the scope of this review to examine the other arguments put forward by \citet{farihi2022}, but we note that some of them have been challenged by other works \citep{bagnulo2022,blouin2022b,obrien2023}. All in all, the case for a binary origin appears inconclusive.

The convective dredge-up scenario nicely describes the atmospheric composition of the classical, cool DQ stars, but can it also explain the existence of other groups of carbon-bearing white dwarfs? Currently, this is not the case of the relatively hot, carbon-polluted DB stars with $T_{\mathrm{eff}} \simeq 25,000$\,K and $\log N_{\mathrm{C}}/N_{\mathrm{He}} \simeq -5.5$ (Figure~\ref{fig:carbon}). At this temperature, the convection zone is still very shallow (see the bottom panel of Figure~\ref{fig:convzone}), hence standard transport simulations do not predict any surface carbon contamination. This suggests that the gravitational settling of carbon must be significantly slower than expected in these objects. However, none of the physical mechanisms usually invoked in the framework of spectral evolution provides a viable explanation for this behaviour. Therefore, the origin of carbon pollution in hot DB white dwarfs remains an open question \citep{fontaine2005,koester2014b}.

As for the massive DQ stars forming a second sequence in the temperature--abundance diagram (Figure~\ref{fig:carbon}), it is clear that the convective dredge-up scenario does not apply in its canonical version, given that these objects originate from white dwarf mergers rather than single-star evolution \citep{dunlap2015,cheng2019,coutu2019,kawka2023}. Nevertheless, it has been demonstrated that the dredge-up process can indeed produce the required carbon-dominated atmospheric composition, provided that the stellar envelope is strongly helium-deficient (in addition to being strongly hydrogen-deficient, of course). More specifically, transport models indicate that the total helium mass must be between $\simeq 10^{-8}\,M$ and $10^{-6}\,M$ \citep{althaus2009a,hollands2020,koester2020}, many orders of magnitude smaller than the value expected from single-star evolution, $\simeq 10^{-2}\,M$ \citep{iben1984,herwig1999,lawlor2006,miller-bertolami2006b,renedo2010,romero2012}. Given such an initial condition, the surface helium layer built by diffusion remains extremely thin, and thus the expansion of the convection zone at $T_{\mathrm{eff}} \simeq 20,000$\,K (see the bottom panel of Figure~\ref{fig:convzone}) suddenly brings a large amount of carbon to the surface \citep{althaus2009a}. The inferred strong helium deficiency represents yet another peculiar property of warm DQ stars (besides their high masses, space velocities, magnetic fields, and rotation rates) and provides valuable insight into the outcome of white dwarf mergers. 

That said, we note that no single transport simulation has so far been able to reproduce the shape of the massive DQ sequence in the temperature--abundance diagram \citep{brassard2007,althaus2009a}. This may be because these objects have a highly non-standard cooling history, an aspect that is not considered in existing models as it has been uncovered only recently. In the Gaia colour--magnitude diagram, most massive DQ stars lie on the so-called Q branch, a nearly horizontal overdensity of white dwarfs at absolute magnitude $M_{G} \simeq 13$ (see Figure\,\ref{fig:bifurcation1}; \citealt{gaia2018b}). This pile-up reveals that some white dwarf merger remnants experience a very long cooling delay of about 10\,Gyr, which is likely due to a chemical distillation process triggered by the crystallisation of their core \citep{cheng2019,blouin2021a,bedard2024}. The warm DQ stars probably all belong to this delayed population, and therefore a given object is presumably stuck at a given effective temperature for billions of years. This means that the massive DQ sequence in the temperature--abundance diagram may not be an evolutionary sequence, and may instead arise from the fact that different objects experience the cooling delay at different temperatures (due to their different masses; \citealt{bedard2024}). The spread in atmospheric composition could then simply reflect a diversity of residual hydrogen and helium contents and/or convection zone depths \citep{koester2020,kilic2024}. In any case, models of element transport over the multi-Gyr cooling delay are needed to shed further light on the chemical structure of merger remnants.

\begin{figure*}[t]
\centering
\includegraphics[width=2.0\columnwidth,clip=true,trim=20.00cm 32.00cm 20.00cm 44.00cm]{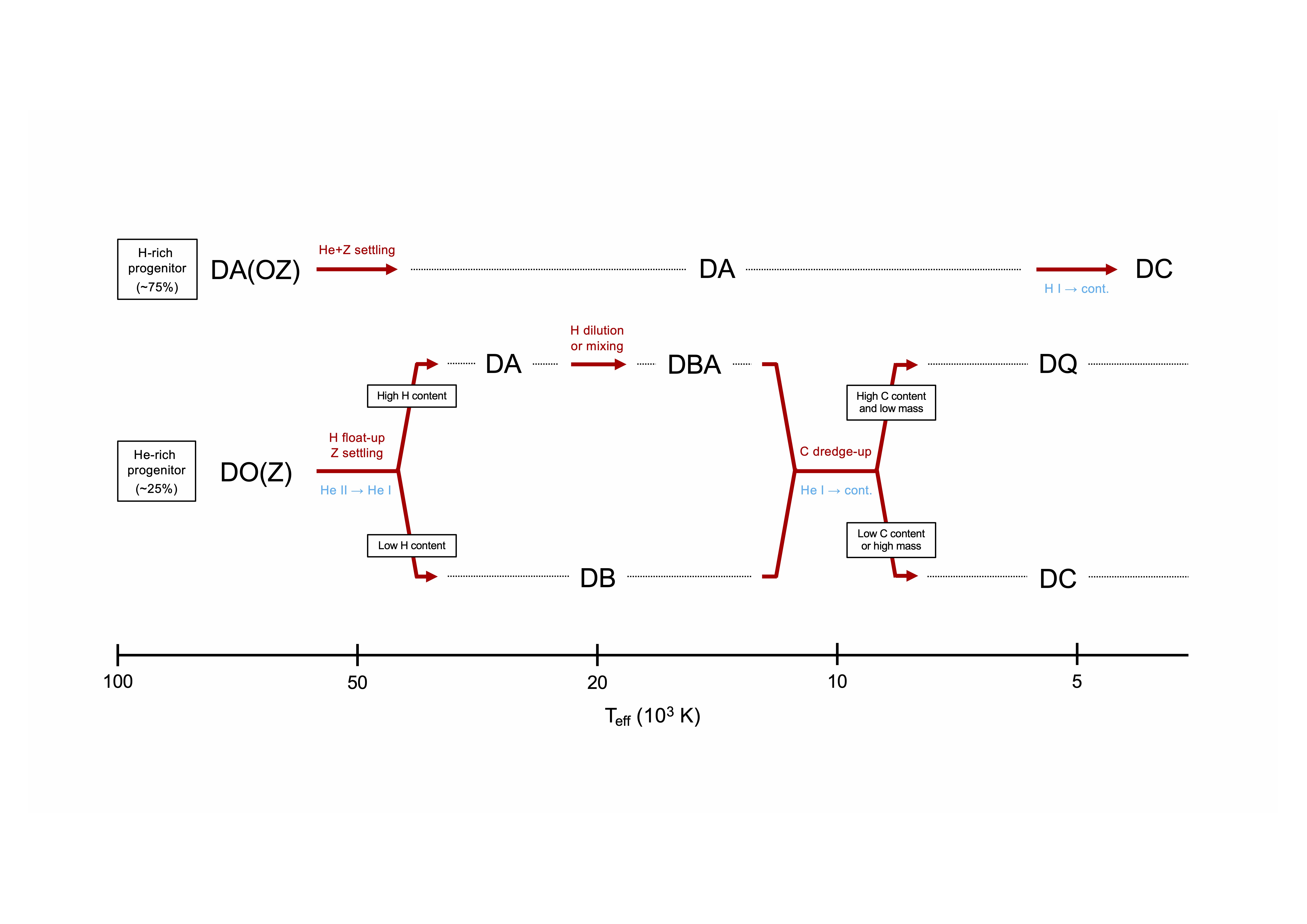}
\caption{Schematic summary of the main white dwarf spectral evolution channels. For clarity, only internal element transport is considered, and it is implied that external accretion may at any time add metals to the atmosphere and thus add a ``Z'' to the spectral type. Some intermediate phases are also omitted for simplicity, most notably the hot stratified white dwarfs (amidst the DO-to-DA transition) and the cool helium-rich DA white dwarfs (amidst the DBA-to-DC/DQ transition).}
\label{fig:summary}
\end{figure*}

In concluding this section, we briefly consider the general (and more realistic) case of a helium-rich progenitor containing both a trace of hydrogen and a significant amount of carbon. The chemical evolution of such an object is simply a combination of the phenomena described in the previous and present sections. The surface composition initially remains unchanged because of the radiative wind, but as gravitational settling starts to operate, the star turns into a helium-atmosphere DO white dwarf due to the sinking of carbon, and then into a hydrogen-atmosphere DA white dwarf due to the floating of hydrogen. Shortly afterwards, the surface hydrogen layer is wiped out by convection, through either the convective dilution or mixing mechanism, thereby producing a DBA or helium-rich DA star. As the white dwarf cools further, the atmospheric hydrogen abundance remains roughly constant but its spectroscopic signature gradually disappears. Meanwhile, the convection zone reaches into the deep reservoir of settling carbon, a small amount of which is thus brought back to the surface. Depending on the magnitude of this dredge-up process, the outcome is either a DQ star if carbon is abundant enough to be spectroscopically detected, or a featureless DC star otherwise. In any case, it is clear that the resulting cool white dwarf does not possess a pure-helium atmosphere, but rather a helium-rich atmosphere contaminated by traces of both hydrogen and carbon \citep{bedard2022b}.

\section{Conclusion}
\label{sec:conclusion}

Overall, our understanding of the spectral evolution of white dwarfs is in a very satisfactory state. On one hand, large-scale photometric and spectroscopic surveys have led to a detailed empirical view of the variations in surface composition along the cooling sequence. On the other hand, sophisticated numerical simulations of element transport have enabled the development of an elaborate theory of spectral evolution, which successfully explains nearly all of the observed features. As a summary, we present in Figure~\ref{fig:summary} a schematic diagram of the main white dwarf spectral evolution channels as currently understood.

However, we still face a number of challenges, the most notable of which are briefly recapitulated here.
\begin{itemize}
\item The inferred fraction of helium-atmosphere white dwarfs remains uncertain, both at high temperature, where only one modern study exists, and at low temperature, where different studies yield different results. Better constraints would help clarify the relative importance of the three main spectral evolution channels among the white dwarf population.
\item The undetected presence of hydrogen and carbon in cool helium-rich DC stars is supported by both observation and theory, but the abundances of individual objects remain elusive, which introduces a certain level of uncertainty in the determination of their atmospheric parameters. Ultraviolet spectroscopy, although currently practicable only for the nearest objects, could allow abundance measurements and hence greatly improve the characterisation of cool helium-dominated white dwarfs.
\item More theoretical work is required to fully understand the transport of carbon in the envelope of white dwarfs. First, despite the overall success of the canonical dredge-up scenario, the quantitative predictions are still subject to uncertainties on the carbon equation of state, which thus needs to be improved. Second, this scenario currently completely fails to account for the carbon pollution seen in DB stars. Third, a holistic explanation for the peculiar atmospheric properties of the massive DQ class has yet to be provided.
\item The spectral evolution of the hottest white dwarfs is believed to be dominated by radiative winds, and yet models describing the generation of such winds are practically nonexistent, forcing the use of very crude approximations in transport simulations. Detailed predictions of the mass-loss rate as a function of the relevant white dwarf parameters would be highly desirable.
\item The predictive power of models of convectively-driven phenomena (namely, convective dilution, convective mixing, and convective dredge-up) is currently severely limited by uncertainties on the extent of overshoot. An expansion of the existing hydrodynamics-based overshoot studies could help reduce these uncertainties. Moreover, we note that the onset of the convective dilution process, where the helium-rich convection zone erodes the overlying hydrogen-rich layer from below, has yet to be modelled in detail. This may become possible with a more accurate description of overshoot.
\end{itemize}

In closing, we note that the advent of the next generation of multi-object spectroscopic surveys, including SDSS-V \citep{kollmeier2017}, DESI \citep{desi2016}, WEAVE \citep{dalton2012}, and 4MOST \citep{dejong2014}, is imminent. This observational revolution will not only significantly increase the number of white dwarfs of all known types, but also possibly reveal entirely new classes of objects with atmospheric properties never seen before. These discoveries will constitute as many new challenges for the theory of white dwarf spectral evolution.

\backmatter

\bmhead{Acknowledgments} AB is grateful to Pierre Bergeron and Pierre Brassard for their guidance and encouragement while supervising his PhD thesis, on which a large part of this review is based. AB thanks the anonymous referee as well as Pierre Bergeron, Simon Blouin, Mairi O'Brien, Ingrid Pelisoli, Pier-Emmanuel Tremblay, and Olivier Vincent for reading the manuscript and providing useful feedback. AB is also grateful to the Executive Committee of the International Astronomical Union (IAU) for awarding him the 2022 IAU PhD Prize (Stars and Stellar Physics Division), which led to this invitation from Springer Nature to contribute to the 2023 Astronomy Prize Awardees Collection. This version of the article has been accepted for publication after peer review, but is not the version of record and does not reflect post-acceptance improvements; the version of record is publicly available online at \url{http://dx.doi.org/10.1007/s10509-024-04307-5}.

\bmhead{Author contributions} AB produced the entirety of the manuscript.

\bmhead{Funding} AB is a Postdoctoral Fellow of the Natural Sciences and Engineering Research Council (NSERC) of Canada and also acknowledges support from the European Research Council (ERC) under the European Union’s Horizon 2020 research and innovation programme (grant agreement no. 101002408).

\bmhead{Competing interests} AB declares no competing interests.

\bibliography{sn-article}

\begin{thebibliography}{323}
\providecommand{\natexlab}[1]{#1}
\providecommand{\url}[1]{{#1}}
\providecommand{\urlprefix}{URL }
\providecommand{\doi}[1]{\url{https://doi.org/#1}}
\providecommand{\eprint}[2][]{\url{#2}}
 \bibcommenthead

\bibitem[{{Alcock} \& {Illarionov}(1980)}]{alcock1980}
{Alcock} C., {Illarionov} A. (1980): {The Surface Chemistry of Stars - Part Two - Fractionated Accretion of Interstellar Matter}. \apj\, 235:541. \doi{10.1086/157657}

\bibitem[{{Althaus} \& {Benvenuto}(1998)}]{althaus1998}
{Althaus} L.~G., {Benvenuto} O.~G. (1998): {Evolution of DA white dwarfs in the context of a new theory of convection}. \mnras\, 296(1):206--216. \doi{10.1046/j.1365-8711.1998.01332.x}, {\href{https://arxiv.org/abs/astro-ph/9812062}{{arXiv:astro-ph/9812062}}} {[astro-ph]}

\bibitem[{{Althaus} \& {C{\'o}rsico}(2004)}]{althaus2004}
{Althaus} L.~G., {C{\'o}rsico} A.~H. (2004): {The double-layered chemical structure in DB white dwarfs}. \aap\, 417:1115--1123. \doi{10.1051/0004-6361:20040021}, {\href{https://arxiv.org/abs/astro-ph/0401321}{{arXiv:astro-ph/0401321}}} {[astro-ph]}

\bibitem[{{Althaus} et~al.(2005{\natexlab{a}}){Althaus}, {Miller Bertolami}, {C{\'o}rsico}, {Garc{\'\i}a-Berro}, \& {Gil-Pons}}]{althaus2005b}
{Althaus} L.~G., {Miller Bertolami} M.~M., {C{\'o}rsico} A.~H., et~al. (2005{\natexlab{a}}): {The formation of DA white dwarfs with thin hydrogen envelopes}. \aap\, 440(1):L1--L4. \doi{10.1051/0004-6361:200500159}, {\href{https://arxiv.org/abs/astro-ph/0507415}{{arXiv:astro-ph/0507415}}} {[astro-ph]}

\bibitem[{{Althaus} et~al.(2005{\natexlab{b}}){Althaus}, {Serenelli}, {Panei}, {C{\'o}rsico}, {Garc{\'\i}a-Berro}, \& {Sc{\'o}ccola}}]{althaus2005a}
{Althaus} L.~G., {Serenelli} A.~M., {Panei} J.~A., et~al. (2005{\natexlab{b}}): {The formation and evolution of hydrogen-deficient post-AGB white dwarfs: The emerging chemical profile and the expectations for the PG 1159-DB-DQ evolutionary connection}. \aap\, 435(2):631--648. \doi{10.1051/0004-6361:20041965}, {\href{https://arxiv.org/abs/astro-ph/0502005}{{arXiv:astro-ph/0502005}}} {[astro-ph]}

\bibitem[{{Althaus} et~al.(2009{\natexlab{a}}){Althaus}, {Garc{\'\i}a-Berro}, {C{\'o}rsico}, {Miller Bertolami}, \& {Romero}}]{althaus2009a}
{Althaus} L.~G., {Garc{\'\i}a-Berro} E., {C{\'o}rsico} A.~H., et~al. (2009{\natexlab{a}}): {On the Formation of Hot DQ White Dwarfs}. \apjl\, 693(1):L23--L26. \doi{10.1088/0004-637X/693/1/L23}, {\href{https://arxiv.org/abs/0901.1836}{{arXiv:0901.1836}}} {[astro-ph.SR]}

\bibitem[{{Althaus} et~al.(2009{\natexlab{b}}){Althaus}, {Panei}, {Miller Bertolami}, {Garc{\'\i}a-Berro}, {C{\'o}rsico}, {Romero}, {Kepler}, \& {Rohrmann}}]{althaus2009b}
{Althaus} L.~G., {Panei} J.~A., {Miller Bertolami} M.~M., et~al. (2009{\natexlab{b}}): {New Evolutionary Sequences for Hot H-Deficient White Dwarfs on the Basis of a Full Account of Progenitor Evolution}. \apj\, 704(2):1605--1615. \doi{10.1088/0004-637X/704/2/1605}, {\href{https://arxiv.org/abs/0909.2689}{{arXiv:0909.2689}}} {[astro-ph.SR]}

\bibitem[{{Althaus} et~al.(2010){Althaus}, {C{\'o}rsico}, {Isern}, \& {Garc{\'\i}a-Berro}}]{althaus2010}
{Althaus} L.~G., {C{\'o}rsico} A.~H., {Isern} J., et~al. (2010): {Evolutionary and pulsational properties of white dwarf stars}. \aapr\, 18(4):471--566. \doi{10.1007/s00159-010-0033-1}, {\href{https://arxiv.org/abs/1007.2659}{{arXiv:1007.2659}}} {[astro-ph.SR]}

\bibitem[{{Althaus} et~al.(2020){Althaus}, {C{\'o}rsico}, {Uzundag}, {Vu{\v{c}}kovi{\'c}}, {Baran}, {Bell}, {Camisassa}, {Calcaferro}, {De Ger{\'o}nimo}, {Kepler}, \& {Silvotti}}]{althaus2020a}
{Althaus} L.~G., {C{\'o}rsico} A.~H., {Uzundag} M., et~al. (2020): {About the existence of warm H-rich pulsating white dwarfs}. \aap\, 633:A20. \doi{10.1051/0004-6361/201936346}, {\href{https://arxiv.org/abs/1911.02442}{{arXiv:1911.02442}}} {[astro-ph.SR]}

\bibitem[{{Arcoragi} \& {Fontaine}(1980)}]{arcoragi1980}
{Arcoragi} J.~P., {Fontaine} G. (1980): {Acoustic fluxes in white dwarfs}. \apj\, 242:1208--1225. \doi{10.1086/158551}

\bibitem[{{Badenas-Agusti} et~al.(2024){Badenas-Agusti}, {Vanderburg}, {Blouin}, {Dufour}, {Via{\~n}a}, {Seager}, \& {Wang}}]{badenas-agusti2024}
{Badenas-Agusti} M., {Vanderburg} A., {Blouin} S., et~al. (2024): {Detection and preliminary characterization of polluted white dwarfs from Gaia EDR3 and LAMOST}. \mnras\, 527(3):4515--4544. \doi{10.1093/mnras/stad3362}, {\href{https://arxiv.org/abs/2310.19790}{{arXiv:2310.19790}}} {[astro-ph.SR]}

\bibitem[{{Baglin} \& {Vauclair}(1973)}]{baglin1973}
{Baglin} A., {Vauclair} G. (1973): {Evolutionary Sequence for DA, DB, DC White Dwarfs}. \aap\, 27:307

\bibitem[{{Bagnulo} \& {Landstreet}(2022)}]{bagnulo2022}
{Bagnulo} S., {Landstreet} J.~D. (2022): {Multiple Channels for the Onset of Magnetism in Isolated White Dwarfs}. \apjl\, 935(1):L12. \doi{10.3847/2041-8213/ac84d3}, {\href{https://arxiv.org/abs/2208.02655}{{arXiv:2208.02655}}} {[astro-ph.SR]}

\bibitem[{{Barstow} \& {Hubeny}(1998)}]{barstow1998a}
{Barstow} M.~A., {Hubeny} I. (1998): {An alternative explanation of the EUV spectrum of the white dwarf G191-B2B invoking a stratified H+He envelope including heavier elements}. \mnras\, 299(2):379--388. \doi{10.1046/j.1365-8711.1998.01711.x}

\bibitem[{{Barstow} et~al.(2003){Barstow}, {Good}, {Holberg}, {Hubeny}, {Bannister}, {Bruhweiler}, {Burleigh}, \& {Napiwotzki}}]{barstow2003a}
{Barstow} M.~A., {Good} S.~A., {Holberg} J.~B., et~al. (2003): {Heavy-element abundance patterns in hot DA white dwarfs}. \mnras\, 341(3):870--890. \doi{10.1046/j.1365-8711.2003.06462.x}, {\href{https://arxiv.org/abs/astro-ph/0301519}{{arXiv:astro-ph/0301519}}} {[astro-ph]}

\bibitem[{{Barstow} et~al.(2014){Barstow}, {Barstow}, {Casewell}, {Holberg}, \& {Hubeny}}]{barstow2014}
{Barstow} M.~A., {Barstow} J.~K., {Casewell} S.~L., et~al. (2014): {Evidence for an external origin of heavy elements in hot DA white dwarfs}. \mnras\, 440(2):1607--1625. \doi{10.1093/mnras/stu216}, {\href{https://arxiv.org/abs/1402.2164}{{arXiv:1402.2164}}} {[astro-ph.SR]}

\bibitem[{{Bauer}(2023)}]{bauer2023}
{Bauer} E.~B. (2023): {Carbon-Oxygen Phase Separation in Modules for Experiments in Stellar Astrophysics (MESA) White Dwarf Models}. \apj\, 950(2):115. \doi{10.3847/1538-4357/acd057}, {\href{https://arxiv.org/abs/2303.10110}{{arXiv:2303.10110}}} {[astro-ph.SR]}

\bibitem[{{Bauer} \& {Bildsten}(2018)}]{bauer2018}
{Bauer} E.~B., {Bildsten} L. (2018): {Increases to Inferred Rates of Planetesimal Accretion due to Thermohaline Mixing in Metal-accreting White Dwarfs}. \apjl\, 859(2):L19. \doi{10.3847/2041-8213/aac492}, {\href{https://arxiv.org/abs/1805.05425}{{arXiv:1805.05425}}} {[astro-ph.SR]}

\bibitem[{{Bauer} \& {Bildsten}(2019)}]{bauer2019}
{Bauer} E.~B., {Bildsten} L. (2019): {Polluted White Dwarfs: Mixing Regions and Diffusion Timescales}. \apj\, 872(1):96. \doi{10.3847/1538-4357/ab0028}, {\href{https://arxiv.org/abs/1812.09602}{{arXiv:1812.09602}}} {[astro-ph.SR]}

\bibitem[{{B{\'e}dard} et~al.(2017){B{\'e}dard}, {Bergeron}, \& {Fontaine}}]{bedard2017}
{B{\'e}dard} A., {Bergeron} P., {Fontaine} G. (2017): {Measurements of Physical Parameters of White Dwarfs: A Test of the Mass-Radius Relation}. \apj\, 848(1):11. \doi{10.3847/1538-4357/aa8bb6}, {\href{https://arxiv.org/abs/1709.02324}{{arXiv:1709.02324}}} {[astro-ph.SR]}

\bibitem[{{B{\'e}dard} et~al.(2020){B{\'e}dard}, {Bergeron}, {Brassard}, \& {Fontaine}}]{bedard2020}
{B{\'e}dard} A., {Bergeron} P., {Brassard} P., et~al. (2020): {On the Spectral Evolution of Hot White Dwarf Stars. I. A Detailed Model Atmosphere Analysis of Hot White Dwarfs from SDSS DR12}. \apj\, 901(2):93. \doi{10.3847/1538-4357/abafbe}, {\href{https://arxiv.org/abs/2008.07469}{{arXiv:2008.07469}}} {[astro-ph.SR]}

\bibitem[{{B{\'e}dard} et~al.(2022{\natexlab{a}}){B{\'e}dard}, {Bergeron}, \& {Brassard}}]{bedard2022b}
{B{\'e}dard} A., {Bergeron} P., {Brassard} P. (2022{\natexlab{a}}): {On the Spectral Evolution of Hot White Dwarf Stars. III. The PG 1159-DO-DB-DQ Evolutionary Channel Revisited}. \apj\, 930(1):8. \doi{10.3847/1538-4357/ac609d}, {\href{https://arxiv.org/abs/2203.12045}{{arXiv:2203.12045}}} {[astro-ph.SR]}

\bibitem[{{B{\'e}dard} et~al.(2022{\natexlab{b}}){B{\'e}dard}, {Brassard}, {Bergeron}, \& {Blouin}}]{bedard2022a}
{B{\'e}dard} A., {Brassard} P., {Bergeron} P., et~al. (2022{\natexlab{b}}): {On the Spectral Evolution of Hot White Dwarf Stars. II. Time-dependent Simulations of Element Transport in Evolving White Dwarfs with STELUM}. \apj\, 927(1):128. \doi{10.3847/1538-4357/ac4497}, {\href{https://arxiv.org/abs/2112.09989}{{arXiv:2112.09989}}} {[astro-ph.SR]}

\bibitem[{{B{\'e}dard} et~al.(2023){B{\'e}dard}, {Bergeron}, \& {Brassard}}]{bedard2023}
{B{\'e}dard} A., {Bergeron} P., {Brassard} P. (2023): {On the Spectral Evolution of Hot White Dwarf Stars. IV. The Diffusion and Mixing of Residual Hydrogen in Helium-rich White Dwarfs}. \apj\, 946(1):24. \doi{10.3847/1538-4357/acbb62}, {\href{https://arxiv.org/abs/2302.05424}{{arXiv:2302.05424}}} {[astro-ph.SR]}

\bibitem[{{B{\'e}dard} et~al.(2024){B{\'e}dard}, {Blouin}, \& {Cheng}}]{bedard2024}
{B{\'e}dard} A., {Blouin} S., {Cheng} S. (2024): {Buoyant crystals halt the cooling of white dwarf stars}. \nat\, 627(8003):286--288. \doi{10.1038/s41586-024-07102-y}

\bibitem[{{Benvenuto} \& {Althaus}(1997)}]{benvenuto1997}
{Benvenuto} O.~G., {Althaus} L.~G. (1997): {DB white dwarf evolution in the frame of the full spectrum turbulence theory}. \mnras\, 288(4):1004--1014. \doi{10.1093/mnras/288.4.1004}

\bibitem[{{Bergeron} \& {Liebert}(2002)}]{bergeron2002}
{Bergeron} P., {Liebert} J. (2002): {Spectroscopic Analysis of the DAB White Dwarf PG 1115+166: An Unresolved DA+DB Degenerate Binary}. \apj\, 566(2):1091--1094. \doi{10.1086/338279}, {\href{https://arxiv.org/abs/astro-ph/0110549}{{arXiv:astro-ph/0110549}}} {[astro-ph]}

\bibitem[{{Bergeron} et~al.(1992){Bergeron}, {Saffer}, \& {Liebert}}]{bergeron1992}
{Bergeron} P., {Saffer} R.~A., {Liebert} J. (1992): {A Spectroscopic Determination of the Mass Distribution of DA White Dwarfs}. \apj\, 394:228. \doi{10.1086/171575}

\bibitem[{{Bergeron} et~al.(1993){Bergeron}, {Wesemael}, {Lamontagne}, \& {Chayer}}]{bergeron1993}
{Bergeron} P., {Wesemael} F., {Lamontagne} R., et~al. (1993): {Metal-Line Blanketing and the Peculiar H beta Line Profile in the DAO Star Feige 55}. \apjl\, 407:L85. \doi{10.1086/186812}

\bibitem[{{Bergeron} et~al.(1994){Bergeron}, {Wesemael}, {Beauchamp}, {Wood}, {Lamontagne}, {Fontaine}, \& {Liebert}}]{bergeron1994}
{Bergeron} P., {Wesemael} F., {Beauchamp} A., et~al. (1994): {A Spectroscopic Analysis of DAO and Hot DA White Dwarfs: The Implications of the Presence of Helium and the Nature of DAO Stars}. \apj\, 432:305. \doi{10.1086/174571}

\bibitem[{{Bergeron} et~al.(1997){Bergeron}, {Ruiz}, \& {Leggett}}]{bergeron1997}
{Bergeron} P., {Ruiz} M.~T., {Leggett} S.~K. (1997): {The Chemical Evolution of Cool White Dwarfs and the Age of the Local Galactic Disk}. \apjs\, 108(1):339--387. \doi{10.1086/312955}

\bibitem[{{Bergeron} et~al.(2001){Bergeron}, {Leggett}, \& {Ruiz}}]{bergeron2001}
{Bergeron} P., {Leggett} S.~K., {Ruiz} M.~T. (2001): {Photometric and Spectroscopic Analysis of Cool White Dwarfs with Trigonometric Parallax Measurements}. \apjs\, 133(2):413--449. \doi{10.1086/320356}, {\href{https://arxiv.org/abs/astro-ph/0011286}{{arXiv:astro-ph/0011286}}} {[astro-ph]}

\bibitem[{{Bergeron} et~al.(2011){Bergeron}, {Wesemael}, {Dufour}, {Beauchamp}, {Hunter}, {Saffer}, {Gianninas}, {Ruiz}, {Limoges}, {Dufour}, {Fontaine}, \& {Liebert}}]{bergeron2011}
{Bergeron} P., {Wesemael} F., {Dufour} P., et~al. (2011): {A Comprehensive Spectroscopic Analysis of DB White Dwarfs}. \apj\, 737(1):28. \doi{10.1088/0004-637X/737/1/28}, {\href{https://arxiv.org/abs/1105.5433}{{arXiv:1105.5433}}} {[astro-ph.SR]}

\bibitem[{{Bergeron} et~al.(2019){Bergeron}, {Dufour}, {Fontaine}, {Coutu}, {Blouin}, {Genest-Beaulieu}, {B{\'e}dard}, \& {Rolland}}]{bergeron2019}
{Bergeron} P., {Dufour} P., {Fontaine} G., et~al. (2019): {On the Measurement of Fundamental Parameters of White Dwarfs in the Gaia Era}. \apj\, 876(1):67. \doi{10.3847/1538-4357/ab153a}, {\href{https://arxiv.org/abs/1904.02022}{{arXiv:1904.02022}}} {[astro-ph.SR]}

\bibitem[{{Bergeron} et~al.(2022){Bergeron}, {Kilic}, {Blouin}, {B{\'e}dard}, {Leggett}, \& {Brown}}]{bergeron2022}
{Bergeron} P., {Kilic} M., {Blouin} S., et~al. (2022): {On the Nature of Ultracool White Dwarfs: Not so Cool after All}. \apj\, 934(1):36. \doi{10.3847/1538-4357/ac76c7}, {\href{https://arxiv.org/abs/2206.03174}{{arXiv:2206.03174}}} {[astro-ph.SR]}

\bibitem[{{Bl{\"o}cker}(2001)}]{blocker2001}
{Bl{\"o}cker} T. (2001): {Evolution on the AGB and beyond: on the formation of H-deficient post-AGB stars}. \apss\, 275:1--14. \doi{10.1023/A:1002777931450}, {\href{https://arxiv.org/abs/astro-ph/0102135}{{arXiv:astro-ph/0102135}}} {[astro-ph]}

\bibitem[{{Blouin}(2022)}]{blouin2022b}
{Blouin} S. (2022): {Missing metals in DQ stars: A simple explanation}. \aap\, 666:L7. \doi{10.1051/0004-6361/202244944}, {\href{https://arxiv.org/abs/2209.11626}{{arXiv:2209.11626}}} {[astro-ph.SR]}

\bibitem[{{Blouin} \& {Dufour}(2019)}]{blouin2019c}
{Blouin} S., {Dufour} P. (2019): {The evolution of carbon-polluted white dwarfs at low effective temperatures}. \mnras\, 490(3):4166--4174. \doi{10.1093/mnras/stz2915}, {\href{https://arxiv.org/abs/1910.06168}{{arXiv:1910.06168}}} {[astro-ph.SR]}

\bibitem[{{Blouin} \& {Xu}(2022)}]{blouin2022a}
{Blouin} S., {Xu} S. (2022): {No evidence for a strong decrease of planetesimal accretion in old white dwarfs}. \mnras\, 510(1):1059--1067. \doi{10.1093/mnras/stab3446}, {\href{https://arxiv.org/abs/2111.12152}{{arXiv:2111.12152}}} {[astro-ph.SR]}

\bibitem[{{Blouin} et~al.(2019{\natexlab{a}}){Blouin}, {Dufour}, {Allard}, {Salim}, {Rich}, \& {Koopmans}}]{blouin2019a}
{Blouin} S., {Dufour} P., {Allard} N.~F., et~al. (2019{\natexlab{a}}): {A New Generation of Cool White Dwarf Atmosphere Models. III. WD J2356-209: Accretion of a Planetesimal with an Unusual Composition}. \apj\, 872(2):188. \doi{10.3847/1538-4357/ab0081}, {\href{https://arxiv.org/abs/1902.03219}{{arXiv:1902.03219}}} {[astro-ph.SR]}

\bibitem[{{Blouin} et~al.(2019{\natexlab{b}}){Blouin}, {Dufour}, {Thibeault}, \& {Allard}}]{blouin2019b}
{Blouin} S., {Dufour} P., {Thibeault} C., et~al. (2019{\natexlab{b}}): {A New Generation of Cool White Dwarf Atmosphere Models. IV. Revisiting the Spectral Evolution of Cool White Dwarfs}. \apj\, 878(1):63. \doi{10.3847/1538-4357/ab1f82}, {\href{https://arxiv.org/abs/1905.02174}{{arXiv:1905.02174}}} {[astro-ph.SR]}

\bibitem[{{Blouin} et~al.(2021){Blouin}, {Daligault}, \& {Saumon}}]{blouin2021a}
{Blouin} S., {Daligault} J., {Saumon} D. (2021): {$^{22}$Ne Phase Separation as a Solution to the Ultramassive White Dwarf Cooling Anomaly}. \apjl\, 911(1):L5. \doi{10.3847/2041-8213/abf14b}, {\href{https://arxiv.org/abs/2103.12892}{{arXiv:2103.12892}}} {[astro-ph.SR]}

\bibitem[{{Blouin} et~al.(2023{\natexlab{a}}){Blouin}, {B{\'e}dard}, \& {Tremblay}}]{blouin2023a}
{Blouin} S., {B{\'e}dard} A., {Tremblay} P.-E. (2023{\natexlab{a}}): {Carbon dredge-up required to explain the Gaia white dwarf colour-magnitude bifurcation}. \mnras\, 523(3):3363--3375. \doi{10.1093/mnras/stad1574}, {\href{https://arxiv.org/abs/2305.02827}{{arXiv:2305.02827}}} {[astro-ph.SR]}

\bibitem[{{Blouin} et~al.(2023{\natexlab{b}}){Blouin}, {Kilic}, {B{\'e}dard}, \& {Tremblay}}]{blouin2023b}
{Blouin} S., {Kilic} M., {B{\'e}dard} A., et~al. (2023{\natexlab{b}}): {The ubiquity of carbon dredge-up in hydrogen-deficient white dwarfs as revealed by GALEX}. \mnras\, 525(1):L112--L116. \doi{10.1093/mnrasl/slad105}, {\href{https://arxiv.org/abs/2307.14295}{{arXiv:2307.14295}}} {[astro-ph.SR]}

\bibitem[{{Bonsor} et~al.(2020){Bonsor}, {Carter}, {Hollands}, {G{\"a}nsicke}, {Leinhardt}, \& {Harrison}}]{bonsor2020}
{Bonsor} A., {Carter} P.~J., {Hollands} M., et~al. (2020): {Are exoplanetesimals differentiated?} \mnras\, 492(2):2683--2697. \doi{10.1093/mnras/stz3603}, {\href{https://arxiv.org/abs/2001.04499}{{arXiv:2001.04499}}} {[astro-ph.EP]}

\bibitem[{{Brassard} et~al.(2007){Brassard}, {Fontaine}, {Dufour}, \& {Bergeron}}]{brassard2007}
{Brassard} P., {Fontaine} G., {Dufour} P., et~al. (2007): {The Origin and Evolution of DQ White Dwarfs: The Carbon Pollution Problem Revisited}. In: {Napiwotzki} R., {Burleigh} M.~R. (eds.) ASP Conf. Ser. 372: 15th European Workshop on White Dwarfs. San Francisco: Astronomical Society of the Pacific, p.~19

\bibitem[{{Buchan} et~al.(2022){Buchan}, {Bonsor}, {Shorttle}, {Wade}, {Harrison}, {Noack}, \& {Koester}}]{buchan2022}
{Buchan} A.~M., {Bonsor} A., {Shorttle} O., et~al. (2022): {Planets or asteroids? A geochemical method to constrain the masses of White Dwarf pollutants}. \mnras\, 510(3):3512--3530. \doi{10.1093/mnras/stab3624}, {\href{https://arxiv.org/abs/2111.08779}{{arXiv:2111.08779}}} {[astro-ph.EP]}

\bibitem[{{Caiazzo} et~al.(2023){Caiazzo}, {Burdge}, {Tremblay}, {Fuller}, {Ferrario}, {G{\"a}nsicke}, {Hermes}, {Heyl}, {Kawka}, {Kulkarni}, {Marsh}, {Mr{\'o}z}, {Prince}, {Richer}, {Rodriguez}, {van Roestel}, {Vanderbosch}, {Vennes}, {Wickramasinghe}, {Dhillon}, {Littlefair}, {Munday}, {Pelisoli}, {Perley}, {Bellm}, {Breedt}, {Brown}, {Dekany}, {Drake}, {Dyer}, {Graham}, {Green}, {Laher}, {Kerry}, {Parsons}, {Riddle}, {Rusholme}, \& {Sahman}}]{caiazzo2023}
{Caiazzo} I., {Burdge} K.~B., {Tremblay} P.-E., et~al. (2023): {A rotating white dwarf shows different compositions on its opposite faces}. \nat\, 620(7972):61--66. \doi{10.1038/s41586-023-06171-9}, {\href{https://arxiv.org/abs/2308.07430}{{arXiv:2308.07430}}} {[astro-ph.SR]}

\bibitem[{{Camisassa} et~al.(2023){Camisassa}, {Torres}, {Hollands}, {Koester}, {Raddi}, {Althaus}, \& {Rebassa-Mansergas}}]{camisassa2023}
{Camisassa} M., {Torres} S., {Hollands} M., et~al. (2023): {A hidden population of white dwarfs with atmospheric carbon traces in the Gaia bifurcation}. \aap\, 674:A213. \doi{10.1051/0004-6361/202346628}, {\href{https://arxiv.org/abs/2305.02110}{{arXiv:2305.02110}}} {[astro-ph.SR]}

\bibitem[{{Camisassa} et~al.(2016){Camisassa}, {Althaus}, {C{\'o}rsico}, {Vinyoles}, {Serenelli}, {Isern}, {Miller Bertolami}, \& {Garc{\'\i}a{\textendash}Berro}}]{camisassa2016}
{Camisassa} M.~E., {Althaus} L.~G., {C{\'o}rsico} A.~H., et~al. (2016): {The Effect of $^{22}$NE Diffusion in the Evolution and Pulsational Properties of White Dwarfs with Solar Metallicity Progenitors}. \apj\, 823(2):158. \doi{10.3847/0004-637X/823/2/158}, {\href{https://arxiv.org/abs/1604.01744}{{arXiv:1604.01744}}} {[astro-ph.SR]}

\bibitem[{{Camisassa} et~al.(2017){Camisassa}, {Althaus}, {Rohrmann}, {Garc{\'\i}a-Berro}, {Torres}, {C{\'o}rsico}, \& {Wachlin}}]{camisassa2017}
{Camisassa} M.~E., {Althaus} L.~G., {Rohrmann} R.~D., et~al. (2017): {Updated Evolutionary Sequences for Hydrogen-deficient White Dwarfs}. \apj\, 839(1):11. \doi{10.3847/1538-4357/aa6797}, {\href{https://arxiv.org/abs/1703.05340}{{arXiv:1703.05340}}} {[astro-ph.SR]}

\bibitem[{{Caron} et~al.(2023){Caron}, {Bergeron}, {Blouin}, \& {Leggett}}]{caron2023}
{Caron} A., {Bergeron} P., {Blouin} S., et~al. (2023): {A spectrophotometric analysis of cool white dwarfs in the Gaia and pan-STARRS footprint}. \mnras\, 519(3):4529--4549. \doi{10.1093/mnras/stac3733}, {\href{https://arxiv.org/abs/2212.08014}{{arXiv:2212.08014}}} {[astro-ph.SR]}

\bibitem[{{Cenarro} et~al.(2019){Cenarro}, {Moles}, {Crist{\'o}bal-Hornillos}, {Mar{\'\i}n-Franch}, {Ederoclite}, {Varela}, {L{\'o}pez-Sanjuan}, {Hern{\'a}ndez-Monteagudo}, {Angulo}, {V{\'a}zquez Rami{\'o}}, {Viironen}, {Bonoli}, {Orsi}, {Hurier}, {San Roman}, {Greisel}, {Vilella-Rojo}, {D{\'\i}az-Garc{\'\i}a}, {Logro{\~n}o-Garc{\'\i}a}, {Gurung-L{\'o}pez}, {Spinoso}, {Izquierdo-Villalba}, {Aguerri}, {Allende Prieto}, {Bonatto}, {Carvano}, {Chies-Santos}, {Daflon}, {Dupke}, {Falc{\'o}n-Barroso}, {Gon{\c{c}}alves}, {Jim{\'e}nez-Teja}, {Molino}, {Placco}, {Solano}, {Whitten}, {Abril}, {Ant{\'o}n}, {Bello}, {Bielsa de Toledo}, {Castillo-Ram{\'\i}rez}, {Chueca}, {Civera}, {D{\'\i}az-Mart{\'\i}n}, {Dom{\'\i}nguez-Mart{\'\i}nez}, {Garzar{\'a}n-Calderaro}, {Hern{\'a}ndez-Fuertes}, {Iglesias-Marzoa}, {I{\~n}iguez}, {Jim{\'e}nez Ruiz}, {Kruuse}, {Lamadrid}, {Lasso-Cabrera}, {L{\'o}pez-Alegre}, {L{\'o}pez-Sainz}, {Ma{\'\i}cas}, {Moreno-Signes}, {Muniesa}, {Rodr{\'\i}guez-Llano}, {Rueda-Teruel}, {Rueda-Teruel},
  {Soriano-Lagu{\'\i}a}, {Tilve}, {Valdivielso}, {Yanes-D{\'\i}az}, {Alcaniz}, {Mendes de Oliveira}, {Sodr{\'e}}, {Coelho}, {Lopes de Oliveira}, {Tamm}, {Xavier}, {Abramo}, {Akras}, {Alfaro}, {Alvarez-Candal}, {Ascaso}, {Beasley}, {Beers}, {Borges Fernandes}, {Bruzual}, {Buzzo}, {Carrasco}, {Cepa}, {Cortesi}, {Costa-Duarte}, {De Pr{\'a}}, {Favole}, {Galarza}, {Galbany}, {Garcia}, {Gonz{\'a}lez Delgado}, {Gonz{\'a}lez-Serrano}, {Guti{\'e}rrez-Soto}, {Hernandez-Jimenez}, {Kanaan}, {Kuncarayakti}, {Landim}, {Laur}, {Licandro}, {Lima Neto}, {Lyman}, {Ma{\'\i}z Apell{\'a}niz}, {Miralda-Escud{\'e}}, {Morate}, {Nogueira-Cavalcante}, {Novais}, {Oncins}, {Oteo}, {Overzier}, {Pereira}, {Rebassa-Mansergas}, {Reis}, {Roig}, {Sako}, {Salvador-Rusi{\~n}ol}, {Sampedro}, {S{\'a}nchez-Bl{\'a}zquez}, {Santos}, {Schmidtobreick}, {Siffert}, {Telles}, \& {Vilchez}}]{cenarro2019}
{Cenarro} A.~J., {Moles} M., {Crist{\'o}bal-Hornillos} D., et~al. (2019): {J-PLUS: The Javalambre Photometric Local Universe Survey}. \aap\, 622:A176. \doi{10.1051/0004-6361/201833036}, {\href{https://arxiv.org/abs/1804.02667}{{arXiv:1804.02667}}} {[astro-ph.GA]}

\bibitem[{{Chambers} et~al.(2016){Chambers}, {Magnier}, {Metcalfe}, {Flewelling}, {Huber}, {Waters}, {Denneau}, {Draper}, {Farrow}, {Finkbeiner}, {Holmberg}, {Koppenhoefer}, {Price}, {Rest}, {Saglia}, {Schlafly}, {Smartt}, {Sweeney}, {Wainscoat}, {Burgett}, {Chastel}, {Grav}, {Heasley}, {Hodapp}, {Jedicke}, {Kaiser}, {Kudritzki}, {Luppino}, {Lupton}, {Monet}, {Morgan}, {Onaka}, {Shiao}, {Stubbs}, {Tonry}, {White}, {Ba{\~n}ados}, {Bell}, {Bender}, {Bernard}, {Boegner}, {Boffi}, {Botticella}, {Calamida}, {Casertano}, {Chen}, {Chen}, {Cole}, {Deacon}, {Frenk}, {Fitzsimmons}, {Gezari}, {Gibbs}, {Goessl}, {Goggia}, {Gourgue}, {Goldman}, {Grant}, {Grebel}, {Hambly}, {Hasinger}, {Heavens}, {Heckman}, {Henderson}, {Henning}, {Holman}, {Hopp}, {Ip}, {Isani}, {Jackson}, {Keyes}, {Koekemoer}, {Kotak}, {Le}, {Liska}, {Long}, {Lucey}, {Liu}, {Martin}, {Masci}, {McLean}, {Mindel}, {Misra}, {Morganson}, {Murphy}, {Obaika}, {Narayan}, {Nieto-Santisteban}, {Norberg}, {Peacock}, {Pier}, {Postman}, {Primak}, {Rae}, {Rai},
  {Riess}, {Riffeser}, {Rix}, {R{\"o}ser}, {Russel}, {Rutz}, {Schilbach}, {Schultz}, {Scolnic}, {Strolger}, {Szalay}, {Seitz}, {Small}, {Smith}, {Soderblom}, {Taylor}, {Thomson}, {Taylor}, {Thakar}, {Thiel}, {Thilker}, {Unger}, {Urata}, {Valenti}, {Wagner}, {Walder}, {Walter}, {Watters}, {Werner}, {Wood-Vasey}, \& {Wyse}}]{chambers2016}
{Chambers} K.~C., {Magnier} E.~A., {Metcalfe} N., et~al. (2016): {The Pan-STARRS1 Surveys}. arXiv e-prints\, arXiv:1612.05560. \doi{10.48550/arXiv.1612.05560}, {\href{https://arxiv.org/abs/1612.05560}{{arXiv:1612.05560}}} {[astro-ph.IM]}

\bibitem[{{Chayer}(2014)}]{chayer2014}
{Chayer} P. (2014): {Radiative levitation of silicon in the atmospheres of two Hyades DA white dwarfs}. \mnras\, 437(1):L95--L99. \doi{10.1093/mnrasl/slt149}, {\href{https://arxiv.org/abs/1310.6245}{{arXiv:1310.6245}}} {[astro-ph.SR]}

\bibitem[{{Chayer} et~al.(1995{\natexlab{a}}){Chayer}, {Fontaine}, \& {Wesemael}}]{chayer1995a}
{Chayer} P., {Fontaine} G., {Wesemael} F. (1995{\natexlab{a}}): {Radiative Levitation in Hot White Dwarfs: Equilibrium Theory}. \apjs\, 99:189. \doi{10.1086/192184}

\bibitem[{{Chayer} et~al.(1995{\natexlab{b}}){Chayer}, {Vennes}, {Pradhan}, {Thejll}, {Beauchamp}, {Fontaine}, \& {Wesemael}}]{chayer1995b}
{Chayer} P., {Vennes} S., {Pradhan} A.~K., et~al. (1995{\natexlab{b}}): {Improved Calculations of the Equilibrium Abundances of Heavy Elements Supported by Radiative Levitation in the Atmospheres of Hot DA White Dwarfs}. \apj\, 454:429. \doi{10.1086/176494}

\bibitem[{{Chayer} et~al.(2005){Chayer}, {Vennes}, {Dupuis}, \& {Kruk}}]{chayer2005}
{Chayer} P., {Vennes} S., {Dupuis} J., et~al. (2005): {Abundance of Elements beyond the Iron Group in Cool DO White Dwarfs}. \apjl\, 630(2):L169--L172. \doi{10.1086/491699}

\bibitem[{{Chayer} et~al.(2023){Chayer}, {Mendoza}, {Mel{\'e}ndez}, {Deprince}, \& {Dupuis}}]{chayer2023}
{Chayer} P., {Mendoza} C., {Mel{\'e}ndez} M., et~al. (2023): {Detection of cesium in the atmosphere of the hot He-rich white dwarf HD 149499B}. \mnras\, 518(1):368--381. \doi{10.1093/mnras/stac3138}, {\href{https://arxiv.org/abs/2211.01868}{{arXiv:2211.01868}}} {[astro-ph.SR]}

\bibitem[{{Chen} \& {Hansen}(2011)}]{chen2011}
{Chen} E.~Y., {Hansen} B. M.~S. (2011): {Cooling curves and chemical evolution curves of convective mixing white dwarf stars}. \mnras\, 413(4):2827--2837. \doi{10.1111/j.1365-2966.2011.18355.x}

\bibitem[{{Chen} \& {Hansen}(2012)}]{chen2012}
{Chen} E.~Y., {Hansen} B. M.~S. (2012): {The Spectral Evolution of Convective Mixing White Dwarfs, the Non-DA Gap, and White Dwarf Cosmochronology}. \apjl\, 753(1):L16. \doi{10.1088/2041-8205/753/1/L16}, {\href{https://arxiv.org/abs/1205.7068}{{arXiv:1205.7068}}} {[astro-ph.SR]}

\bibitem[{{Chen} et~al.(2021){Chen}, {Ferraro}, {Cadelano}, {Salaris}, {Lanzoni}, {Pallanca}, {Althaus}, \& {Dalessandro}}]{chen2021}
{Chen} J., {Ferraro} F.~R., {Cadelano} M., et~al. (2021): {Slowly cooling white dwarfs in M13 from stable hydrogen burning}. Nature Astronomy\, 5:1170--1177. \doi{10.1038/s41550-021-01445-6}, {\href{https://arxiv.org/abs/2109.02306}{{arXiv:2109.02306}}} {[astro-ph.GA]}

\bibitem[{{Cheng} et~al.(2019){Cheng}, {Cummings}, \& {M{\'e}nard}}]{cheng2019}
{Cheng} S., {Cummings} J.~D., {M{\'e}nard} B. (2019): {A Cooling Anomaly of High-mass White Dwarfs}. \apj\, 886(2):100. \doi{10.3847/1538-4357/ab4989}, {\href{https://arxiv.org/abs/1905.12710}{{arXiv:1905.12710}}} {[astro-ph.SR]}

\bibitem[{{Coutu} et~al.(2019){Coutu}, {Dufour}, {Bergeron}, {Blouin}, {Loranger}, {Allard}, \& {Dunlap}}]{coutu2019}
{Coutu} S., {Dufour} P., {Bergeron} P., et~al. (2019): {Analysis of Helium-rich White Dwarfs Polluted by Heavy Elements in the Gaia Era}. \apj\, 885(1):74. \doi{10.3847/1538-4357/ab46b9}, {\href{https://arxiv.org/abs/1907.05932}{{arXiv:1907.05932}}} {[astro-ph.SR]}

\bibitem[{{Cukanovaite} et~al.(2018){Cukanovaite}, {Tremblay}, {Freytag}, {Ludwig}, \& {Bergeron}}]{cukanovaite2018}
{Cukanovaite} E., {Tremblay} P.~E., {Freytag} B., et~al. (2018): {Pure-helium 3D model atmospheres of white dwarfs}. \mnras\, 481(2):1522--1537. \doi{10.1093/mnras/sty2383}, {\href{https://arxiv.org/abs/1809.00590}{{arXiv:1809.00590}}} {[astro-ph.SR]}

\bibitem[{{Cukanovaite} et~al.(2019){Cukanovaite}, {Tremblay}, {Freytag}, {Ludwig}, {Fontaine}, {Brassard}, {Toloza}, \& {Koester}}]{cukanovaite2019}
{Cukanovaite} E., {Tremblay} P.~E., {Freytag} B., et~al. (2019): {Calibration of the mixing-length theory for structures of helium-dominated atmosphere white dwarfs}. \mnras\, 490(1):1010--1025. \doi{10.1093/mnras/stz2656}, {\href{https://arxiv.org/abs/1909.10532}{{arXiv:1909.10532}}} {[astro-ph.SR]}

\bibitem[{{Cukanovaite} et~al.(2021){Cukanovaite}, {Tremblay}, {Bergeron}, {Freytag}, {Ludwig}, \& {Steffen}}]{cukanovaite2021}
{Cukanovaite} E., {Tremblay} P.-E., {Bergeron} P., et~al. (2021): {3D spectroscopic analysis of helium-line white dwarfs}. \mnras\, 501(4):5274--5293. \doi{10.1093/mnras/staa3684}, {\href{https://arxiv.org/abs/2011.12693}{{arXiv:2011.12693}}} {[astro-ph.SR]}

\bibitem[{{Cukanovaite} et~al.(2023){Cukanovaite}, {Tremblay}, {Toonen}, {Temmink}, {Manser}, {O'Brien}, \& {McCleery}}]{cukanovaite2023}
{Cukanovaite} E., {Tremblay} P.~E., {Toonen} S., et~al. (2023): {Local stellar formation history from the 40 pc white dwarf sample}. \mnras\, 522(2):1643--1661. \doi{10.1093/mnras/stad1020}, {\href{https://arxiv.org/abs/2209.13919}{{arXiv:2209.13919}}} {[astro-ph.SR]}

\bibitem[{{Cummings} et~al.(2018){Cummings}, {Kalirai}, {Tremblay}, {Ramirez-Ruiz}, \& {Choi}}]{cummings2018}
{Cummings} J.~D., {Kalirai} J.~S., {Tremblay} P.~E., et~al. (2018): {The White Dwarf Initial-Final Mass Relation for Progenitor Stars from 0.85 to 7.5 M $_{{\ensuremath{\odot}}}$}. \apj\, 866(1):21. \doi{10.3847/1538-4357/aadfd6}, {\href{https://arxiv.org/abs/1809.01673}{{arXiv:1809.01673}}} {[astro-ph.SR]}

\bibitem[{{Cunningham} et~al.(2019){Cunningham}, {Tremblay}, {Freytag}, {Ludwig}, \& {Koester}}]{cunningham2019}
{Cunningham} T., {Tremblay} P.-E., {Freytag} B., et~al. (2019): {Convective overshoot and macroscopic diffusion in pure-hydrogen-atmosphere white dwarfs}. \mnras\, 488(2):2503--2522. \doi{10.1093/mnras/stz1759}, {\href{https://arxiv.org/abs/1906.11252}{{arXiv:1906.11252}}} {[astro-ph.SR]}

\bibitem[{{Cunningham} et~al.(2020){Cunningham}, {Tremblay}, {Gentile Fusillo}, {Hollands}, \& {Cukanovaite}}]{cunningham2020}
{Cunningham} T., {Tremblay} P.-E., {Gentile Fusillo} N.~P., et~al. (2020): {From hydrogen to helium: the spectral evolution of white dwarfs as evidence for convective mixing}. \mnras\, 492(3):3540--3552. \doi{10.1093/mnras/stz3638}, {\href{https://arxiv.org/abs/1911.00014}{{arXiv:1911.00014}}} {[astro-ph.SR]}

\bibitem[{{Dalton} et~al.(2012){Dalton}, {Trager}, {Abrams}, {Carter}, {Bonifacio}, {Aguerri}, {MacIntosh}, {Evans}, {Lewis}, {Navarro}, {Agocs}, {Dee}, {Rousset}, {Tosh}, {Middleton}, {Pragt}, {Terrett}, {Brock}, {Benn}, {Verheijen}, {Cano Infantes}, {Bevil}, {Steele}, {Mottram}, {Bates}, {Gribbin}, {Rey}, {Rodriguez}, {Delgado}, {Guinouard}, {Walton}, {Irwin}, {Jagourel}, {Stuik}, {Gerlofsma}, {Roelfsma}, {Skillen}, {Ridings}, {Balcells}, {Daban}, {Gouvret}, {Venema}, \& {Girard}}]{dalton2012}
{Dalton} G., {Trager} S.~C., {Abrams} D.~C., et~al. (2012): {WEAVE: the next generation wide-field spectroscopy facility for the William Herschel Telescope}. In: {McLean} I.~S., {Ramsay} S.~K., {Takami} H. (eds.) Ground-based and Airborne Instrumentation for Astronomy IV, vol. 8446. Society of Photo-Optical Instrumentation Engineers, p. 84460P

\bibitem[{{D'Antona} \& {Mazzitelli}(1989)}]{dantona1989}
{D'Antona} F., {Mazzitelli} I. (1989): {The Fastest Evolving White Dwarfs}. \apj\, 347:934. \doi{10.1086/168185}

\bibitem[{{de Jong} et~al.(2014){de Jong}, {Barden}, {Bellido-Tirado}, {Brynnel}, {Chiappini}, {Depagne}, {Haynes}, {Johl}, {Phillips}, {Schnurr}, {Schwope}, {Walcher}, {Bauer}, {Cescutti}, {Cioni}, {Dionies}, {Enke}, {Haynes}, {Kelz}, {Kitaura}, {Lamer}, {Minchev}, {M{\"u}ller}, {Nuza}, {Olaya}, {Piffl}, {Popow}, {Saviauk}, {Steinmetz}, {Ural}, {Valentini}, {Winkler}, {Wisotzki}, {Ansorge}, {Banerji}, {Gonzalez Solares}, {Irwin}, {Kennicutt}, {King}, {McMahon}, {Koposov}, {Parry}, {Sun}, {Walton}, {Finger}, {Iwert}, {Krumpe}, {Lizon}, {Mainieri}, {Amans}, {Bonifacio}, {Cohen}, {Fran{\c{c}}ois}, {Jagourel}, {Mignot}, {Royer}, {Sartoretti}, {Bender}, {Hess}, {Lang-Bardl}, {Muschielok}, {Schlichter}, {B{\"o}hringer}, {Boller}, {Bongiorno}, {Brusa}, {Dwelly}, {Merloni}, {Nandra}, {Salvato}, {Pragt}, {Navarro}, {Gerlofsma}, {Roelfsema}, {Dalton}, {Middleton}, {Tosh}, {Boeche}, {Caffau}, {Christlieb}, {Grebel}, {Hansen}, {Koch}, {Ludwig}, {Mandel}, {Quirrenbach}, {Sbordone}, {Seifert}, {Thimm}, {Helmi}, {trager},
  {Bensby}, {Feltzing}, {Ruchti}, {Edvardsson}, {Korn}, {Lind}, {Boland}, {Colless}, {Frost}, {Gilbert}, {Gillingham}, {Lawrence}, {Legg}, {Saunders}, {Sheinis}, {Driver}, {Robotham}, {Bacon}, {Caillier}, {Kosmalski}, {Laurent}, \& {Richard}}]{dejong2014}
{de Jong} R.~S., {Barden} S., {Bellido-Tirado} O., et~al. (2014): {4MOST: 4-metre Multi-Object Spectroscopic Telescope}. In: {Ramsay} S.~K., {McLean} I.~S., {Takami} H. (eds.) Ground-based and Airborne Instrumentation for Astronomy V, vol. 9147. Society of Photo-Optical Instrumentation Engineers, p. 91470M

\bibitem[{{Deal} et~al.(2013){Deal}, {Deheuvels}, {Vauclair}, {Vauclair}, \& {Wachlin}}]{deal2013}
{Deal} M., {Deheuvels} S., {Vauclair} G., et~al. (2013): {Accretion from debris disks onto white dwarfs. Fingering (thermohaline) instability and derived accretion rates}. \aap\, 557:L12. \doi{10.1051/0004-6361/201322206}, {\href{https://arxiv.org/abs/1308.5406}{{arXiv:1308.5406}}} {[astro-ph.SR]}

\bibitem[{{Debes} \& {Sigurdsson}(2002)}]{debes2002}
{Debes} J.~H., {Sigurdsson} S. (2002): {Are There Unstable Planetary Systems around White Dwarfs?} \apj\, 572(1):556--565. \doi{10.1086/340291}, {\href{https://arxiv.org/abs/astro-ph/0202273}{{arXiv:astro-ph/0202273}}} {[astro-ph]}

\bibitem[{{Dehner} \& {Kawaler}(1995)}]{dehner1995}
{Dehner} B.~T., {Kawaler} S.~D. (1995): {Thick to Thin: The Evolutionary Connection between PG 1159 Stars and the Thin Helium--enveloped Pulsating White Dwarf GD 358}. \apjl\, 445:L141. \doi{10.1086/187909}, {\href{https://arxiv.org/abs/astro-ph/9503099}{{arXiv:astro-ph/9503099}}} {[astro-ph]}

\bibitem[{{DESI Collaboration} et~al.(2016){DESI Collaboration}, {Aghamousa}, {Aguilar}, {Ahlen}, {Alam}, {Allen}, {Allende Prieto}, {Annis}, {Bailey}, {Balland}, {Ballester}, {Baltay}, {Beaufore}, {Bebek}, {Beers}, {Bell}, {Bernal}, {Besuner}, {Beutler}, {Blake}, {Bleuler}, {Blomqvist}, {Blum}, {Bolton}, {Briceno}, {Brooks}, {Brownstein}, {Buckley-Geer}, {Burden}, {Burtin}, {Busca}, {Cahn}, {Cai}, {Cardiel-Sas}, {Carlberg}, {Carton}, {Casas}, {Castander}, {Cervantes-Cota}, {Claybaugh}, {Close}, {Coker}, {Cole}, {Comparat}, {Cooper}, {Cousinou}, {Crocce}, {Cuby}, {Cunningham}, {Davis}, {Dawson}, {de la Macorra}, {De Vicente}, {Delubac}, {Derwent}, {Dey}, {Dhungana}, {Ding}, {Doel}, {Duan}, {Ealet}, {Edelstein}, {Eftekharzadeh}, {Eisenstein}, {Elliott}, {Escoffier}, {Evatt}, {Fagrelius}, {Fan}, {Fanning}, {Farahi}, {Farihi}, {Favole}, {Feng}, {Fernandez}, {Findlay}, {Finkbeiner}, {Fitzpatrick}, {Flaugher}, {Flender}, {Font-Ribera}, {Forero-Romero}, {Fosalba}, {Frenk}, {Fumagalli}, {Gaensicke}, {Gallo},
  {Garcia-Bellido}, {Gaztanaga}, {Pietro Gentile Fusillo}, {Gerard}, {Gershkovich}, {Giannantonio}, {Gillet}, {Gonzalez-de-Rivera}, {Gonzalez-Perez}, {Gott}, {Graur}, {Gutierrez}, {Guy}, {Habib}, {Heetderks}, {Heetderks}, {Heitmann}, {Hellwing}, {Herrera}, {Ho}, {Holland}, {Honscheid}, {Huff}, {Hutchinson}, {Huterer}, {Hwang}, {Illa Laguna}, {Ishikawa}, {Jacobs}, {Jeffrey}, {Jelinsky}, {Jennings}, {Jiang}, {Jimenez}, {Johnson}, {Joyce}, {Jullo}, {Juneau}, {Kama}, {Karcher}, {Karkar}, {Kehoe}, {Kennamer}, {Kent}, {Kilbinger}, {Kim}, {Kirkby}, {Kisner}, {Kitanidis}, {Kneib}, {Koposov}, {Kovacs}, {Koyama}, {Kremin}, {Kron}, {Kronig}, {Kueter-Young}, {Lacey}, {Lafever}, {Lahav}, {Lambert}, {Lampton}, {Landriau}, {Lang}, {Lauer}, {Le Goff}, {Le Guillou}, {Le Van Suu}, {Lee}, {Lee}, {Leitner}, {Lesser}, {Levi}, {L'Huillier}, {Li}, {Liang}, {Lin}, {Linder}, {Loebman}, {Luki{\'c}}, {Ma}, {MacCrann}, {Magneville}, {Makarem}, {Manera}, {Manser}, {Marshall}, {Martini}, {Massey}, {Matheson}, {McCauley}, {McDonald},
  {McGreer}, {Meisner}, {Metcalfe}, {Miller}, {Miquel}, {Moustakas}, {Myers}, {Naik}, {Newman}, {Nichol}, {Nicola}, {Nicolati da Costa}, {Nie}, {Niz}, {Norberg}, {Nord}, {Norman}, {Nugent}, {O'Brien}, {Oh}, {Olsen}, {Padilla}, {Padmanabhan}, {Padmanabhan}, {Palanque-Delabrouille}, {Palmese}, {Pappalardo}, {P{\^a}ris}, {Park}, {Patej}, {Peacock}, {Peiris}, {Peng}, {Percival}, {Perruchot}, {Pieri}, {Pogge}, {Pollack}, {Poppett}, {Prada}, {Prakash}, {Probst}, {Rabinowitz}, {Raichoor}, {Ree}, {Refregier}, {Regal}, {Reid}, {Reil}, {Rezaie}, {Rockosi}, {Roe}, {Ronayette}, {Roodman}, {Ross}, {Ross}, {Rossi}, {Rozo}, {Ruhlmann-Kleider}, {Rykoff}, {Sabiu}, {Samushia}, {Sanchez}, {Sanchez}, {Schlegel}, {Schneider}, {Schubnell}, {Secroun}, {Seljak}, {Seo}, {Serrano}, {Shafieloo}, {Shan}, {Sharples}, {Sholl}, {Shourt}, {Silber}, {Silva}, {Sirk}, {Slosar}, {Smith}, {Smoot}, {Som}, {Song}, {Sprayberry}, {Staten}, {Stefanik}, {Tarle}, {Sien Tie}, {Tinker}, {Tojeiro}, {Valdes}, {Valenzuela}, {Valluri}, {Vargas-Magana},
  {Verde}, {Walker}, {Wang}, {Wang}, {Weaver}, {Weaverdyck}, {Wechsler}, {Weinberg}, {White}, {Yang}, {Yeche}, {Zhang}, {Zhao}, {Zheng}, {Zhou}, {Zhou}, {Zhu}, {Zou}, \& {Zu}}]{desi2016}
{DESI Collaboration}, {Aghamousa} A., {Aguilar} J., et~al. (2016): {The DESI Experiment Part I: Science,Targeting, and Survey Design}. arXiv e-prints\, arXiv:1611.00036. {\href{https://arxiv.org/abs/1611.00036}{{arXiv:1611.00036}}} {[astro-ph.IM]}

\bibitem[{{Doyle} et~al.(2023){Doyle}, {Klein}, {Dufour}, {Melis}, {Zuckerman}, {Xu}, {Weinberger}, {Trierweiler}, {Monson}, {Jura}, \& {Young}}]{doyle2023}
{Doyle} A.~E., {Klein} B.~L., {Dufour} P., et~al. (2023): {New Chondritic Bodies Identified in Eight Oxygen-bearing White Dwarfs}. \apj\, 950(2):93. \doi{10.3847/1538-4357/acbd44}, {\href{https://arxiv.org/abs/2303.00063}{{arXiv:2303.00063}}} {[astro-ph.SR]}

\bibitem[{{Dreizler}(1999)}]{dreizler1999b}
{Dreizler} S. (1999): {Hubble Space Telescope spectroscopy of hot helium-rich white dwarfs: metal abundances along the cooling sequence}. \aap\, 352:632--644

\bibitem[{{Dreizler} \& {Heber}(1998)}]{dreizler1998}
{Dreizler} S., {Heber} U. (1998): {Spectral analyses of PG 1159 star: constraints on the GW Virginis pulsations from HST observations}. \aap\, 334:618--632

\bibitem[{{Dreizler} \& {Werner}(1996)}]{dreizler1996}
{Dreizler} S., {Werner} K. (1996): {Spectral analysis of hot helium-rich white dwarfs.} \aap\, 314:217--232

\bibitem[{{Dreizler} \& {Wolff}(1999)}]{dreizler1999a}
{Dreizler} S., {Wolff} B. (1999): {Analysis of ultraviolet and extreme-ultraviolet spectra of the DA white dwarf G 191-B2B using self-consistent diffusion models}. \aap\, 348:189--197

\bibitem[{{Dufour}(2011)}]{dufour2011}
{Dufour} P. (2011): {Stars with Unusual Compositions: Carbon and Oxygen in Cool White Dwarfs}. In: {Hoard} D.~W. (ed.) White Dwarf Atmospheres and Circumstellar Environments. New York: Wiley, p. 53--88

\bibitem[{{Dufour} et~al.(2005){Dufour}, {Bergeron}, \& {Fontaine}}]{dufour2005}
{Dufour} P., {Bergeron} P., {Fontaine} G. (2005): {Detailed Spectroscopic and Photometric Analysis of DQ White Dwarfs}. \apj\, 627(1):404--417. \doi{10.1086/430373}, {\href{https://arxiv.org/abs/astro-ph/0503112}{{arXiv:astro-ph/0503112}}} {[astro-ph]}

\bibitem[{{Dufour} et~al.(2007{\natexlab{a}}){Dufour}, {Bergeron}, {Liebert}, {Harris}, {Knapp}, {Anderson}, {Hall}, {Strauss}, {Collinge}, \& {Edwards}}]{dufour2007a}
{Dufour} P., {Bergeron} P., {Liebert} J., et~al. (2007{\natexlab{a}}): {On the Spectral Evolution of Cool, Helium-Atmosphere White Dwarfs: Detailed Spectroscopic and Photometric Analysis of DZ Stars}. \apj\, 663(2):1291--1308. \doi{10.1086/518468}, {\href{https://arxiv.org/abs/astro-ph/0703758}{{arXiv:astro-ph/0703758}}} {[astro-ph]}

\bibitem[{{Dufour} et~al.(2007{\natexlab{b}}){Dufour}, {Liebert}, {Fontaine}, \& {Behara}}]{dufour2007b}
{Dufour} P., {Liebert} J., {Fontaine} G., et~al. (2007{\natexlab{b}}): {White dwarf stars with carbon atmospheres}. \nat\, 450(7169):522--524. \doi{10.1038/nature06318}, {\href{https://arxiv.org/abs/0711.3227}{{arXiv:0711.3227}}} {[astro-ph]}

\bibitem[{{Dufour} et~al.(2008){Dufour}, {Fontaine}, {Liebert}, {Schmidt}, \& {Behara}}]{dufour2008}
{Dufour} P., {Fontaine} G., {Liebert} J., et~al. (2008): {Hot DQ White Dwarfs: Something Different}. \apj\, 683(2):978--989. \doi{10.1086/589855}, {\href{https://arxiv.org/abs/0805.0331}{{arXiv:0805.0331}}} {[astro-ph]}

\bibitem[{{Dufour} et~al.(2010{\natexlab{a}}){Dufour}, {Desharnais}, {Wesemael}, {Chayer}, {Lanz}, {Bergeron}, {Fontaine}, {Beauchamp}, {Saffer}, {Kruk}, \& {Limoges}}]{dufour2010a}
{Dufour} P., {Desharnais} S., {Wesemael} F., et~al. (2010{\natexlab{a}}): {Multiwavelength Observations of the Hot DB Star PG 0112+104}. \apj\, 718(2):647--656. \doi{10.1088/0004-637X/718/2/647}, {\href{https://arxiv.org/abs/1006.0365}{{arXiv:1006.0365}}} {[astro-ph.SR]}

\bibitem[{{Dufour} et~al.(2010{\natexlab{b}}){Dufour}, {Kilic}, {Fontaine}, {Bergeron}, {Lachapelle}, {Kleinman}, \& {Leggett}}]{dufour2010b}
{Dufour} P., {Kilic} M., {Fontaine} G., et~al. (2010{\natexlab{b}}): {The Discovery of the Most Metal-rich White Dwarf: Composition of a Tidally Disrupted Extrasolar Dwarf Planet}. \apj\, 719(1):803--809. \doi{10.1088/0004-637X/719/1/803}, {\href{https://arxiv.org/abs/1006.3710}{{arXiv:1006.3710}}} {[astro-ph.SR]}

\bibitem[{{Dufour} et~al.(2012){Dufour}, {Kilic}, {Fontaine}, {Bergeron}, {Melis}, \& {Bochanski}}]{dufour2012}
{Dufour} P., {Kilic} M., {Fontaine} G., et~al. (2012): {Detailed Compositional Analysis of the Heavily Polluted DBZ White Dwarf SDSS J073842.56+183509.06: A Window on Planet Formation?} \apj\, 749(1):6. \doi{10.1088/0004-637X/749/1/6}, {\href{https://arxiv.org/abs/1201.6252}{{arXiv:1201.6252}}} {[astro-ph.SR]}

\bibitem[{{Dufour} et~al.(2013){Dufour}, {Vornanen}, {Bergeron}, \& {Fontaine}}]{dufour2013}
{Dufour} P., {Vornanen} T., {Bergeron} P., et~al. (2013): {White Dwarfs with Carbon Dominated Atmosphere: New Observations and Analysis}. In: {Krzesi{\'n}ski} J., {Stachowski} G., {Moskalik} P., et~al. (eds.) ASP Conf. Ser. 469: 18th European White Dwarf Workshop. San Francisco: Astronomical Society of the Pacific, p. 167

\bibitem[{{Dufour} et~al.(2017){Dufour}, {Blouin}, {Coutu}, {Fortin-Archambault}, {Thibeault}, {Bergeron}, \& {Fontaine}}]{dufour2017}
{Dufour} P., {Blouin} S., {Coutu} S., et~al. (2017): {The Montreal White Dwarf Database: A Tool for the Community}. In: {Tremblay} P.~E., {G{\"a}nsicke} B., {Marsh} T. (eds.) ASP Conf. Ser. 509: 20th European Workshop on White Dwarfs. San Francisco: Astronomical Society of the Pacific, p.~3, \eprint{1610.00986}

\bibitem[{{Dunlap} \& {Clemens}(2015)}]{dunlap2015}
{Dunlap} B.~H., {Clemens} J.~C. (2015): {Hot DQ White Dwarf Stars as Failed Type Ia Supernovae}. In: {Dufour} P., {Bergeron} P., {Fontaine} G. (eds.) ASP Conf. Ser. 493: 19th European Workshop on White Dwarfs. San Francisco: Astronomical Society of the Pacific, p. 547

\bibitem[{{Dupuis} et~al.(1992){Dupuis}, {Fontaine}, {Pelletier}, \& {Wesemael}}]{dupuis1992}
{Dupuis} J., {Fontaine} G., {Pelletier} C., et~al. (1992): {A Study of Metal Abundance Patterns in Cool White Dwarfs. I. Time-dependent Calculations of Gravitational Settling}. \apjs\, 82:505. \doi{10.1086/191728}

\bibitem[{{Dupuis} et~al.(1993{\natexlab{a}}){Dupuis}, {Fontaine}, {Pelletier}, \& {Wesemael}}]{dupuis1993a}
{Dupuis} J., {Fontaine} G., {Pelletier} C., et~al. (1993{\natexlab{a}}): {A Study of Metal Abundance Patterns in Cool White Dwarfs. II. Simulations of Accretion Episodes}. \apjs\, 84:73. \doi{10.1086/191746}

\bibitem[{{Dupuis} et~al.(1993{\natexlab{b}}){Dupuis}, {Fontaine}, \& {Wesemael}}]{dupuis1993b}
{Dupuis} J., {Fontaine} G., {Wesemael} F. (1993{\natexlab{b}}): {A Study of Metal Abundance Patterns in Cool White Dwarfs. III. Comparison of the Predictions of the Two-Phase Accretion Model with the Observations}. \apjs\, 87:345. \doi{10.1086/191808}

\bibitem[{{Dwomoh} \& {Bauer}(2023)}]{dwomoh2023}
{Dwomoh} A.~M., {Bauer} E.~B. (2023): {Reinterpreting the Polluted White Dwarf SDSS J122859.93+104032.9 in Light of Thermohaline Mixing Models: More Polluting Material from a Larger Orbiting Solid Body}. \apj\, 952(2):95. \doi{10.3847/1538-4357/acdb69}, {\href{https://arxiv.org/abs/2306.03864}{{arXiv:2306.03864}}} {[astro-ph.SR]}

\bibitem[{{Eisenstein} et~al.(2006){Eisenstein}, {Liebert}, {Koester}, {Kleinmann}, {Nitta}, {Smith}, {Barentine}, {Brewington}, {Brinkmann}, {Harvanek}, {Krzesi{\'n}ski}, {Neilsen}, {Long}, {Schneider}, \& {Snedden}}]{eisenstein2006a}
{Eisenstein} D.~J., {Liebert} J., {Koester} D., et~al. (2006): {Hot DB White Dwarfs from the Sloan Digital Sky Survey}. \aj\, 132(2):676--691. \doi{10.1086/504424}, {\href{https://arxiv.org/abs/astro-ph/0606702}{{arXiv:astro-ph/0606702}}} {[astro-ph]}

\bibitem[{{El-Badry} et~al.(2018){El-Badry}, {Rix}, \& {Weisz}}]{elbadry2018}
{El-Badry} K., {Rix} H.-W., {Weisz} D.~R. (2018): {An Empirical Measurement of the Initial-Final Mass Relation with Gaia White Dwarfs}. \apjl\, 860(2):L17. \doi{10.3847/2041-8213/aaca9c}, {\href{https://arxiv.org/abs/1805.05849}{{arXiv:1805.05849}}} {[astro-ph.SR]}

\bibitem[{{Elms} et~al.(2022){Elms}, {Tremblay}, {G{\"a}nsicke}, {Koester}, {Hollands}, {Gentile Fusillo}, {Cunningham}, \& {Apps}}]{elms2022}
{Elms} A.~K., {Tremblay} P.-E., {G{\"a}nsicke} B.~T., et~al. (2022): {Spectral analysis of ultra-cool white dwarfs polluted by planetary debris}. \mnras\, 517(3):4557--4574. \doi{10.1093/mnras/stac2908}, {\href{https://arxiv.org/abs/2206.05258}{{arXiv:2206.05258}}} {[astro-ph.SR]}

\bibitem[{{Fantin} et~al.(2019){Fantin}, {C{\^o}t{\'e}}, {McConnachie}, {Bergeron}, {Cuillandre}, {Gwyn}, {Ibata}, {Thomas}, {Carlberg}, {Fabbro}, {Haywood}, {Lan{\c{c}}on}, {Lewis}, {Malhan}, {Martin}, {Navarro}, {Scott}, \& {Starkenburg}}]{fantin2019}
{Fantin} N.~J., {C{\^o}t{\'e}} P., {McConnachie} A.~W., et~al. (2019): {The Canada-France Imaging Survey: Reconstructing the Milky Way Star Formation History from Its White Dwarf Population}. \apj\, 887(2):148. \doi{10.3847/1538-4357/ab5521}, {\href{https://arxiv.org/abs/1911.02576}{{arXiv:1911.02576}}} {[astro-ph.GA]}

\bibitem[{{Farihi}(2016)}]{farihi2016a}
{Farihi} J. (2016): {Circumstellar debris and pollution at white dwarf stars}. \nar\, 71:9--34. \doi{10.1016/j.newar.2016.03.001}, {\href{https://arxiv.org/abs/1604.03092}{{arXiv:1604.03092}}} {[astro-ph.EP]}

\bibitem[{{Farihi} et~al.(2010){Farihi}, {Barstow}, {Redfield}, {Dufour}, \& {Hambly}}]{farihi2010}
{Farihi} J., {Barstow} M.~A., {Redfield} S., et~al. (2010): {Rocky planetesimals as the origin of metals in DZ stars}. \mnras\, 404(4):2123--2135. \doi{10.1111/j.1365-2966.2010.16426.x}, {\href{https://arxiv.org/abs/1001.5025}{{arXiv:1001.5025}}} {[astro-ph.EP]}

\bibitem[{{Farihi} et~al.(2011){Farihi}, {Brinkworth}, {G{\"a}nsicke}, {Marsh}, {Girven}, {Hoard}, {Klein}, \& {Koester}}]{farihi2011}
{Farihi} J., {Brinkworth} C.~S., {G{\"a}nsicke} B.~T., et~al. (2011): {Possible Signs of Water and Differentiation in a Rocky Exoplanetary Body}. \apjl\, 728(1):L8. \doi{10.1088/2041-8205/728/1/L8}, {\href{https://arxiv.org/abs/1101.0158}{{arXiv:1101.0158}}} {[astro-ph.EP]}

\bibitem[{{Farihi} et~al.(2013){Farihi}, {G{\"a}nsicke}, \& {Koester}}]{farihi2013}
{Farihi} J., {G{\"a}nsicke} B.~T., {Koester} D. (2013): {Evidence for Water in the Rocky Debris of a Disrupted Extrasolar Minor Planet}. Science\, 342(6155):218--220. \doi{10.1126/science.1239447}, {\href{https://arxiv.org/abs/1310.3269}{{arXiv:1310.3269}}} {[astro-ph.EP]}

\bibitem[{{Farihi} et~al.(2016){Farihi}, {Koester}, {Zuckerman}, {Vican}, {G{\"a}nsicke}, {Smith}, {Walth}, \& {Breedt}}]{farihi2016b}
{Farihi} J., {Koester} D., {Zuckerman} B., et~al. (2016): {Solar abundances of rock-forming elements, extreme oxygen and hydrogen in a young polluted white dwarf}. \mnras\, 463(3):3186--3192. \doi{10.1093/mnras/stw2182}, {\href{https://arxiv.org/abs/1608.07278}{{arXiv:1608.07278}}} {[astro-ph.EP]}

\bibitem[{{Farihi} et~al.(2022){Farihi}, {Dufour}, \& {Wilson}}]{farihi2022}
{Farihi} J., {Dufour} P., {Wilson} T.~G. (2022): {Missing Metals in DQ Stars; a Compelling Clue to their Origin}. arXiv e-prints\, arXiv:2208.05990. \doi{10.48550/arXiv.2208.05990}, {\href{https://arxiv.org/abs/2208.05990}{{arXiv:2208.05990}}} {[astro-ph.SR]}

\bibitem[{{Finley} et~al.(1997){Finley}, {Koester}, \& {Basri}}]{finley1997}
{Finley} D.~S., {Koester} D., {Basri} G. (1997): {The Temperature Scale and Mass Distribution of Hot DA White Dwarfs}. \apj\, 488(1):375--396. \doi{10.1086/304668}

\bibitem[{{Fleming} et~al.(1986){Fleming}, {Liebert}, \& {Green}}]{fleming1986}
{Fleming} T.~A., {Liebert} J., {Green} R.~F. (1986): {The Luminosity Function of DA White Dwarfs}. \apj\, 308:176. \doi{10.1086/164488}

\bibitem[{{Fontaine} \& {Brassard}(2002)}]{fontaine2002}
{Fontaine} G., {Brassard} P. (2002): {Can White Dwarf Asteroseismology Really Constrain the $^{12}$C({\ensuremath{\alpha}}, {\ensuremath{\gamma}})$^{16}$O Reaction Rate?} \apjl\, 581(1):L33--L37. \doi{10.1086/345787}

\bibitem[{{Fontaine} \& {Brassard}(2005)}]{fontaine2005}
{Fontaine} G., {Brassard} P. (2005): {Carbon in Hot DB White Dwarfs: A New Challenge for the Theory of the Spectral Evolution of White Dwarfs}. In: {Koester} D., {Moehler} S. (eds.) ASP Conf. Ser. 334: 14th European Workshop on White Dwarfs. San Francisco: Astronomical Society of the Pacific, p.~49

\bibitem[{{Fontaine} \& {Michaud}(1979)}]{fontaine1979}
{Fontaine} G., {Michaud} G. (1979): {Diffusion time scales in white dwarfs.} \apj\, 231:826--840. \doi{10.1086/157247}

\bibitem[{{Fontaine} \& {van Horn}(1976)}]{fontaine1976}
{Fontaine} G., {van Horn} H.~M. (1976): {Convective white-dwarf envelope model grids for H-, He-, and C-rich compositions.} \apjs\, 31:467--487. \doi{10.1086/190388}

\bibitem[{{Fontaine} \& {Wesemael}(1987)}]{fontaine1987}
{Fontaine} G., {Wesemael} F. (1987): {Recent advances in the theory of white dwarf spectral evolution.} In: {Philip} A.~G.~D., {Hayes} D.~S., {Liebert} J.~W. (eds.) IAU Colloq. 95: Second Conference on Faint Blue Stars. Schenectady: Davis Press, pp. 319--326

\bibitem[{{Fontaine} et~al.(1984){Fontaine}, {Villeneuve}, {Wesemael}, \& {Wegner}}]{fontaine1984}
{Fontaine} G., {Villeneuve} B., {Wesemael} F., et~al. (1984): {Carbon in the cool DC and C2 white dwarfs - Dredge-up in compositionally stratified envelopes}. \apjl\, 277:L61--L64. \doi{10.1086/184203}

\bibitem[{{Fontaine} et~al.(2001){Fontaine}, {Brassard}, \& {Bergeron}}]{fontaine2001}
{Fontaine} G., {Brassard} P., {Bergeron} P. (2001): {The Potential of White Dwarf Cosmochronology}. \pasp\, 113(782):409--435. \doi{10.1086/319535}

\bibitem[{{Fortin-Archambault} et~al.(2020){Fortin-Archambault}, {Dufour}, \& {Xu}}]{fortin-archambault2020}
{Fortin-Archambault} M., {Dufour} P., {Xu} S. (2020): {Modeling of the Variable Circumstellar Absorption Features of WD 1145+017}. \apj\, 888(1):47. \doi{10.3847/1538-4357/ab585a}, {\href{https://arxiv.org/abs/1911.05690}{{arXiv:1911.05690}}} {[astro-ph.SR]}

\bibitem[{{Gaia Collaboration} et~al.(2016){Gaia Collaboration}, {Prusti}, {de Bruijne}, {Brown}, {Vallenari}, {Babusiaux}, {Bailer-Jones}, {Bastian}, {Biermann}, {Evans}, {Eyer}, {Jansen}, {Jordi}, {Klioner}, {Lammers}, {Lindegren}, {Luri}, {Mignard}, {Milligan}, {Panem}, {Poinsignon}, {Pourbaix}, {Randich}, {Sarri}, {Sartoretti}, {Siddiqui}, {Soubiran}, {Valette}, {van Leeuwen}, {Walton}, {Aerts}, {Arenou}, {Cropper}, {Drimmel}, {H{\o}g}, {Katz}, {Lattanzi}, {O'Mullane}, {Grebel}, {Holland}, {Huc}, {Passot}, {Bramante}, {Cacciari}, {Casta{\~n}eda}, {Chaoul}, {Cheek}, {De Angeli}, {Fabricius}, {Guerra}, {Hern{\'a}ndez}, {Jean-Antoine-Piccolo}, {Masana}, {Messineo}, {Mowlavi}, {Nienartowicz}, {Ord{\'o}{\~n}ez-Blanco}, {Panuzzo}, {Portell}, {Richards}, {Riello}, {Seabroke}, {Tanga}, {Th{\'e}venin}, {Torra}, {Els}, {Gracia-Abril}, {Comoretto}, {Garcia-Reinaldos}, {Lock}, {Mercier}, {Altmann}, {Andrae}, {Astraatmadja}, {Bellas-Velidis}, {Benson}, {Berthier}, {Blomme}, {Busso}, {Carry}, {Cellino}, {Clementini},
  {Cowell}, {Creevey}, {Cuypers}, {Davidson}, {De Ridder}, {de Torres}, {Delchambre}, {Dell'Oro}, {Ducourant}, {Fr{\'e}mat}, {Garc{\'\i}a-Torres}, {Gosset}, {Halbwachs}, {Hambly}, {Harrison}, {Hauser}, {Hestroffer}, {Hodgkin}, {Huckle}, {Hutton}, {Jasniewicz}, {Jordan}, {Kontizas}, {Korn}, {Lanzafame}, {Manteiga}, {Moitinho}, {Muinonen}, {Osinde}, {Pancino}, {Pauwels}, {Petit}, {Recio-Blanco}, {Robin}, {Sarro}, {Siopis}, {Smith}, {Smith}, {Sozzetti}, {Thuillot}, {van Reeven}, {Viala}, {Abbas}, {Abreu Aramburu}, {Accart}, {Aguado}, {Allan}, {Allasia}, {Altavilla}, {{\'A}lvarez}, {Alves}, {Anderson}, {Andrei}, {Anglada Varela}, {Antiche}, {Antoja}, {Ant{\'o}n}, {Arcay}, {Atzei}, {Ayache}, {Bach}, {Baker}, {Balaguer-N{\'u}{\~n}ez}, {Barache}, {Barata}, {Barbier}, {Barblan}, {Baroni}, {Barrado y Navascu{\'e}s}, {Barros}, {Barstow}, {Becciani}, {Bellazzini}, {Bellei}, {Bello Garc{\'\i}a}, {Belokurov}, {Bendjoya}, {Berihuete}, {Bianchi}, {Bienaym{\'e}}, {Billebaud}, {Blagorodnova}, {Blanco-Cuaresma}, {Boch},
  {Bombrun}, {Borrachero}, {Bouquillon}, {Bourda}, {Bouy}, {Bragaglia}, {Breddels}, {Brouillet}, {Br{\"u}semeister}, {Bucciarelli}, {Budnik}, {Burgess}, {Burgon}, {Burlacu}, {Busonero}, {Buzzi}, {Caffau}, {Cambras}, {Campbell}, {Cancelliere}, {Cantat-Gaudin}, {Carlucci}, {Carrasco}, {Castellani}, {Charlot}, {Charnas}, {Charvet}, {Chassat}, {Chiavassa}, {Clotet}, {Cocozza}, {Collins}, {Collins}, {Costigan}, {Crifo}, {Cross}, {Crosta}, {Crowley}, {Dafonte}, {Damerdji}, {Dapergolas}, {David}, {David}, {De Cat}, {de Felice}, {de Laverny}, {De Luise}, {De March}, {de Martino}, {de Souza}, {Debosscher}, {del Pozo}, {Delbo}, {Delgado}, {Delgado}, {di Marco}, {Di Matteo}, {Diakite}, {Distefano}, {Dolding}, {Dos Anjos}, {Drazinos}, {Dur{\'a}n}, {Dzigan}, {Ecale}, {Edvardsson}, {Enke}, {Erdmann}, {Escolar}, {Espina}, {Evans}, {Eynard Bontemps}, {Fabre}, {Fabrizio}, {Faigler}, {Falc{\~a}o}, {Farr{\`a}s Casas}, {Faye}, {Federici}, {Fedorets}, {Fern{\'a}ndez-Hern{\'a}ndez}, {Fernique}, {Fienga}, {Figueras}, {Filippi},
  {Findeisen}, {Fonti}, {Fouesneau}, {Fraile}, {Fraser}, {Fuchs}, {Furnell}, {Gai}, {Galleti}, {Galluccio}, {Garabato}, {Garc{\'\i}a-Sedano}, {Gar{\'e}}, {Garofalo}, {Garralda}, {Gavras}, {Gerssen}, {Geyer}, {Gilmore}, {Girona}, {Giuffrida}, {Gomes}, {Gonz{\'a}lez-Marcos}, {Gonz{\'a}lez-N{\'u}{\~n}ez}, {Gonz{\'a}lez-Vidal}, {Granvik}, {Guerrier}, {Guillout}, {Guiraud}, {G{\'u}rpide}, {Guti{\'e}rrez-S{\'a}nchez}, {Guy}, {Haigron}, {Hatzidimitriou}, {Haywood}, {Heiter}, {Helmi}, {Hobbs}, {Hofmann}, {Holl}, {Holland}, {Hunt}, {Hypki}, {Icardi}, {Irwin}, {Jevardat de Fombelle}, {Jofr{\'e}}, {Jonker}, {Jorissen}, {Julbe}, {Karampelas}, {Kochoska}, {Kohley}, {Kolenberg}, {Kontizas}, {Koposov}, {Kordopatis}, {Koubsky}, {Kowalczyk}, {Krone-Martins}, {Kudryashova}, {Kull}, {Bachchan}, {Lacoste-Seris}, {Lanza}, {Lavigne}, {Le Poncin-Lafitte}, {Lebreton}, {Lebzelter}, {Leccia}, {Leclerc}, {Lecoeur-Taibi}, {Lemaitre}, {Lenhardt}, {Leroux}, {Liao}, {Licata}, {Lindstr{\o}m}, {Lister}, {Livanou}, {Lobel}, {L{\"o}ffler},
  {L{\'o}pez}, {Lopez-Lozano}, {Lorenz}, {Loureiro}, {MacDonald}, {Magalh{\~a}es Fernandes}, {Managau}, {Mann}, {Mantelet}, {Marchal}, {Marchant}, {Marconi}, {Marie}, {Marinoni}, {Marrese}, {Marschalk{\'o}}, {Marshall}, {Mart{\'\i}n-Fleitas}, {Martino}, {Mary}, {Matijevi{\v{c}}}, {Mazeh}, {McMillan}, {Messina}, {Mestre}, {Michalik}, {Millar}, {Miranda}, {Molina}, {Molinaro}, {Molinaro}, {Moln{\'a}r}, {Moniez}, {Montegriffo}, {Monteiro}, {Mor}, {Mora}, {Morbidelli}, {Morel}, {Morgenthaler}, {Morley}, {Morris}, {Mulone}, {Muraveva}, {Musella}, {Narbonne}, {Nelemans}, {Nicastro}, {Noval}, {Ord{\'e}novic}, {Ordieres-Mer{\'e}}, {Osborne}, {Pagani}, {Pagano}, {Pailler}, {Palacin}, {Palaversa}, {Parsons}, {Paulsen}, {Pecoraro}, {Pedrosa}, {Pentik{\"a}inen}, {Pereira}, {Pichon}, {Piersimoni}, {Pineau}, {Plachy}, {Plum}, {Poujoulet}, {Pr{\v{s}}a}, {Pulone}, {Ragaini}, {Rago}, {Rambaux}, {Ramos-Lerate}, {Ranalli}, {Rauw}, {Read}, {Regibo}, {Renk}, {Reyl{\'e}}, {Ribeiro}, {Rimoldini}, {Ripepi}, {Riva}, {Rixon},
  {Roelens}, {Romero-G{\'o}mez}, {Rowell}, {Royer}, {Rudolph}, {Ruiz-Dern}, {Sadowski}, {Sagrist{\`a} Sell{\'e}s}, {Sahlmann}, {Salgado}, {Salguero}, {Sarasso}, {Savietto}, {Schnorhk}, {Schultheis}, {Sciacca}, {Segol}, {Segovia}, {Segransan}, {Serpell}, {Shih}, {Smareglia}, {Smart}, {Smith}, {Solano}, {Solitro}, {Sordo}, {Soria Nieto}, {Souchay}, {Spagna}, {Spoto}, {Stampa}, {Steele}, {Steidelm{\"u}ller}, {Stephenson}, {Stoev}, {Suess}, {S{\"u}veges}, {Surdej}, {Szabados}, {Szegedi-Elek}, {Tapiador}, {Taris}, {Tauran}, {Taylor}, {Teixeira}, {Terrett}, {Tingley}, {Trager}, {Turon}, {Ulla}, {Utrilla}, {Valentini}, {van Elteren}, {Van Hemelryck}, {van Leeuwen}, {Varadi}, {Vecchiato}, {Veljanoski}, {Via}, {Vicente}, {Vogt}, {Voss}, {Votruba}, {Voutsinas}, {Walmsley}, {Weiler}, {Weingrill}, {Werner}, {Wevers}, {Whitehead}, {Wyrzykowski}, {Yoldas}, {{\v{Z}}erjal}, {Zucker}, {Zurbach}, {Zwitter}, {Alecu}, {Allen}, {Allende Prieto}, {Amorim}, {Anglada-Escud{\'e}}, {Arsenijevic}, {Azaz}, {Balm}, {Beck}, {Bernstein},
  {Bigot}, {Bijaoui}, {Blasco}, {Bonfigli}, {Bono}, {Boudreault}, {Bressan}, {Brown}, {Brunet}, {Bunclark}, {Buonanno}, {Butkevich}, {Carret}, {Carrion}, {Chemin}, {Ch{\'e}reau}, {Corcione}, {Darmigny}, {de Boer}, {de Teodoro}, {de Zeeuw}, {Delle Luche}, {Domingues}, {Dubath}, {Fodor}, {Fr{\'e}zouls}, {Fries}, {Fustes}, {Fyfe}, {Gallardo}, {Gallegos}, {Gardiol}, {Gebran}, {Gomboc}, {G{\'o}mez}, {Grux}, {Gueguen}, {Heyrovsky}, {Hoar}, {Iannicola}, {Isasi Parache}, {Janotto}, {Joliet}, {Jonckheere}, {Keil}, {Kim}, {Klagyivik}, {Klar}, {Knude}, {Kochukhov}, {Kolka}, {Kos}, {Kutka}, {Lainey}, {LeBouquin}, {Liu}, {Loreggia}, {Makarov}, {Marseille}, {Martayan}, {Martinez-Rubi}, {Massart}, {Meynadier}, {Mignot}, {Munari}, {Nguyen}, {Nordlander}, {Ocvirk}, {O'Flaherty}, {Olias Sanz}, {Ortiz}, {Osorio}, {Oszkiewicz}, {Ouzounis}, {Palmer}, {Park}, {Pasquato}, {Peltzer}, {Peralta}, {P{\'e}turaud}, {Pieniluoma}, {Pigozzi}, {Poels}, {Prat}, {Prod'homme}, {Raison}, {Rebordao}, {Risquez}, {Rocca-Volmerange}, {Rosen},
  {Ruiz-Fuertes}, {Russo}, {Sembay}, {Serraller Vizcaino}, {Short}, {Siebert}, {Silva}, {Sinachopoulos}, {Slezak}, {Soffel}, {Sosnowska}, {Strai{\v{z}}ys}, {ter Linden}, {Terrell}, {Theil}, {Tiede}, {Troisi}, {Tsalmantza}, {Tur}, {Vaccari}, {Vachier}, {Valles}, {Van Hamme}, {Veltz}, {Virtanen}, {Wallut}, {Wichmann}, {Wilkinson}, {Ziaeepour}, \& {Zschocke}}]{gaia2016}
{Gaia Collaboration}, {Prusti} T., {de Bruijne} J.~H.~J., et~al. (2016): {The Gaia mission}. \aap\, 595:A1. \doi{10.1051/0004-6361/201629272}, {\href{https://arxiv.org/abs/1609.04153}{{arXiv:1609.04153}}} {[astro-ph.IM]}

\bibitem[{{Gaia Collaboration} et~al.(2018{\natexlab{a}}){Gaia Collaboration}, {Babusiaux}, {van Leeuwen}, {Barstow}, {Jordi}, {Vallenari}, {Bossini}, {Bressan}, {Cantat-Gaudin}, {van Leeuwen}, {Brown}, {Prusti}, {de Bruijne}, {Bailer-Jones}, {Biermann}, {Evans}, {Eyer}, {Jansen}, {Klioner}, {Lammers}, {Lindegren}, {Luri}, {Mignard}, {Panem}, {Pourbaix}, {Randich}, {Sartoretti}, {Siddiqui}, {Soubiran}, {Walton}, {Arenou}, {Bastian}, {Cropper}, {Drimmel}, {Katz}, {Lattanzi}, {Bakker}, {Cacciari}, {Casta{\~n}eda}, {Chaoul}, {Cheek}, {De Angeli}, {Fabricius}, {Guerra}, {Holl}, {Masana}, {Messineo}, {Mowlavi}, {Nienartowicz}, {Panuzzo}, {Portell}, {Riello}, {Seabroke}, {Tanga}, {Th{\'e}venin}, {Gracia-Abril}, {Comoretto}, {Garcia-Reinaldos}, {Teyssier}, {Altmann}, {Andrae}, {Audard}, {Bellas-Velidis}, {Benson}, {Berthier}, {Blomme}, {Burgess}, {Busso}, {Carry}, {Cellino}, {Clementini}, {Clotet}, {Creevey}, {Davidson}, {De Ridder}, {Delchambre}, {Dell'Oro}, {Ducourant}, {Fern{\'a}ndez-Hern{\'a}ndez}, {Fouesneau},
  {Fr{\'e}mat}, {Galluccio}, {Garc{\'\i}a-Torres}, {Gonz{\'a}lez-N{\'u}{\~n}ez}, {Gonz{\'a}lez-Vidal}, {Gosset}, {Guy}, {Halbwachs}, {Hambly}, {Harrison}, {Hern{\'a}ndez}, {Hestroffer}, {Hodgkin}, {Hutton}, {Jasniewicz}, {Jean-Antoine-Piccolo}, {Jordan}, {Korn}, {Krone-Martins}, {Lanzafame}, {Lebzelter}, {L{\"o}ffler}, {Manteiga}, {Marrese}, {Mart{\'\i}n-Fleitas}, {Moitinho}, {Mora}, {Muinonen}, {Osinde}, {Pancino}, {Pauwels}, {Petit}, {Recio-Blanco}, {Richards}, {Rimoldini}, {Robin}, {Sarro}, {Siopis}, {Smith}, {Sozzetti}, {S{\"u}veges}, {Torra}, {van Reeven}, {Abbas}, {Abreu Aramburu}, {Accart}, {Aerts}, {Altavilla}, {{\'A}lvarez}, {Alvarez}, {Alves}, {Anderson}, {Andrei}, {Anglada Varela}, {Antiche}, {Antoja}, {Arcay}, {Astraatmadja}, {Bach}, {Baker}, {Balaguer-N{\'u}{\~n}ez}, {Balm}, {Barache}, {Barata}, {Barbato}, {Barblan}, {Barklem}, {Barrado}, {Barros}, {Bartholom{\'e} Mu{\~n}oz}, {Bassilana}, {Becciani}, {Bellazzini}, {Berihuete}, {Bertone}, {Bianchi}, {Bienaym{\'e}}, {Blanco-Cuaresma}, {Boch},
  {Boeche}, {Bombrun}, {Borrachero}, {Bouquillon}, {Bourda}, {Bragaglia}, {Bramante}, {Breddels}, {Brouillet}, {Br{\"u}semeister}, {Brugaletta}, {Bucciarelli}, {Burlacu}, {Busonero}, {Butkevich}, {Buzzi}, {Caffau}, {Cancelliere}, {Cannizzaro}, {Carballo}, {Carlucci}, {Carrasco}, {Casamiquela}, {Castellani}, {Castro-Ginard}, {Charlot}, {Chemin}, {Chiavassa}, {Cocozza}, {Costigan}, {Cowell}, {Crifo}, {Crosta}, {Crowley}, {Cuypers}, {Dafonte}, {Damerdji}, {Dapergolas}, {David}, {David}, {de Laverny}, {De Luise}, {De March}, {de Martino}, {de Souza}, {de Torres}, {Debosscher}, {del Pozo}, {Delbo}, {Delgado}, {Delgado}, {Diakite}, {Diener}, {Distefano}, {Dolding}, {Drazinos}, {Dur{\'a}n}, {Edvardsson}, {Enke}, {Eriksson}, {Esquej}, {Eynard Bontemps}, {Fabre}, {Fabrizio}, {Faigler}, {Falc{\~a}o}, {Farr{\`a}s Casas}, {Federici}, {Fedorets}, {Fernique}, {Figueras}, {Filippi}, {Findeisen}, {Fonti}, {Fraile}, {Fraser}, {Fr{\'e}zouls}, {Gai}, {Galleti}, {Garabato}, {Garc{\'\i}a-Sedano}, {Garofalo}, {Garralda}, {Gavel},
  {Gavras}, {Gerssen}, {Geyer}, {Giacobbe}, {Gilmore}, {Girona}, {Giuffrida}, {Glass}, {Gomes}, {Granvik}, {Gueguen}, {Guerrier}, {Guiraud}, {Guti{\'e}}, {Haigron}, {Hatzidimitriou}, {Hauser}, {Haywood}, {Heiter}, {Helmi}, {Heu}, {Hilger}, {Hobbs}, {Hofmann}, {Holland}, {Huckle}, {Hypki}, {Icardi}, {Jan{\ss}en}, {Jevardat de Fombelle}, {Jonker}, {Juh{\'a}sz}, {Julbe}, {Karampelas}, {Kewley}, {Klar}, {Kochoska}, {Kohley}, {Kolenberg}, {Kontizas}, {Kontizas}, {Koposov}, {Kordopatis}, {Kostrzewa-Rutkowska}, {Koubsky}, {Lambert}, {Lanza}, {Lasne}, {Lavigne}, {Le Fustec}, {Le Poncin-Lafitte}, {Lebreton}, {Leccia}, {Leclerc}, {Lecoeur-Taibi}, {Lenhardt}, {Leroux}, {Liao}, {Licata}, {Lindstr{\o}m}, {Lister}, {Livanou}, {Lobel}, {L{\'o}pez}, {Managau}, {Mann}, {Mantelet}, {Marchal}, {Marchant}, {Marconi}, {Marinoni}, {Marschalk{\'o}}, {Marshall}, {Martino}, {Marton}, {Mary}, {Massari}, {Matijevi{\v{c}}}, {Mazeh}, {McMillan}, {Messina}, {Michalik}, {Millar}, {Molina}, {Molinaro}, {Moln{\'a}r}, {Montegriffo}, {Mor},
  {Morbidelli}, {Morel}, {Morris}, {Mulone}, {Muraveva}, {Musella}, {Nelemans}, {Nicastro}, {Noval}, {O'Mullane}, {Ord{\'e}novic}, {Ord{\'o}{\~n}ez-Blanco}, {Osborne}, {Pagani}, {Pagano}, {Pailler}, {Palacin}, {Palaversa}, {Panahi}, {Pawlak}, {Piersimoni}, {Pineau}, {Plachy}, {Plum}, {Poggio}, {Poujoulet}, {Pr{\v{s}}a}, {Pulone}, {Racero}, {Ragaini}, {Rambaux}, {Ramos-Lerate}, {Regibo}, {Reyl{\'e}}, {Riclet}, {Ripepi}, {Riva}, {Rivard}, {Rixon}, {Roegiers}, {Roelens}, {Romero-G{\'o}mez}, {Rowell}, {Royer}, {Ruiz-Dern}, {Sadowski}, {Sagrist{\`a} Sell{\'e}s}, {Sahlmann}, {Salgado}, {Salguero}, {Sanna}, {Santana-Ros}, {Sarasso}, {Savietto}, {Schultheis}, {Sciacca}, {Segol}, {Segovia}, {S{\'e}gransan}, {Shih}, {Siltala}, {Silva}, {Smart}, {Smith}, {Solano}, {Solitro}, {Sordo}, {Soria Nieto}, {Souchay}, {Spagna}, {Spoto}, {Stampa}, {Steele}, {Steidelm{\"u}ller}, {Stephenson}, {Stoev}, {Suess}, {Surdej}, {Szabados}, {Szegedi-Elek}, {Tapiador}, {Taris}, {Tauran}, {Taylor}, {Teixeira}, {Terrett}, {Teyssandier},
  {Thuillot}, {Titarenko}, {Torra Clotet}, {Turon}, {Ulla}, {Utrilla}, {Uzzi}, {Vaillant}, {Valentini}, {Valette}, {van Elteren}, {Van Hemelryck}, {Vaschetto}, {Vecchiato}, {Veljanoski}, {Viala}, {Vicente}, {Vogt}, {von Essen}, {Voss}, {Votruba}, {Voutsinas}, {Walmsley}, {Weiler}, {Wertz}, {Wevers}, {Wyrzykowski}, {Yoldas}, {{\v{Z}}erjal}, {Ziaeepour}, {Zorec}, {Zschocke}, {Zucker}, {Zurbach}, \& {Zwitter}}]{gaia2018b}
{Gaia Collaboration}, {Babusiaux} C., {van Leeuwen} F., et~al. (2018{\natexlab{a}}): {Gaia Data Release 2. Observational Hertzsprung-Russell diagrams}. \aap\, 616:A10. \doi{10.1051/0004-6361/201832843}, {\href{https://arxiv.org/abs/1804.09378}{{arXiv:1804.09378}}} {[astro-ph.SR]}

\bibitem[{{Gaia Collaboration} et~al.(2018{\natexlab{b}}){Gaia Collaboration}, {Brown}, {Vallenari}, {Prusti}, {de Bruijne}, {Babusiaux}, {Bailer-Jones}, {Biermann}, {Evans}, {Eyer}, {Jansen}, {Jordi}, {Klioner}, {Lammers}, {Lindegren}, {Luri}, {Mignard}, {Panem}, {Pourbaix}, {Randich}, {Sartoretti}, {Siddiqui}, {Soubiran}, {van Leeuwen}, {Walton}, {Arenou}, {Bastian}, {Cropper}, {Drimmel}, {Katz}, {Lattanzi}, {Bakker}, {Cacciari}, {Casta{\~n}eda}, {Chaoul}, {Cheek}, {De Angeli}, {Fabricius}, {Guerra}, {Holl}, {Masana}, {Messineo}, {Mowlavi}, {Nienartowicz}, {Panuzzo}, {Portell}, {Riello}, {Seabroke}, {Tanga}, {Th{\'e}venin}, {Gracia-Abril}, {Comoretto}, {Garcia-Reinaldos}, {Teyssier}, {Altmann}, {Andrae}, {Audard}, {Bellas-Velidis}, {Benson}, {Berthier}, {Blomme}, {Burgess}, {Busso}, {Carry}, {Cellino}, {Clementini}, {Clotet}, {Creevey}, {Davidson}, {De Ridder}, {Delchambre}, {Dell'Oro}, {Ducourant}, {Fern{\'a}ndez-Hern{\'a}ndez}, {Fouesneau}, {Fr{\'e}mat}, {Galluccio}, {Garc{\'\i}a-Torres},
  {Gonz{\'a}lez-N{\'u}{\~n}ez}, {Gonz{\'a}lez-Vidal}, {Gosset}, {Guy}, {Halbwachs}, {Hambly}, {Harrison}, {Hern{\'a}ndez}, {Hestroffer}, {Hodgkin}, {Hutton}, {Jasniewicz}, {Jean-Antoine-Piccolo}, {Jordan}, {Korn}, {Krone-Martins}, {Lanzafame}, {Lebzelter}, {L{\"o}ffler}, {Manteiga}, {Marrese}, {Mart{\'\i}n-Fleitas}, {Moitinho}, {Mora}, {Muinonen}, {Osinde}, {Pancino}, {Pauwels}, {Petit}, {Recio-Blanco}, {Richards}, {Rimoldini}, {Robin}, {Sarro}, {Siopis}, {Smith}, {Sozzetti}, {S{\"u}veges}, {Torra}, {van Reeven}, {Abbas}, {Abreu Aramburu}, {Accart}, {Aerts}, {Altavilla}, {{\'A}lvarez}, {Alvarez}, {Alves}, {Anderson}, {Andrei}, {Anglada Varela}, {Antiche}, {Antoja}, {Arcay}, {Astraatmadja}, {Bach}, {Baker}, {Balaguer-N{\'u}{\~n}ez}, {Balm}, {Barache}, {Barata}, {Barbato}, {Barblan}, {Barklem}, {Barrado}, {Barros}, {Barstow}, {Bartholom{\'e} Mu{\~n}oz}, {Bassilana}, {Becciani}, {Bellazzini}, {Berihuete}, {Bertone}, {Bianchi}, {Bienaym{\'e}}, {Blanco-Cuaresma}, {Boch}, {Boeche}, {Bombrun}, {Borrachero},
  {Bossini}, {Bouquillon}, {Bourda}, {Bragaglia}, {Bramante}, {Breddels}, {Bressan}, {Brouillet}, {Br{\"u}semeister}, {Brugaletta}, {Bucciarelli}, {Burlacu}, {Busonero}, {Butkevich}, {Buzzi}, {Caffau}, {Cancelliere}, {Cannizzaro}, {Cantat-Gaudin}, {Carballo}, {Carlucci}, {Carrasco}, {Casamiquela}, {Castellani}, {Castro-Ginard}, {Charlot}, {Chemin}, {Chiavassa}, {Cocozza}, {Costigan}, {Cowell}, {Crifo}, {Crosta}, {Crowley}, {Cuypers}, {Dafonte}, {Damerdji}, {Dapergolas}, {David}, {David}, {de Laverny}, {De Luise}, {De March}, {de Martino}, {de Souza}, {de Torres}, {Debosscher}, {del Pozo}, {Delbo}, {Delgado}, {Delgado}, {Di Matteo}, {Diakite}, {Diener}, {Distefano}, {Dolding}, {Drazinos}, {Dur{\'a}n}, {Edvardsson}, {Enke}, {Eriksson}, {Esquej}, {Eynard Bontemps}, {Fabre}, {Fabrizio}, {Faigler}, {Falc{\~a}o}, {Farr{\`a}s Casas}, {Federici}, {Fedorets}, {Fernique}, {Figueras}, {Filippi}, {Findeisen}, {Fonti}, {Fraile}, {Fraser}, {Fr{\'e}zouls}, {Gai}, {Galleti}, {Garabato}, {Garc{\'\i}a-Sedano}, {Garofalo},
  {Garralda}, {Gavel}, {Gavras}, {Gerssen}, {Geyer}, {Giacobbe}, {Gilmore}, {Girona}, {Giuffrida}, {Glass}, {Gomes}, {Granvik}, {Gueguen}, {Guerrier}, {Guiraud}, {Guti{\'e}rrez-S{\'a}nchez}, {Haigron}, {Hatzidimitriou}, {Hauser}, {Haywood}, {Heiter}, {Helmi}, {Heu}, {Hilger}, {Hobbs}, {Hofmann}, {Holland}, {Huckle}, {Hypki}, {Icardi}, {Jan{\ss}en}, {Jevardat de Fombelle}, {Jonker}, {Juh{\'a}sz}, {Julbe}, {Karampelas}, {Kewley}, {Klar}, {Kochoska}, {Kohley}, {Kolenberg}, {Kontizas}, {Kontizas}, {Koposov}, {Kordopatis}, {Kostrzewa-Rutkowska}, {Koubsky}, {Lambert}, {Lanza}, {Lasne}, {Lavigne}, {Le Fustec}, {Le Poncin-Lafitte}, {Lebreton}, {Leccia}, {Leclerc}, {Lecoeur-Taibi}, {Lenhardt}, {Leroux}, {Liao}, {Licata}, {Lindstr{\o}m}, {Lister}, {Livanou}, {Lobel}, {L{\'o}pez}, {Managau}, {Mann}, {Mantelet}, {Marchal}, {Marchant}, {Marconi}, {Marinoni}, {Marschalk{\'o}}, {Marshall}, {Martino}, {Marton}, {Mary}, {Massari}, {Matijevi{\v{c}}}, {Mazeh}, {McMillan}, {Messina}, {Michalik}, {Millar}, {Molina}, {Molinaro},
  {Moln{\'a}r}, {Montegriffo}, {Mor}, {Morbidelli}, {Morel}, {Morris}, {Mulone}, {Muraveva}, {Musella}, {Nelemans}, {Nicastro}, {Noval}, {O'Mullane}, {Ord{\'e}novic}, {Ord{\'o}{\~n}ez-Blanco}, {Osborne}, {Pagani}, {Pagano}, {Pailler}, {Palacin}, {Palaversa}, {Panahi}, {Pawlak}, {Piersimoni}, {Pineau}, {Plachy}, {Plum}, {Poggio}, {Poujoulet}, {Pr{\v{s}}a}, {Pulone}, {Racero}, {Ragaini}, {Rambaux}, {Ramos-Lerate}, {Regibo}, {Reyl{\'e}}, {Riclet}, {Ripepi}, {Riva}, {Rivard}, {Rixon}, {Roegiers}, {Roelens}, {Romero-G{\'o}mez}, {Rowell}, {Royer}, {Ruiz-Dern}, {Sadowski}, {Sagrist{\`a} Sell{\'e}s}, {Sahlmann}, {Salgado}, {Salguero}, {Sanna}, {Santana-Ros}, {Sarasso}, {Savietto}, {Schultheis}, {Sciacca}, {Segol}, {Segovia}, {S{\'e}gransan}, {Shih}, {Siltala}, {Silva}, {Smart}, {Smith}, {Solano}, {Solitro}, {Sordo}, {Soria Nieto}, {Souchay}, {Spagna}, {Spoto}, {Stampa}, {Steele}, {Steidelm{\"u}ller}, {Stephenson}, {Stoev}, {Suess}, {Surdej}, {Szabados}, {Szegedi-Elek}, {Tapiador}, {Taris}, {Tauran}, {Taylor},
  {Teixeira}, {Terrett}, {Teyssandier}, {Thuillot}, {Titarenko}, {Torra Clotet}, {Turon}, {Ulla}, {Utrilla}, {Uzzi}, {Vaillant}, {Valentini}, {Valette}, {van Elteren}, {Van Hemelryck}, {van Leeuwen}, {Vaschetto}, {Vecchiato}, {Veljanoski}, {Viala}, {Vicente}, {Vogt}, {von Essen}, {Voss}, {Votruba}, {Voutsinas}, {Walmsley}, {Weiler}, {Wertz}, {Wevers}, {Wyrzykowski}, {Yoldas}, {{\v{Z}}erjal}, {Ziaeepour}, {Zorec}, {Zschocke}, {Zucker}, {Zurbach}, \& {Zwitter}}]{gaia2018a}
{Gaia Collaboration}, {Brown} A.~G.~A., {Vallenari} A., et~al. (2018{\natexlab{b}}): {Gaia Data Release 2. Summary of the contents and survey properties}. \aap\, 616:A1. \doi{10.1051/0004-6361/201833051}, {\href{https://arxiv.org/abs/1804.09365}{{arXiv:1804.09365}}} {[astro-ph.GA]}

\bibitem[{{Gaia Collaboration} et~al.(2021){Gaia Collaboration}, {Brown}, {Vallenari}, {Prusti}, {de Bruijne}, {Babusiaux}, {Biermann}, {Creevey}, {Evans}, {Eyer}, {Hutton}, {Jansen}, {Jordi}, {Klioner}, {Lammers}, {Lindegren}, {Luri}, {Mignard}, {Panem}, {Pourbaix}, {Randich}, {Sartoretti}, {Soubiran}, {Walton}, {Arenou}, {Bailer-Jones}, {Bastian}, {Cropper}, {Drimmel}, {Katz}, {Lattanzi}, {van Leeuwen}, {Bakker}, {Cacciari}, {Casta{\~n}eda}, {De Angeli}, {Ducourant}, {Fabricius}, {Fouesneau}, {Fr{\'e}mat}, {Guerra}, {Guerrier}, {Guiraud}, {Jean-Antoine Piccolo}, {Masana}, {Messineo}, {Mowlavi}, {Nicolas}, {Nienartowicz}, {Pailler}, {Panuzzo}, {Riclet}, {Roux}, {Seabroke}, {Sordo}, {Tanga}, {Th{\'e}venin}, {Gracia-Abril}, {Portell}, {Teyssier}, {Altmann}, {Andrae}, {Bellas-Velidis}, {Benson}, {Berthier}, {Blomme}, {Brugaletta}, {Burgess}, {Busso}, {Carry}, {Cellino}, {Cheek}, {Clementini}, {Damerdji}, {Davidson}, {Delchambre}, {Dell'Oro}, {Fern{\'a}ndez-Hern{\'a}ndez}, {Galluccio}, {Garc{\'\i}a-Lario},
  {Garcia-Reinaldos}, {Gonz{\'a}lez-N{\'u}{\~n}ez}, {Gosset}, {Haigron}, {Halbwachs}, {Hambly}, {Harrison}, {Hatzidimitriou}, {Heiter}, {Hern{\'a}ndez}, {Hestroffer}, {Hodgkin}, {Holl}, {Jan{\ss}en}, {Jevardat de Fombelle}, {Jordan}, {Krone-Martins}, {Lanzafame}, {L{\"o}ffler}, {Lorca}, {Manteiga}, {Marchal}, {Marrese}, {Moitinho}, {Mora}, {Muinonen}, {Osborne}, {Pancino}, {Pauwels}, {Petit}, {Recio-Blanco}, {Richards}, {Riello}, {Rimoldini}, {Robin}, {Roegiers}, {Rybizki}, {Sarro}, {Siopis}, {Smith}, {Sozzetti}, {Ulla}, {Utrilla}, {van Leeuwen}, {van Reeven}, {Abbas}, {Abreu Aramburu}, {Accart}, {Aerts}, {Aguado}, {Ajaj}, {Altavilla}, {{\'A}lvarez}, {{\'A}lvarez Cid-Fuentes}, {Alves}, {Anderson}, {Anglada Varela}, {Antoja}, {Audard}, {Baines}, {Baker}, {Balaguer-N{\'u}{\~n}ez}, {Balbinot}, {Balog}, {Barache}, {Barbato}, {Barros}, {Barstow}, {Bartolom{\'e}}, {Bassilana}, {Bauchet}, {Baudesson-Stella}, {Becciani}, {Bellazzini}, {Bernet}, {Bertone}, {Bianchi}, {Blanco-Cuaresma}, {Boch}, {Bombrun}, {Bossini},
  {Bouquillon}, {Bragaglia}, {Bramante}, {Breedt}, {Bressan}, {Brouillet}, {Bucciarelli}, {Burlacu}, {Busonero}, {Butkevich}, {Buzzi}, {Caffau}, {Cancelliere}, {C{\'a}novas}, {Cantat-Gaudin}, {Carballo}, {Carlucci}, {Carnerero}, {Carrasco}, {Casamiquela}, {Castellani}, {Castro-Ginard}, {Castro Sampol}, {Chaoul}, {Charlot}, {Chemin}, {Chiavassa}, {Cioni}, {Comoretto}, {Cooper}, {Cornez}, {Cowell}, {Crifo}, {Crosta}, {Crowley}, {Dafonte}, {Dapergolas}, {David}, {David}, {de Laverny}, {De Luise}, {De March}, {De Ridder}, {de Souza}, {de Teodoro}, {de Torres}, {del Peloso}, {del Pozo}, {Delbo}, {Delgado}, {Delgado}, {Delisle}, {Di Matteo}, {Diakite}, {Diener}, {Distefano}, {Dolding}, {Eappachen}, {Edvardsson}, {Enke}, {Esquej}, {Fabre}, {Fabrizio}, {Faigler}, {Fedorets}, {Fernique}, {Fienga}, {Figueras}, {Fouron}, {Fragkoudi}, {Fraile}, {Franke}, {Gai}, {Garabato}, {Garcia-Gutierrez}, {Garc{\'\i}a-Torres}, {Garofalo}, {Gavras}, {Gerlach}, {Geyer}, {Giacobbe}, {Gilmore}, {Girona}, {Giuffrida}, {Gomel}, {Gomez},
  {Gonzalez-Santamaria}, {Gonz{\'a}lez-Vidal}, {Granvik}, {Guti{\'e}rrez-S{\'a}nchez}, {Guy}, {Hauser}, {Haywood}, {Helmi}, {Hidalgo}, {Hilger}, {H{\l}adczuk}, {Hobbs}, {Holland}, {Huckle}, {Jasniewicz}, {Jonker}, {Juaristi Campillo}, {Julbe}, {Karbevska}, {Kervella}, {Khanna}, {Kochoska}, {Kontizas}, {Kordopatis}, {Korn}, {Kostrzewa-Rutkowska}, {Kruszy{\'n}ska}, {Lambert}, {Lanza}, {Lasne}, {Le Campion}, {Le Fustec}, {Lebreton}, {Lebzelter}, {Leccia}, {Leclerc}, {Lecoeur-Taibi}, {Liao}, {Licata}, {Lindstr{\o}m}, {Lister}, {Livanou}, {Lobel}, {Madrero Pardo}, {Managau}, {Mann}, {Marchant}, {Marconi}, {Marcos Santos}, {Marinoni}, {Marocco}, {Marshall}, {Martin Polo}, {Mart{\'\i}n-Fleitas}, {Masip}, {Massari}, {Mastrobuono-Battisti}, {Mazeh}, {McMillan}, {Messina}, {Michalik}, {Millar}, {Mints}, {Molina}, {Molinaro}, {Moln{\'a}r}, {Montegriffo}, {Mor}, {Morbidelli}, {Morel}, {Morris}, {Mulone}, {Munoz}, {Muraveva}, {Murphy}, {Musella}, {Noval}, {Ord{\'e}novic}, {Orr{\`u}}, {Osinde}, {Pagani}, {Pagano},
  {Palaversa}, {Palicio}, {Panahi}, {Pawlak}, {Pe{\~n}alosa Esteller}, {Penttil{\"a}}, {Piersimoni}, {Pineau}, {Plachy}, {Plum}, {Poggio}, {Poretti}, {Poujoulet}, {Pr{\v{s}}a}, {Pulone}, {Racero}, {Ragaini}, {Rainer}, {Raiteri}, {Rambaux}, {Ramos}, {Ramos-Lerate}, {Re Fiorentin}, {Regibo}, {Reyl{\'e}}, {Ripepi}, {Riva}, {Rixon}, {Robichon}, {Robin}, {Roelens}, {Rohrbasser}, {Romero-G{\'o}mez}, {Rowell}, {Royer}, {Rybicki}, {Sadowski}, {Sagrist{\`a} Sell{\'e}s}, {Sahlmann}, {Salgado}, {Salguero}, {Samaras}, {Sanchez Gimenez}, {Sanna}, {Santove{\~n}a}, {Sarasso}, {Schultheis}, {Sciacca}, {Segol}, {Segovia}, {S{\'e}gransan}, {Semeux}, {Shahaf}, {Siddiqui}, {Siebert}, {Siltala}, {Slezak}, {Smart}, {Solano}, {Solitro}, {Souami}, {Souchay}, {Spagna}, {Spoto}, {Steele}, {Steidelm{\"u}ller}, {Stephenson}, {S{\"u}veges}, {Szabados}, {Szegedi-Elek}, {Taris}, {Tauran}, {Taylor}, {Teixeira}, {Thuillot}, {Tonello}, {Torra}, {Torra}, {Turon}, {Unger}, {Vaillant}, {van Dillen}, {Vanel}, {Vecchiato}, {Viala}, {Vicente},
  {Voutsinas}, {Weiler}, {Wevers}, {Wyrzykowski}, {Yoldas}, {Yvard}, {Zhao}, {Zorec}, {Zucker}, {Zurbach}, \& {Zwitter}}]{gaia2021}
{Gaia Collaboration}, {Brown} A.~G.~A., {Vallenari} A., et~al. (2021): {Gaia Early Data Release 3. Summary of the contents and survey properties}. \aap\, 649:A1. \doi{10.1051/0004-6361/202039657}, {\href{https://arxiv.org/abs/2012.01533}{{arXiv:2012.01533}}} {[astro-ph.GA]}

\bibitem[{{Gaia Collaboration} et~al.(2023{\natexlab{a}}){Gaia Collaboration}, {Montegriffo}, {Bellazzini}, {De Angeli}, {Andrae}, {Barstow}, {Bossini}, {Bragaglia}, {Burgess}, {Cacciari}, {Carrasco}, {Chornay}, {Delchambre}, {Evans}, {Fouesneau}, {Fr{\'e}mat}, {Garabato}, {Jordi}, {Manteiga}, {Massari}, {Palaversa}, {Pancino}, {Riello}, {Ruz Mieres}, {Sanna}, {Santove{\~n}a}, {Sordo}, {Vallenari}, {Walton}, {Brown}, {Prusti}, {de Bruijne}, {Arenou}, {Babusiaux}, {Biermann}, {Creevey}, {Ducourant}, {Eyer}, {Guerra}, {Hutton}, {Klioner}, {Lammers}, {Lindegren}, {Luri}, {Mignard}, {Panem}, {Pourbaix}, {Randich}, {Sartoretti}, {Soubiran}, {Tanga}, {Bailer-Jones}, {Bastian}, {Drimmel}, {Jansen}, {Katz}, {Lattanzi}, {van Leeuwen}, {Bakker}, {Casta{\~n}eda}, {Fabricius}, {Galluccio}, {Guerrier}, {Heiter}, {Masana}, {Messineo}, {Mowlavi}, {Nicolas}, {Nienartowicz}, {Pailler}, {Panuzzo}, {Riclet}, {Roux}, {Seabroke}, {Th{\'e}venin}, {Gracia-Abril}, {Portell}, {Teyssier}, {Altmann}, {Audard}, {Bellas-Velidis},
  {Benson}, {Berthier}, {Blomme}, {Busonero}, {Busso}, {C{\'a}novas}, {Carry}, {Cellino}, {Cheek}, {Clementini}, {Damerdji}, {Davidson}, {de Teodoro}, {Nu{\~n}ez Campos}, {Dell'Oro}, {Esquej}, {Fern{\'a}ndez-Hern{\'a}ndez}, {Fraile}, {Garc{\'\i}a-Lario}, {Gosset}, {Haigron}, {Halbwachs}, {Hambly}, {Harrison}, {Hern{\'a}ndez}, {Hestroffer}, {Hodgkin}, {Holl}, {Jan{\ss}en}, {Jevardat de Fombelle}, {Jordan}, {Krone-Martins}, {Lanzafame}, {L{\"o}ffler}, {Marchal}, {Marrese}, {Moitinho}, {Muinonen}, {Osborne}, {Pauwels}, {Recio-Blanco}, {Reyl{\'e}}, {Rimoldini}, {Roegiers}, {Rybizki}, {Sarro}, {Siopis}, {Smith}, {Sozzetti}, {Utrilla}, {van Leeuwen}, {Abbas}, {{\'A}brah{\'a}m}, {Abreu Aramburu}, {Aerts}, {Aguado}, {Ajaj}, {Aldea-Montero}, {Altavilla}, {{\'A}lvarez}, {Alves}, {Anderson}, {Anglada Varela}, {Antoja}, {Baines}, {Baker}, {Balaguer-N{\'u}{\~n}ez}, {Balbinot}, {Balog}, {Barache}, {Barbato}, {Barros}, {Bartolom{\'e}}, {Bassilana}, {Bauchet}, {Becciani}, {Berihuete}, {Bernet}, {Bertone}, {Bianchi},
  {Binnenfeld}, {Blanco-Cuaresma}, {Boch}, {Bombrun}, {Bouquillon}, {Bramante}, {Breedt}, {Bressan}, {Brouillet}, {Brugaletta}, {Bucciarelli}, {Burlacu}, {Butkevich}, {Buzzi}, {Caffau}, {Cancelliere}, {Cantat-Gaudin}, {Carballo}, {Carlucci}, {Carnerero}, {Casamiquela}, {Castellani}, {Castro-Ginard}, {Chaoul}, {Charlot}, {Chemin}, {Chiaramida}, {Chiavassa}, {Comoretto}, {Contursi}, {Cooper}, {Cornez}, {Cowell}, {Crifo}, {Cropper}, {Crosta}, {Crowley}, {Dafonte}, {Dapergolas}, {David}, {de Laverny}, {De Luise}, {De March}, {De Ridder}, {de Souza}, {de Torres}, {del Peloso}, {del Pozo}, {Delbo}, {Delgado}, {Delisle}, {Demouchy}, {Dharmawardena}, {Diakite}, {Diener}, {Distefano}, {Dolding}, {Enke}, {Fabre}, {Fabrizio}, {Faigler}, {Fedorets}, {Fernique}, {Figueras}, {Fournier}, {Fouron}, {Fragkoudi}, {Gai}, {Garcia-Gutierrez}, {Garcia-Reinaldos}, {Garc{\'\i}a-Torres}, {Garofalo}, {Gavel}, {Gavras}, {Gerlach}, {Geyer}, {Giacobbe}, {Gilmore}, {Girona}, {Giuffrida}, {Gomel}, {Gomez}, {Gonz{\'a}lez-N{\'u}{\~n}ez},
  {Gonz{\'a}lez-Santamar{\'\i}a}, {Gonz{\'a}lez-Vidal}, {Granvik}, {Guillout}, {Guiraud}, {Guti{\'e}rrez-S{\'a}nchez}, {Guy}, {Hatzidimitriou}, {Hauser}, {Haywood}, {Helmer}, {Helmi}, {Sarmiento}, {Hidalgo}, {H{\l}adczuk}, {Hobbs}, {Holland}, {Huckle}, {Jardine}, {Jasniewicz}, {Jean-Antoine Piccolo}, {Jim{\'e}nez-Arranz}, {Juaristi Campillo}, {Julbe}, {Karbevska}, {Kervella}, {Khanna}, {Kordopatis}, {Korn}, {K{\'o}sp{\'a}l}, {Kostrzewa-Rutkowska}, {Kruszy{\'n}ska}, {Kun}, {Laizeau}, {Lambert}, {Lanza}, {Lasne}, {Le Campion}, {Lebreton}, {Lebzelter}, {Leccia}, {Leclerc}, {Lecoeur-Taibi}, {Liao}, {Licata}, {Lindstr{\'o}m}, {Lister}, {Livanou}, {Lobel}, {Lorca}, {Loup}, {Madrero Pardo}, {Magdaleno Romeo}, {Managau}, {Mann}, {Marchant}, {Marconi}, {Marcos}, {Marcos Santos}, {Mar{\'\i}n Pina}, {Marinoni}, {Marocco}, {Marshall}, {Martin Polo}, {Mart{\'\i}n-Fleitas}, {Marton}, {Mary}, {Masip}, {Mastrobuono-Battisti}, {Mazeh}, {McMillan}, {Messina}, {Michalik}, {Millar}, {Mints}, {Molina}, {Molinaro}, {Moln{\'a}r},
  {Monari}, {Mongui{\'o}}, {Montero}, {Mor}, {Mora}, {Morbidelli}, {Morel}, {Morris}, {Muraveva}, {Murphy}, {Musella}, {Nagy}, {Noval}, {Oca{\~n}a}, {Ogden}, {Ordenovic}, {Osinde}, {Pagani}, {Pagano}, {Palicio}, {Pallas-Quintela}, {Panahi}, {Payne-Wardenaar}, {Pe{\~n}alosa Esteller}, {Penttil{\"a}}, {Pichon}, {Piersimoni}, {Pineau}, {Plachy}, {Plum}, {Poggio}, {Pr{\v{s}}a}, {Pulone}, {Racero}, {Ragaini}, {Rainer}, {Raiteri}, {Ramos}, {Ramos-Lerate}, {Re Fiorentin}, {Regibo}, {Richards}, {Rios Diaz}, {Ripepi}, {Riva}, {Rix}, {Rixon}, {Robichon}, {Robin}, {Robin}, {Roelens}, {Rogues}, {Rohrbasser}, {Romero-G{\'o}mez}, {Rowell}, {Royer}, {Rybicki}, {Sadowski}, {S{\'a}ez N{\'u}{\~n}ez}, {Sagrist{\`a} Sell{\'e}s}, {Sahlmann}, {Salguero}, {Samaras}, {Sanchez Gimenez}, {Sarasso}, {Schultheis}, {Sciacca}, {Segol}, {Segovia}, {S{\'e}gransan}, {Semeux}, {Shahaf}, {Siddiqui}, {Siebert}, {Siltala}, {Silvelo}, {Slezak}, {Slezak}, {Smart}, {Snaith}, {Solano}, {Solitro}, {Souami}, {Souchay}, {Spagna}, {Spina}, {Spoto},
  {Steele}, {Steidelm{\"u}ller}, {Stephenson}, {S{\"u}veges}, {Surdej}, {Szabados}, {Szegedi-Elek}, {Taris}, {Taylor}, {Teixeira}, {Tolomei}, {Tonello}, {Torra}, {Torra}, {Torralba Elipe}, {Trabucchi}, {Tsounis}, {Turon}, {Ulla}, {Unger}, {Vaillant}, {van Dillen}, {van Reeven}, {Vanel}, {Vecchiato}, {Viala}, {Vicente}, {Voutsinas}, {Wevers}, {Wyrzykowski}, {Yoldas}, {Yvard}, {Zhao}, {Zorec}, {Zucker}, \& {Zwitter}}]{gaia2023b}
{Gaia Collaboration}, {Montegriffo} P., {Bellazzini} M., et~al. (2023{\natexlab{a}}): {Gaia Data Release 3. The Galaxy in your preferred colours: Synthetic photometry from Gaia low-resolution spectra}. \aap\, 674:A33. \doi{10.1051/0004-6361/202243709}, {\href{https://arxiv.org/abs/2206.06215}{{arXiv:2206.06215}}} {[astro-ph.SR]}

\bibitem[{{Gaia Collaboration} et~al.(2023{\natexlab{b}}){Gaia Collaboration}, {Vallenari}, {Brown}, {Prusti}, {de Bruijne}, {Arenou}, {Babusiaux}, {Biermann}, {Creevey}, {Ducourant}, {Evans}, {Eyer}, {Guerra}, {Hutton}, {Jordi}, {Klioner}, {Lammers}, {Lindegren}, {Luri}, {Mignard}, {Panem}, {Pourbaix}, {Randich}, {Sartoretti}, {Soubiran}, {Tanga}, {Walton}, {Bailer-Jones}, {Bastian}, {Drimmel}, {Jansen}, {Katz}, {Lattanzi}, {van Leeuwen}, {Bakker}, {Cacciari}, {Casta{\~n}eda}, {De Angeli}, {Fabricius}, {Fouesneau}, {Fr{\'e}mat}, {Galluccio}, {Guerrier}, {Heiter}, {Masana}, {Messineo}, {Mowlavi}, {Nicolas}, {Nienartowicz}, {Pailler}, {Panuzzo}, {Riclet}, {Roux}, {Seabroke}, {Sordo}, {Th{\'e}venin}, {Gracia-Abril}, {Portell}, {Teyssier}, {Altmann}, {Andrae}, {Audard}, {Bellas-Velidis}, {Benson}, {Berthier}, {Blomme}, {Burgess}, {Busonero}, {Busso}, {C{\'a}novas}, {Carry}, {Cellino}, {Cheek}, {Clementini}, {Damerdji}, {Davidson}, {de Teodoro}, {Nu{\~n}ez Campos}, {Delchambre}, {Dell'Oro}, {Esquej},
  {Fern{\'a}ndez-Hern{\'a}ndez}, {Fraile}, {Garabato}, {Garc{\'\i}a-Lario}, {Gosset}, {Haigron}, {Halbwachs}, {Hambly}, {Harrison}, {Hern{\'a}ndez}, {Hestroffer}, {Hodgkin}, {Holl}, {Jan{\ss}en}, {Jevardat de Fombelle}, {Jordan}, {Krone-Martins}, {Lanzafame}, {L{\"o}ffler}, {Marchal}, {Marrese}, {Moitinho}, {Muinonen}, {Osborne}, {Pancino}, {Pauwels}, {Recio-Blanco}, {Reyl{\'e}}, {Riello}, {Rimoldini}, {Roegiers}, {Rybizki}, {Sarro}, {Siopis}, {Smith}, {Sozzetti}, {Utrilla}, {van Leeuwen}, {Abbas}, {{\'A}brah{\'a}m}, {Abreu Aramburu}, {Aerts}, {Aguado}, {Ajaj}, {Aldea-Montero}, {Altavilla}, {{\'A}lvarez}, {Alves}, {Anders}, {Anderson}, {Anglada Varela}, {Antoja}, {Baines}, {Baker}, {Balaguer-N{\'u}{\~n}ez}, {Balbinot}, {Balog}, {Barache}, {Barbato}, {Barros}, {Barstow}, {Bartolom{\'e}}, {Bassilana}, {Bauchet}, {Becciani}, {Bellazzini}, {Berihuete}, {Bernet}, {Bertone}, {Bianchi}, {Binnenfeld}, {Blanco-Cuaresma}, {Blazere}, {Boch}, {Bombrun}, {Bossini}, {Bouquillon}, {Bragaglia}, {Bramante}, {Breedt},
  {Bressan}, {Brouillet}, {Brugaletta}, {Bucciarelli}, {Burlacu}, {Butkevich}, {Buzzi}, {Caffau}, {Cancelliere}, {Cantat-Gaudin}, {Carballo}, {Carlucci}, {Carnerero}, {Carrasco}, {Casamiquela}, {Castellani}, {Castro-Ginard}, {Chaoul}, {Charlot}, {Chemin}, {Chiaramida}, {Chiavassa}, {Chornay}, {Comoretto}, {Contursi}, {Cooper}, {Cornez}, {Cowell}, {Crifo}, {Cropper}, {Crosta}, {Crowley}, {Dafonte}, {Dapergolas}, {David}, {David}, {de Laverny}, {De Luise}, {De March}, {De Ridder}, {de Souza}, {de Torres}, {del Peloso}, {del Pozo}, {Delbo}, {Delgado}, {Delisle}, {Demouchy}, {Dharmawardena}, {Di Matteo}, {Diakite}, {Diener}, {Distefano}, {Dolding}, {Edvardsson}, {Enke}, {Fabre}, {Fabrizio}, {Faigler}, {Fedorets}, {Fernique}, {Fienga}, {Figueras}, {Fournier}, {Fouron}, {Fragkoudi}, {Gai}, {Garcia-Gutierrez}, {Garcia-Reinaldos}, {Garc{\'\i}a-Torres}, {Garofalo}, {Gavel}, {Gavras}, {Gerlach}, {Geyer}, {Giacobbe}, {Gilmore}, {Girona}, {Giuffrida}, {Gomel}, {Gomez}, {Gonz{\'a}lez-N{\'u}{\~n}ez},
  {Gonz{\'a}lez-Santamar{\'\i}a}, {Gonz{\'a}lez-Vidal}, {Granvik}, {Guillout}, {Guiraud}, {Guti{\'e}rrez-S{\'a}nchez}, {Guy}, {Hatzidimitriou}, {Hauser}, {Haywood}, {Helmer}, {Helmi}, {Sarmiento}, {Hidalgo}, {Hilger}, {H{\l}adczuk}, {Hobbs}, {Holland}, {Huckle}, {Jardine}, {Jasniewicz}, {Jean-Antoine Piccolo}, {Jim{\'e}nez-Arranz}, {Jorissen}, {Juaristi Campillo}, {Julbe}, {Karbevska}, {Kervella}, {Khanna}, {Kontizas}, {Kordopatis}, {Korn}, {K{\'o}sp{\'a}l}, {Kostrzewa-Rutkowska}, {Kruszy{\'n}ska}, {Kun}, {Laizeau}, {Lambert}, {Lanza}, {Lasne}, {Le Campion}, {Lebreton}, {Lebzelter}, {Leccia}, {Leclerc}, {Lecoeur-Taibi}, {Liao}, {Licata}, {Lindstr{\o}m}, {Lister}, {Livanou}, {Lobel}, {Lorca}, {Loup}, {Madrero Pardo}, {Magdaleno Romeo}, {Managau}, {Mann}, {Manteiga}, {Marchant}, {Marconi}, {Marcos}, {Marcos Santos}, {Mar{\'\i}n Pina}, {Marinoni}, {Marocco}, {Marshall}, {Martin Polo}, {Mart{\'\i}n-Fleitas}, {Marton}, {Mary}, {Masip}, {Massari}, {Mastrobuono-Battisti}, {Mazeh}, {McMillan}, {Messina}, {Michalik},
  {Millar}, {Mints}, {Molina}, {Molinaro}, {Moln{\'a}r}, {Monari}, {Mongui{\'o}}, {Montegriffo}, {Montero}, {Mor}, {Mora}, {Morbidelli}, {Morel}, {Morris}, {Muraveva}, {Murphy}, {Musella}, {Nagy}, {Noval}, {Oca{\~n}a}, {Ogden}, {Ordenovic}, {Osinde}, {Pagani}, {Pagano}, {Palaversa}, {Palicio}, {Pallas-Quintela}, {Panahi}, {Payne-Wardenaar}, {Pe{\~n}alosa Esteller}, {Penttil{\"a}}, {Pichon}, {Piersimoni}, {Pineau}, {Plachy}, {Plum}, {Poggio}, {Pr{\v{s}}a}, {Pulone}, {Racero}, {Ragaini}, {Rainer}, {Raiteri}, {Rambaux}, {Ramos}, {Ramos-Lerate}, {Re Fiorentin}, {Regibo}, {Richards}, {Rios Diaz}, {Ripepi}, {Riva}, {Rix}, {Rixon}, {Robichon}, {Robin}, {Robin}, {Roelens}, {Rogues}, {Rohrbasser}, {Romero-G{\'o}mez}, {Rowell}, {Royer}, {Ruz Mieres}, {Rybicki}, {Sadowski}, {S{\'a}ez N{\'u}{\~n}ez}, {Sagrist{\`a} Sell{\'e}s}, {Sahlmann}, {Salguero}, {Samaras}, {Sanchez Gimenez}, {Sanna}, {Santove{\~n}a}, {Sarasso}, {Schultheis}, {Sciacca}, {Segol}, {Segovia}, {S{\'e}gransan}, {Semeux}, {Shahaf}, {Siddiqui}, {Siebert},
  {Siltala}, {Silvelo}, {Slezak}, {Slezak}, {Smart}, {Snaith}, {Solano}, {Solitro}, {Souami}, {Souchay}, {Spagna}, {Spina}, {Spoto}, {Steele}, {Steidelm{\"u}ller}, {Stephenson}, {S{\"u}veges}, {Surdej}, {Szabados}, {Szegedi-Elek}, {Taris}, {Taylor}, {Teixeira}, {Tolomei}, {Tonello}, {Torra}, {Torra}, {Torralba Elipe}, {Trabucchi}, {Tsounis}, {Turon}, {Ulla}, {Unger}, {Vaillant}, {van Dillen}, {van Reeven}, {Vanel}, {Vecchiato}, {Viala}, {Vicente}, {Voutsinas}, {Weiler}, {Wevers}, {Wyrzykowski}, {Yoldas}, {Yvard}, {Zhao}, {Zorec}, {Zucker}, \& {Zwitter}}]{gaia2023a}
{Gaia Collaboration}, {Vallenari} A., {Brown} A.~G.~A., et~al. (2023{\natexlab{b}}): {Gaia Data Release 3. Summary of the content and survey properties}. \aap\, 674:A1. \doi{10.1051/0004-6361/202243940}, {\href{https://arxiv.org/abs/2208.00211}{{arXiv:2208.00211}}} {[astro-ph.GA]}

\bibitem[{{G{\"a}nsicke} et~al.(2012){G{\"a}nsicke}, {Koester}, {Farihi}, {Girven}, {Parsons}, \& {Breedt}}]{gansicke2012}
{G{\"a}nsicke} B.~T., {Koester} D., {Farihi} J., et~al. (2012): {The chemical diversity of exo-terrestrial planetary debris around white dwarfs}. \mnras\, 424(1):333--347. \doi{10.1111/j.1365-2966.2012.21201.x}, {\href{https://arxiv.org/abs/1205.0167}{{arXiv:1205.0167}}} {[astro-ph.EP]}

\bibitem[{{Garc{\'\i}a-Berro} et~al.(2010){Garc{\'\i}a-Berro}, {Torres}, {Althaus}, {Renedo}, {Lor{\'e}n-Aguilar}, {C{\'o}rsico}, {Rohrmann}, {Salaris}, \& {Isern}}]{garcia-berro2010}
{Garc{\'\i}a-Berro} E., {Torres} S., {Althaus} L.~G., et~al. (2010): {A white dwarf cooling age of 8Gyr for NGC 6791 from physical separation processes}. \nat\, 465(7295):194--196. \doi{10.1038/nature09045}, {\href{https://arxiv.org/abs/1005.2272}{{arXiv:1005.2272}}} {[astro-ph.SR]}

\bibitem[{{Garc{\'\i}a-Zamora} et~al.(2023){Garc{\'\i}a-Zamora}, {Torres}, \& {Rebassa-Mansergas}}]{garcia-zamora2023}
{Garc{\'\i}a-Zamora} E.~M., {Torres} S., {Rebassa-Mansergas} A. (2023): {White dwarf Random Forest classification through Gaia spectral coefficients}. \aap\, 679:A127. \doi{10.1051/0004-6361/202347601}, {\href{https://arxiv.org/abs/2308.07090}{{arXiv:2308.07090}}} {[astro-ph.SR]}

\bibitem[{{Genest-Beaulieu} \& {Bergeron}(2019{\natexlab{a}})}]{genest-beaulieu2019a}
{Genest-Beaulieu} C., {Bergeron} P. (2019{\natexlab{a}}): {A Comprehensive Spectroscopic and Photometric Analysis of DA and DB White Dwarfs from SDSS and Gaia}. \apj\, 871(2):169. \doi{10.3847/1538-4357/aafac6}

\bibitem[{{Genest-Beaulieu} \& {Bergeron}(2019{\natexlab{b}})}]{genest-beaulieu2019b}
{Genest-Beaulieu} C., {Bergeron} P. (2019{\natexlab{b}}): {A Photometric and Spectroscopic Investigation of the DB White Dwarf Population Using SDSS and Gaia Data}. \apj\, 882(2):106. \doi{10.3847/1538-4357/ab379e}, {\href{https://arxiv.org/abs/1908.01728}{{arXiv:1908.01728}}} {[astro-ph.SR]}

\bibitem[{{Gentile Fusillo} et~al.(2017){Gentile Fusillo}, {G{\"a}nsicke}, {Farihi}, {Koester}, {Schreiber}, \& {Pala}}]{gentile-fusillo2017}
{Gentile Fusillo} N.~P., {G{\"a}nsicke} B.~T., {Farihi} J., et~al. (2017): {Trace hydrogen in helium atmosphere white dwarfs as a possible signature of water accretion}. \mnras\, 468(1):971--980. \doi{10.1093/mnras/stx468}, {\href{https://arxiv.org/abs/1702.06542}{{arXiv:1702.06542}}} {[astro-ph.SR]}

\bibitem[{{Gentile Fusillo} et~al.(2019){Gentile Fusillo}, {Tremblay}, {G{\"a}nsicke}, {Manser}, {Cunningham}, {Cukanovaite}, {Hollands}, {Marsh}, {Raddi}, {Jordan}, {Toonen}, {Geier}, {Barstow}, \& {Cummings}}]{gentile-fusillo2019}
{Gentile Fusillo} N.~P., {Tremblay} P.-E., {G{\"a}nsicke} B.~T., et~al. (2019): {A Gaia Data Release 2 catalogue of white dwarfs and a comparison with SDSS}. \mnras\, 482(4):4570--4591. \doi{10.1093/mnras/sty3016}, {\href{https://arxiv.org/abs/1807.03315}{{arXiv:1807.03315}}} {[astro-ph.SR]}

\bibitem[{{Gentile Fusillo} et~al.(2020){Gentile Fusillo}, {Tremblay}, {Bohlin}, {Deustua}, \& {Kalirai}}]{gentile-fusillo2020}
{Gentile Fusillo} N.~P., {Tremblay} P.-E., {Bohlin} R.~C., et~al. (2020): {Cool white dwarfs as standards for infrared observations}. \mnras\, 491(3):3613--3623. \doi{10.1093/mnras/stz2984}, {\href{https://arxiv.org/abs/1910.08087}{{arXiv:1910.08087}}} {[astro-ph.SR]}

\bibitem[{{Gentile Fusillo} et~al.(2021){Gentile Fusillo}, {Tremblay}, {Cukanovaite}, {Vorontseva}, {Lallement}, {Hollands}, {G{\"a}nsicke}, {Burdge}, {McCleery}, \& {Jordan}}]{gentile-fusillo2021}
{Gentile Fusillo} N.~P., {Tremblay} P.~E., {Cukanovaite} E., et~al. (2021): {A catalogue of white dwarfs in Gaia EDR3}. \mnras\, 508(3):3877--3896. \doi{10.1093/mnras/stab2672}, {\href{https://arxiv.org/abs/2106.07669}{{arXiv:2106.07669}}} {[astro-ph.SR]}

\bibitem[{{Giammichele} et~al.(2012){Giammichele}, {Bergeron}, \& {Dufour}}]{giammichele2012}
{Giammichele} N., {Bergeron} P., {Dufour} P. (2012): {Know Your Neighborhood: A Detailed Model Atmosphere Analysis of Nearby White Dwarfs}. \apjs\, 199(2):29. \doi{10.1088/0067-0049/199/2/29}, {\href{https://arxiv.org/abs/1202.5581}{{arXiv:1202.5581}}} {[astro-ph.SR]}

\bibitem[{{Gianninas} et~al.(2010){Gianninas}, {Bergeron}, {Dupuis}, \& {Ruiz}}]{gianninas2010}
{Gianninas} A., {Bergeron} P., {Dupuis} J., et~al. (2010): {Spectroscopic Analysis of Hot, Hydrogen-rich White Dwarfs: The Presence of Metals and the Balmer-line Problem}. \apj\, 720(1):581--602. \doi{10.1088/0004-637X/720/1/581}

\bibitem[{{Gianninas} et~al.(2011){Gianninas}, {Bergeron}, \& {Ruiz}}]{gianninas2011}
{Gianninas} A., {Bergeron} P., {Ruiz} M.~T. (2011): {A Spectroscopic Survey and Analysis of Bright, Hydrogen-rich White Dwarfs}. \apj\, 743(2):138. \doi{10.1088/0004-637X/743/2/138}, {\href{https://arxiv.org/abs/1109.3171}{{arXiv:1109.3171}}} {[astro-ph.SR]}

\bibitem[{{Good} et~al.(2004){Good}, {Barstow}, {Holberg}, {Sing}, {Burleigh}, \& {Dobbie}}]{good2004}
{Good} S.~A., {Barstow} M.~A., {Holberg} J.~B., et~al. (2004): {Comparison of the effective temperatures, gravities and helium abundances of DAO white dwarfs from Balmer and Lyman line studies}. \mnras\, 355(3):1031--1040. \doi{10.1111/j.1365-2966.2004.08406.x}

\bibitem[{{Good} et~al.(2005){Good}, {Barstow}, {Burleigh}, {Dobbie}, {Holberg}, \& {Hubeny}}]{good2005}
{Good} S.~A., {Barstow} M.~A., {Burleigh} M.~R., et~al. (2005): {Heavy element abundances in DAO white dwarfs measured from FUSE data}. \mnras\, 363(1):183--196. \doi{10.1111/j.1365-2966.2005.09428.x}, {\href{https://arxiv.org/abs/astro-ph/0507341}{{arXiv:astro-ph/0507341}}} {[astro-ph]}

\bibitem[{{Greenstein}(1986)}]{greenstein1986}
{Greenstein} J.~L. (1986): {The Frequency of Hydrogen White Dwarfs as Observed at High Signal to Noise}. \apj\, 304:334. \doi{10.1086/164168}

\bibitem[{{Hansen} et~al.(2007){Hansen}, {Anderson}, {Brewer}, {Dotter}, {Fahlman}, {Hurley}, {Kalirai}, {King}, {Reitzel}, {Richer}, {Rich}, {Shara}, \& {Stetson}}]{hansen2007}
{Hansen} B. M.~S., {Anderson} J., {Brewer} J., et~al. (2007): {The White Dwarf Cooling Sequence of NGC 6397}. \apj\, 671(1):380--401. \doi{10.1086/522567}, {\href{https://arxiv.org/abs/astro-ph/0701738}{{arXiv:astro-ph/0701738}}} {[astro-ph]}

\bibitem[{{Harrison} et~al.(2018){Harrison}, {Bonsor}, \& {Madhusudhan}}]{harrison2018}
{Harrison} J. H.~D., {Bonsor} A., {Madhusudhan} N. (2018): {Polluted white dwarfs: constraints on the origin and geology of exoplanetary material}. \mnras\, 479(3):3814--3841. \doi{10.1093/mnras/sty1700}, {\href{https://arxiv.org/abs/1806.09917}{{arXiv:1806.09917}}} {[astro-ph.EP]}

\bibitem[{{Harrison} et~al.(2021){Harrison}, {Bonsor}, {Kama}, {Buchan}, {Blouin}, \& {Koester}}]{harrison2021}
{Harrison} J. H.~D., {Bonsor} A., {Kama} M., et~al. (2021): {Bayesian constraints on the origin and geology of exoplanetary material using a population of externally polluted white dwarfs}. \mnras\, 504(2):2853--2867. \doi{10.1093/mnras/stab736}, {\href{https://arxiv.org/abs/2103.05713}{{arXiv:2103.05713}}} {[astro-ph.EP]}

\bibitem[{{Heber} et~al.(1997){Heber}, {Napiwotzki}, {Lemke}, \& {Edelmann}}]{heber1997}
{Heber} U., {Napiwotzki} R., {Lemke} M., et~al. (1997): {Helium line profile variations in the DAB white dwarf HS 0209+0832.} \aap\, 324:L53--L56

\bibitem[{{Heinonen} et~al.(2020){Heinonen}, {Saumon}, {Daligault}, {Starrett}, {Baalrud}, \& {Fontaine}}]{heinonen2020}
{Heinonen} R.~A., {Saumon} D., {Daligault} J., et~al. (2020): {Diffusion Coefficients in the Envelopes of White Dwarfs}. \apj\, 896(1):2. \doi{10.3847/1538-4357/ab91ad}, {\href{https://arxiv.org/abs/2005.05891}{{arXiv:2005.05891}}} {[astro-ph.SR]}

\bibitem[{{Heintz} et~al.(2022){Heintz}, {Hermes}, {El-Badry}, {Walsh}, {van Saders}, {Fields}, \& {Koester}}]{heintz2022}
{Heintz} T.~M., {Hermes} J.~J., {El-Badry} K., et~al. (2022): {Testing White Dwarf Age Estimates Using Wide Double White Dwarf Binaries from Gaia EDR3}. \apj\, 934(2):148. \doi{10.3847/1538-4357/ac78d9}, {\href{https://arxiv.org/abs/2206.00025}{{arXiv:2206.00025}}} {[astro-ph.SR]}

\bibitem[{{Herwig} et~al.(1999){Herwig}, {Bl{\"o}cker}, {Langer}, \& {Driebe}}]{herwig1999}
{Herwig} F., {Bl{\"o}cker} T., {Langer} N., et~al. (1999): {On the formation of hydrogen-deficient post-AGB stars}. \aap\, 349:L5--L8. {\href{https://arxiv.org/abs/astro-ph/9908108}{{arXiv:astro-ph/9908108}}} {[astro-ph]}

\bibitem[{{Holberg} et~al.(1993){Holberg}, {Barstow}, {Buckley}, {Chen}, {Dreizler}, {Marsh}, {O'Donoghue}, {Sion}, {Tweedy}, {Vauclair}, \& {Werner}}]{holberg1993}
{Holberg} J.~B., {Barstow} M.~A., {Buckley} D.~A.~H., et~al. (1993): {Two New Extremely Iron-rich Hot DA White Dwarfs and the Nature of the EUV Opacity}. \apj\, 416:806. \doi{10.1086/173278}

\bibitem[{{Hollands} et~al.(2017){Hollands}, {Koester}, {Alekseev}, {Herbert}, \& {G{\"a}nsicke}}]{hollands2017}
{Hollands} M.~A., {Koester} D., {Alekseev} V., et~al. (2017): {Cool DZ white dwarfs - I. Identification and spectral analysis}. \mnras\, 467(4):4970--5000. \doi{10.1093/mnras/stx250}, {\href{https://arxiv.org/abs/1701.07827}{{arXiv:1701.07827}}} {[astro-ph.SR]}

\bibitem[{{Hollands} et~al.(2018{\natexlab{a}}){Hollands}, {G{\"a}nsicke}, \& {Koester}}]{hollands2018a}
{Hollands} M.~A., {G{\"a}nsicke} B.~T., {Koester} D. (2018{\natexlab{a}}): {Cool DZ white dwarfs II: compositions and evolution of old remnant planetary systems}. \mnras\, 477(1):93--111. \doi{10.1093/mnras/sty592}, {\href{https://arxiv.org/abs/1801.07714}{{arXiv:1801.07714}}} {[astro-ph.SR]}

\bibitem[{{Hollands} et~al.(2018{\natexlab{b}}){Hollands}, {Tremblay}, {G{\"a}nsicke}, {Gentile-Fusillo}, \& {Toonen}}]{hollands2018b}
{Hollands} M.~A., {Tremblay} P.~E., {G{\"a}nsicke} B.~T., et~al. (2018{\natexlab{b}}): {The Gaia 20 pc white dwarf sample}. \mnras\, 480(3):3942--3961. \doi{10.1093/mnras/sty2057}, {\href{https://arxiv.org/abs/1805.12590}{{arXiv:1805.12590}}} {[astro-ph.SR]}

\bibitem[{{Hollands} et~al.(2020){Hollands}, {Tremblay}, {G{\"a}nsicke}, {Camisassa}, {Koester}, {Aungwerojwit}, {Chote}, {C{\'o}rsico}, {Dhillon}, {Gentile-Fusillo}, {Hoskin}, {Izquierdo}, {Marsh}, \& {Steeghs}}]{hollands2020}
{Hollands} M.~A., {Tremblay} P.~E., {G{\"a}nsicke} B.~T., et~al. (2020): {An ultra-massive white dwarf with a mixed hydrogen-carbon atmosphere as a likely merger remnant}. Nature Astronomy\, 4:663--669. \doi{10.1038/s41550-020-1028-0}, {\href{https://arxiv.org/abs/2003.00028}{{arXiv:2003.00028}}} {[astro-ph.SR]}

\bibitem[{{Hollands} et~al.(2021){Hollands}, {Tremblay}, {G{\"a}nsicke}, {Koester}, \& {Gentile-Fusillo}}]{hollands2021}
{Hollands} M.~A., {Tremblay} P.-E., {G{\"a}nsicke} B.~T., et~al. (2021): {Alkali metals in white dwarf atmospheres as tracers of ancient planetary crusts}. Nature Astronomy\, 5:451--459. \doi{10.1038/s41550-020-01296-7}, {\href{https://arxiv.org/abs/2101.01225}{{arXiv:2101.01225}}} {[astro-ph.EP]}

\bibitem[{{Hollands} et~al.(2022){Hollands}, {Tremblay}, {G{\"a}nsicke}, \& {Koester}}]{hollands2022}
{Hollands} M.~A., {Tremblay} P.~E., {G{\"a}nsicke} B.~T., et~al. (2022): {Spectral analysis of cool white dwarfs accreting from planetary systems: from the ultraviolet to the optical}. \mnras\, 511(1):71--82. \doi{10.1093/mnras/stab3696}, {\href{https://arxiv.org/abs/2112.08887}{{arXiv:2112.08887}}} {[astro-ph.SR]}

\bibitem[{{Hoskin} et~al.(2020){Hoskin}, {Toloza}, {G{\"a}nsicke}, {Raddi}, {Koester}, {Pala}, {Manser}, {Farihi}, {Belmonte}, {Hollands}, {Gentile Fusillo}, \& {Swan}}]{hoskin2020}
{Hoskin} M.~J., {Toloza} O., {G{\"a}nsicke} B.~T., et~al. (2020): {White dwarf pollution by hydrated planetary remnants: hydrogen and metals in WD J204713.76-125908.9}. \mnras\, 499(1):171--182. \doi{10.1093/mnras/staa2717}, {\href{https://arxiv.org/abs/2009.05053}{{arXiv:2009.05053}}} {[astro-ph.EP]}

\bibitem[{{Hoyer} et~al.(2017){Hoyer}, {Rauch}, {Werner}, {Kruk}, \& {Quinet}}]{hoyer2017}
{Hoyer} D., {Rauch} T., {Werner} K., et~al. (2017): {Complete spectral energy distribution of the hot, helium-rich white dwarf RX J0503.9-2854}. \aap\, 598:A135. \doi{10.1051/0004-6361/201629869}, {\href{https://arxiv.org/abs/1610.09177}{{arXiv:1610.09177}}} {[astro-ph.SR]}

\bibitem[{{Hoyer} et~al.(2018){Hoyer}, {Rauch}, {Werner}, \& {Kruk}}]{hoyer2018}
{Hoyer} D., {Rauch} T., {Werner} K., et~al. (2018): {Search for trans-iron elements in hot, helium-rich white dwarfs with the HST Cosmic Origins Spectrograph}. \aap\, 612:A62. \doi{10.1051/0004-6361/201732401}, {\href{https://arxiv.org/abs/1801.02414}{{arXiv:1801.02414}}} {[astro-ph.SR]}

\bibitem[{{H{\"u}gelmeyer} et~al.(2005){H{\"u}gelmeyer}, {Dreizler}, {Werner}, {Krzesi{\'n}ski}, {Nitta}, \& {Kleinman}}]{hugelmeyer2005}
{H{\"u}gelmeyer} S.~D., {Dreizler} S., {Werner} K., et~al. (2005): {Spectral analyses of DO white dwarfs and PG 1159 stars from the Sloan Digital Sky Survey}. \aap\, 442(1):309--314. \doi{10.1051/0004-6361:20053280}, {\href{https://arxiv.org/abs/astro-ph/0508101}{{arXiv:astro-ph/0508101}}} {[astro-ph]}

\bibitem[{{H{\"u}gelmeyer} et~al.(2006){H{\"u}gelmeyer}, {Dreizler}, {Homeier}, {Krzesi{\'n}ski}, {Werner}, {Nitta}, \& {Kleinman}}]{hugelmeyer2006}
{H{\"u}gelmeyer} S.~D., {Dreizler} S., {Homeier} D., et~al. (2006): {Spectral analyses of eighteen hot H-deficient (pre-) white dwarfs from the Sloan Digital Sky Survey Data Release 4}. \aap\, 454(2):617--624. \doi{10.1051/0004-6361:20064869}, {\href{https://arxiv.org/abs/astro-ph/0605551}{{arXiv:astro-ph/0605551}}} {[astro-ph]}

\bibitem[{{Iben} \& {Tutukov}(1984)}]{iben1984}
{Iben} J.I., {Tutukov} A.~V. (1984): {Cooling of low-mass carbon-oxygen dwarfs from the planetary nucleus stage through the crystallization stage}. \apj\, 282:615--630. \doi{10.1086/162241}

\bibitem[{{Iben} et~al.(1983){Iben}, {Kaler}, {Truran}, \& {Renzini}}]{iben1983}
{Iben} J.I., {Kaler} J.~B., {Truran} J.~W., et~al. (1983): {On the evolution of those nuclei of planetary nebulae that experiencea final helium shell flash.} \apj\, 264:605--612. \doi{10.1086/160631}

\bibitem[{{Isern}(2019)}]{isern2019}
{Isern} J. (2019): {The Star Formation History in the Solar Neighborhood as Told by Massive White Dwarfs}. \apjl\, 878(1):L11. \doi{10.3847/2041-8213/ab238e}, {\href{https://arxiv.org/abs/1905.10779}{{arXiv:1905.10779}}} {[astro-ph.GA]}

\bibitem[{{Izquierdo} et~al.(2021){Izquierdo}, {Toloza}, {G{\"a}nsicke}, {Rodr{\'\i}guez-Gil}, {Farihi}, {Koester}, {Guo}, \& {Redfield}}]{izquierdo2021}
{Izquierdo} P., {Toloza} O., {G{\"a}nsicke} B.~T., et~al. (2021): {GD 424 - a helium-atmosphere white dwarf with a large amount of trace hydrogen in the process of digesting a rocky planetesimal}. \mnras\, 501(3):4276--4288. \doi{10.1093/mnras/staa3987}, {\href{https://arxiv.org/abs/2012.12957}{{arXiv:2012.12957}}} {[astro-ph.EP]}

\bibitem[{{Jim{\'e}nez-Esteban} et~al.(2018){Jim{\'e}nez-Esteban}, {Torres}, {Rebassa-Mansergas}, {Skorobogatov}, {Solano}, {Cantero}, \& {Rodrigo}}]{jimenez-esteban2018}
{Jim{\'e}nez-Esteban} F.~M., {Torres} S., {Rebassa-Mansergas} A., et~al. (2018): {A white dwarf catalogue from Gaia-DR2 and the Virtual Observatory}. \mnras\, 480(4):4505--4518. \doi{10.1093/mnras/sty2120}, {\href{https://arxiv.org/abs/1807.02559}{{arXiv:1807.02559}}} {[astro-ph.SR]}

\bibitem[{{Jim{\'e}nez-Esteban} et~al.(2023){Jim{\'e}nez-Esteban}, {Torres}, {Rebassa-Mansergas}, {Cruz}, {Murillo-Ojeda}, {Solano}, {Rodrigo}, \& {Camisassa}}]{jimenez-esteban2023}
{Jim{\'e}nez-Esteban} F.~M., {Torres} S., {Rebassa-Mansergas} A., et~al. (2023): {Spectral classification of the 100 pc white dwarf population from Gaia-DR3 and the virtual observatory}. \mnras\, 518(4):5106--5122. \doi{10.1093/mnras/stac3382}, {\href{https://arxiv.org/abs/2211.08852}{{arXiv:2211.08852}}} {[astro-ph.SR]}

\bibitem[{{Johnson} et~al.(2022){Johnson}, {Klein}, {Koester}, {Melis}, {Zuckerman}, \& {Jura}}]{johnson2022}
{Johnson} T.~M., {Klein} B.~L., {Koester} D., et~al. (2022): {Unusual Abundances from Planetary System Material Polluting the White Dwarf G238-44}. \apj\, 941(2):113. \doi{10.3847/1538-4357/aca089}, {\href{https://arxiv.org/abs/2211.02673}{{arXiv:2211.02673}}} {[astro-ph.EP]}

\bibitem[{{Jordan} \& {Koester}(1986)}]{jordan1986}
{Jordan} S., {Koester} D. (1986): {Model atmospheres and synthetic spectra for white dwarfs with chemically stratified atmospheres.} \aaps\, 65:367--377

\bibitem[{{Jura} \& {Xu}(2010)}]{jura2010}
{Jura} M., {Xu} S. (2010): {The Survival of Water Within Extrasolar Minor Planets}. \aj\, 140(5):1129--1136. \doi{10.1088/0004-6256/140/5/1129}, {\href{https://arxiv.org/abs/1001.2595}{{arXiv:1001.2595}}} {[astro-ph.EP]}

\bibitem[{{Jura} \& {Young}(2014)}]{jura2014}
{Jura} M., {Young} E.~D. (2014): {Extrasolar Cosmochemistry}. \areps\, 42(1):45--67. \doi{10.1146/annurev-earth-060313-054740}

\bibitem[{{Jura} et~al.(2012){Jura}, {Xu}, {Klein}, {Koester}, \& {Zuckerman}}]{jura2012}
{Jura} M., {Xu} S., {Klein} B., et~al. (2012): {Two Extrasolar Asteroids with Low Volatile-element Mass Fractions}. \apj\, 750(1):69. \doi{10.1088/0004-637X/750/1/69}, {\href{https://arxiv.org/abs/1203.2885}{{arXiv:1203.2885}}} {[astro-ph.EP]}

\bibitem[{{Kaiser} et~al.(2021){Kaiser}, {Clemens}, {Blouin}, {Dufour}, {Hegedus}, {Reding}, \& {B{\'e}dard}}]{kaiser2021}
{Kaiser} B.~C., {Clemens} J.~C., {Blouin} S., et~al. (2021): {Lithium pollution of a white dwarf records the accretion of an extrasolar planetesimal}. Science\, 371(6525):168--172. \doi{10.1126/science.abd1714}, {\href{https://arxiv.org/abs/2012.12900}{{arXiv:2012.12900}}} {[astro-ph.EP]}

\bibitem[{{Kalirai}(2012)}]{kalirai2012}
{Kalirai} J.~S. (2012): {The age of the Milky Way inner halo}. \nat\, 486(7401):90--92. \doi{10.1038/nature11062}, {\href{https://arxiv.org/abs/1205.6802}{{arXiv:1205.6802}}} {[astro-ph.GA]}

\bibitem[{{Kawka} et~al.(2023){Kawka}, {Ferrario}, \& {Vennes}}]{kawka2023}
{Kawka} A., {Ferrario} L., {Vennes} S. (2023): {The non-explosive stellar merging origin of the ultra-massive carbon-rich white dwarfs}. \mnras\, 520(4):6299--6311. \doi{10.1093/mnras/stad553}, {\href{https://arxiv.org/abs/2302.11118}{{arXiv:2302.11118}}} {[astro-ph.SR]}

\bibitem[{{Kepler} et~al.(2019){Kepler}, {Pelisoli}, {Koester}, {Reindl}, {Geier}, {Romero}, {Ourique}, {Oliveira}, \& {Amaral}}]{kepler2019}
{Kepler} S.~O., {Pelisoli} I., {Koester} D., et~al. (2019): {White dwarf and subdwarf stars in the Sloan Digital Sky Survey Data Release 14}. \mnras\, 486(2):2169--2183. \doi{10.1093/mnras/stz960}, {\href{https://arxiv.org/abs/1904.01626}{{arXiv:1904.01626}}} {[astro-ph.SR]}

\bibitem[{{Kepler} et~al.(2021){Kepler}, {Koester}, {Pelisoli}, {Romero}, \& {Ourique}}]{kepler2021}
{Kepler} S.~O., {Koester} D., {Pelisoli} I., et~al. (2021): {White dwarf and subdwarf stars in the Sloan Digital Sky Survey Data Release 16}. \mnras\, 507(3):4646--4660. \doi{10.1093/mnras/stab2411}, {\href{https://arxiv.org/abs/2108.10915}{{arXiv:2108.10915}}} {[astro-ph.SR]}

\bibitem[{{Kilic} et~al.(2017){Kilic}, {Munn}, {Harris}, {von Hippel}, {Liebert}, {Williams}, {Jeffery}, \& {DeGennaro}}]{kilic2017}
{Kilic} M., {Munn} J.~A., {Harris} H.~C., et~al. (2017): {The Ages of the Thin Disk, Thick Disk, and the Halo from Nearby White Dwarfs}. \apj\, 837(2):162. \doi{10.3847/1538-4357/aa62a5}, {\href{https://arxiv.org/abs/1702.06984}{{arXiv:1702.06984}}} {[astro-ph.SR]}

\bibitem[{{Kilic} et~al.(2018){Kilic}, {Hambly}, {Bergeron}, {Genest-Beaulieu}, \& {Rowell}}]{kilic2018}
{Kilic} M., {Hambly} N.~C., {Bergeron} P., et~al. (2018): {Gaia reveals evidence for merged white dwarfs}. \mnras\, 479(1):L113--L117. \doi{10.1093/mnrasl/sly110}, {\href{https://arxiv.org/abs/1805.01227}{{arXiv:1805.01227}}} {[astro-ph.SR]}

\bibitem[{{Kilic} et~al.(2020){Kilic}, {Bergeron}, {Kosakowski}, {Brown}, {Ag{\"u}eros}, \& {Blouin}}]{kilic2020}
{Kilic} M., {Bergeron} P., {Kosakowski} A., et~al. (2020): {The 100 pc White Dwarf Sample in the SDSS Footprint}. \apj\, 898(1):84. \doi{10.3847/1538-4357/ab9b8d}, {\href{https://arxiv.org/abs/2006.00323}{{arXiv:2006.00323}}} {[astro-ph.SR]}

\bibitem[{{Kilic} et~al.(2024){Kilic}, {Bergeron}, {Blouin}, {Jewett}, {Brown}, \& {Moss}}]{kilic2024}
{Kilic} M., {Bergeron} P., {Blouin} S., et~al. (2024): {White Dwarf Merger Remnants: The DAQ Subclass}. arXiv e-prints\, arXiv:2403.08878. \doi{10.48550/arXiv.2403.08878}, {\href{https://arxiv.org/abs/2403.08878}{{arXiv:2403.08878}}} {[astro-ph.SR]}

\bibitem[{{Klein} et~al.(2010){Klein}, {Jura}, {Koester}, {Zuckerman}, \& {Melis}}]{klein2010}
{Klein} B., {Jura} M., {Koester} D., et~al. (2010): {Chemical Abundances in the Externally Polluted White Dwarf GD 40: Evidence of a Rocky Extrasolar Minor Planet}. \apj\, 709(2):950--962. \doi{10.1088/0004-637X/709/2/950}, {\href{https://arxiv.org/abs/0912.1422}{{arXiv:0912.1422}}} {[astro-ph.EP]}

\bibitem[{{Klein} et~al.(2011){Klein}, {Jura}, {Koester}, \& {Zuckerman}}]{klein2011}
{Klein} B., {Jura} M., {Koester} D., et~al. (2011): {Rocky Extrasolar Planetary Compositions Derived from Externally Polluted White Dwarfs}. \apj\, 741(1):64. \doi{10.1088/0004-637X/741/1/64}, {\href{https://arxiv.org/abs/1108.1565}{{arXiv:1108.1565}}} {[astro-ph.EP]}

\bibitem[{{Klein} et~al.(2021){Klein}, {Doyle}, {Zuckerman}, {Dufour}, {Blouin}, {Melis}, {Weinberger}, \& {Young}}]{klein2021}
{Klein} B.~L., {Doyle} A.~E., {Zuckerman} B., et~al. (2021): {Discovery of Beryllium in White Dwarfs Polluted by Planetesimal Accretion}. \apj\, 914(1):61. \doi{10.3847/1538-4357/abe40b}, {\href{https://arxiv.org/abs/2102.01834}{{arXiv:2102.01834}}} {[astro-ph.SR]}

\bibitem[{{Koester}(1976)}]{koester1976}
{Koester} D. (1976): {Convective Mixing and Accretion in White Dwarfs}. \aap\, 52:415

\bibitem[{{Koester}(2009)}]{koester2009}
{Koester} D. (2009): {Accretion and diffusion in white dwarfs. New diffusion timescales and applications to GD 362 and G 29-38}. \aap\, 498(2):517--525. \doi{10.1051/0004-6361/200811468}, {\href{https://arxiv.org/abs/0903.1499}{{arXiv:0903.1499}}} {[astro-ph.SR]}

\bibitem[{{Koester}(2015)}]{koester2015a}
{Koester} D. (2015): {On Thermohaline Mixing in Accreting White Dwarfs}. In: {Dufour} P., {Bergeron} P., {Fontaine} G. (eds.) ASP Conf. Ser. 493: 19th European Workshop on White Dwarfs. San Francisco: Astronomical Society of the Pacific, p. 129, \doi{10.48550/arXiv.1408.6934}, \eprint{1408.6934}

\bibitem[{{Koester} \& {Kepler}(2015)}]{koester2015b}
{Koester} D., {Kepler} S.~O. (2015): {DB white dwarfs in the Sloan Digital Sky Survey data release 10 and 12}. \aap\, 583:A86. \doi{10.1051/0004-6361/201527169}, {\href{https://arxiv.org/abs/1509.08244}{{arXiv:1509.08244}}} {[astro-ph.SR]}

\bibitem[{{Koester} \& {Kepler}(2019)}]{koester2019}
{Koester} D., {Kepler} S.~O. (2019): {Carbon-rich (DQ) white dwarfs in the Sloan Digital Sky Survey}. \aap\, 628:A102. \doi{10.1051/0004-6361/201935946}, {\href{https://arxiv.org/abs/1905.11174}{{arXiv:1905.11174}}} {[astro-ph.SR]}

\bibitem[{{Koester} \& {Knist}(2006)}]{koester2006b}
{Koester} D., {Knist} S. (2006): {New DQ white dwarfs in the Sloan Digital Sky Survey DR4: confirmation of two sequences}. \aap\, 454(3):951--956. \doi{10.1051/0004-6361:20065287}, {\href{https://arxiv.org/abs/astro-ph/0603734}{{arXiv:astro-ph/0603734}}} {[astro-ph]}

\bibitem[{{Koester} \& {Wilken}(2006)}]{koester2006a}
{Koester} D., {Wilken} D. (2006): {The accretion-diffusion scenario for metals in cool white dwarfs}. \aap\, 453(3):1051--1057. \doi{10.1051/0004-6361:20064843}, {\href{https://arxiv.org/abs/astro-ph/0603185}{{arXiv:astro-ph/0603185}}} {[astro-ph]}

\bibitem[{{Koester} et~al.(1982){Koester}, {Weidemann}, \& {Zeidler}}]{koester1982}
{Koester} D., {Weidemann} V., {Zeidler} E.~M. (1982): {Atmospheric parameters and carbon abundance of white dwarfs of spectral types C2 and DC.} \aap\, 116:147--157

\bibitem[{{Koester} et~al.(1994){Koester}, {Liebert}, \& {Saffer}}]{koester1994}
{Koester} D., {Liebert} J., {Saffer} R.~A. (1994): {GD 323: New Observations and Analysis of the Prototype DAB White Dwarf}. \apj\, 422:783. \doi{10.1086/173770}

\bibitem[{{Koester} et~al.(2005){Koester}, {Rollenhagen}, {Napiwotzki}, {Voss}, {Christlieb}, {Homeier}, \& {Reimers}}]{koester2005}
{Koester} D., {Rollenhagen} K., {Napiwotzki} R., et~al. (2005): {Metal traces in white dwarfs of the SPY (ESO Supernova Ia Progenitor Survey) sample}. \aap\, 432(3):1025--1032. \doi{10.1051/0004-6361:20041927}

\bibitem[{{Koester} et~al.(2011){Koester}, {Girven}, {G{\"a}nsicke}, \& {Dufour}}]{koester2011}
{Koester} D., {Girven} J., {G{\"a}nsicke} B.~T., et~al. (2011): {Cool DZ white dwarfs in the SDSS}. \aap\, 530:A114. \doi{10.1051/0004-6361/201116816}, {\href{https://arxiv.org/abs/1105.0268}{{arXiv:1105.0268}}} {[astro-ph.SR]}

\bibitem[{{Koester} et~al.(2014{\natexlab{a}}){Koester}, {G{\"a}nsicke}, \& {Farihi}}]{koester2014a}
{Koester} D., {G{\"a}nsicke} B.~T., {Farihi} J. (2014{\natexlab{a}}): {The frequency of planetary debris around young white dwarfs}. \aap\, 566:A34. \doi{10.1051/0004-6361/201423691}, {\href{https://arxiv.org/abs/1404.2617}{{arXiv:1404.2617}}} {[astro-ph.SR]}

\bibitem[{{Koester} et~al.(2014{\natexlab{b}}){Koester}, {Provencal}, \& {G{\"a}nsicke}}]{koester2014b}
{Koester} D., {Provencal} J., {G{\"a}nsicke} B.~T. (2014{\natexlab{b}}): {Atmospheric parameters and carbon abundance for hot DB white dwarfs}. \aap\, 568:A118. \doi{10.1051/0004-6361/201424231}, {\href{https://arxiv.org/abs/1407.6157}{{arXiv:1407.6157}}} {[astro-ph.SR]}

\bibitem[{{Koester} et~al.(2020){Koester}, {Kepler}, \& {Irwin}}]{koester2020}
{Koester} D., {Kepler} S.~O., {Irwin} A.~W. (2020): {New white dwarf envelope models and diffusion. Application to DQ white dwarfs}. \aap\, 635:A103. \doi{10.1051/0004-6361/202037530}, {\href{https://arxiv.org/abs/2002.10170}{{arXiv:2002.10170}}} {[astro-ph.SR]}

\bibitem[{{Kollmeier} et~al.(2017){Kollmeier}, {Zasowski}, {Rix}, {Johns}, {Anderson}, {Drory}, {Johnson}, {Pogge}, {Bird}, {Blanc}, {Brownstein}, {Crane}, {De Lee}, {Klaene}, {Kreckel}, {MacDonald}, {Merloni}, {Ness}, {O'Brien}, {Sanchez-Gallego}, {Sayres}, {Shen}, {Thakar}, {Tkachenko}, {Aerts}, {Blanton}, {Eisenstein}, {Holtzman}, {Maoz}, {Nandra}, {Rockosi}, {Weinberg}, {Bovy}, {Casey}, {Chaname}, {Clerc}, {Conroy}, {Eracleous}, {G{\"a}nsicke}, {Hekker}, {Horne}, {Kauffmann}, {McQuinn}, {Pellegrini}, {Schinnerer}, {Schlafly}, {Schwope}, {Seibert}, {Teske}, \& {van Saders}}]{kollmeier2017}
{Kollmeier} J.~A., {Zasowski} G., {Rix} H.-W., et~al. (2017): {SDSS-V: Pioneering Panoptic Spectroscopy}. arXiv e-prints\, arXiv:1711.03234. \doi{10.48550/arXiv.1711.03234}, {\href{https://arxiv.org/abs/1711.03234}{{arXiv:1711.03234}}} {[astro-ph.GA]}

\bibitem[{{Krzesinski} et~al.(2009){Krzesinski}, {Kleinman}, {Nitta}, {H{\"u}gelmeyer}, {Dreizler}, {Liebert}, \& {Harris}}]{krzesinski2009}
{Krzesinski} J., {Kleinman} S.~J., {Nitta} A., et~al. (2009): {A hot white dwarf luminosity function from the Sloan Digital Sky Survey}. \aap\, 508(1):339--344. \doi{10.1051/0004-6361/200912094}

\bibitem[{{Kudritzki} \& {Puls}(2000)}]{kudritzki2000}
{Kudritzki} R.-P., {Puls} J. (2000): {Winds from Hot Stars}. \araa\, 38:613--666. \doi{10.1146/annurev.astro.38.1.613}

\bibitem[{{Kupka} et~al.(2018){Kupka}, {Zaussinger}, \& {Montgomery}}]{kupka2018}
{Kupka} F., {Zaussinger} F., {Montgomery} M.~H. (2018): {Mixing and overshooting in surface convection zones of DA white dwarfs: first results from ANTARES}. \mnras\, 474(4):4660--4671. \doi{10.1093/mnras/stx3119}, {\href{https://arxiv.org/abs/1712.00641}{{arXiv:1712.00641}}} {[astro-ph.SR]}

\bibitem[{{Lawlor} \& {MacDonald}(2006)}]{lawlor2006}
{Lawlor} T.~M., {MacDonald} J. (2006): {The mass of helium in white dwarf stars and the formation and evolution of hydrogen-deficient post-AGB stars}. \mnras\, 371(1):263--282. \doi{10.1111/j.1365-2966.2006.10641.x}, {\href{https://arxiv.org/abs/astro-ph/0605747}{{arXiv:astro-ph/0605747}}} {[astro-ph]}

\bibitem[{{Leggett} et~al.(1998){Leggett}, {Ruiz}, \& {Bergeron}}]{leggett1998}
{Leggett} S.~K., {Ruiz} M.~T., {Bergeron} P. (1998): {The Cool White Dwarf Luminosity Function and the Age of the Galactic Disk}. \apj\, 497(1):294--302. \doi{10.1086/305463}

\bibitem[{{Liebert}(1983)}]{liebert1983}
{Liebert} J. (1983): {G35-26: carbon in a peculiar DA white dwarf.} \pasp\, 95:878--882. \doi{10.1086/131264}

\bibitem[{{Liebert} et~al.(2005){Liebert}, {Bergeron}, \& {Holberg}}]{liebert2005}
{Liebert} J., {Bergeron} P., {Holberg} J.~B. (2005): {The Formation Rate and Mass and Luminosity Functions of DA White Dwarfs from the Palomar Green Survey}. \apjs\, 156(1):47--68. \doi{10.1086/425738}, {\href{https://arxiv.org/abs/astro-ph/0406657}{{arXiv:astro-ph/0406657}}} {[astro-ph]}

\bibitem[{{Limoges} \& {Bergeron}(2010)}]{limoges2010}
{Limoges} M.~M., {Bergeron} P. (2010): {A Spectroscopic Analysis of White Dwarfs in the Kiso Survey}. \apj\, 714(2):1037--1051. \doi{10.1088/0004-637X/714/2/1037}, {\href{https://arxiv.org/abs/1003.4313}{{arXiv:1003.4313}}} {[astro-ph.SR]}

\bibitem[{{Limoges} et~al.(2009){Limoges}, {Bergeron}, \& {Dufour}}]{limoges2009}
{Limoges} M.~M., {Bergeron} P., {Dufour} P. (2009): {Spectroscopic Analysis of the White Dwarf KUV 02196+2816: A New Unresolved DA+DB Degenerate Binary}. \apj\, 696(2):1461--1465. \doi{10.1088/0004-637X/696/2/1461}, {\href{https://arxiv.org/abs/0902.3640}{{arXiv:0902.3640}}} {[astro-ph.SR]}

\bibitem[{{Limoges} et~al.(2015){Limoges}, {Bergeron}, \& {L{\'e}pine}}]{limoges2015}
{Limoges} M.~M., {Bergeron} P., {L{\'e}pine} S. (2015): {Physical Properties of the Current Census of Northern White Dwarfs within 40 pc of the Sun}. \apjs\, 219(2):19. \doi{10.1088/0067-0049/219/2/19}, {\href{https://arxiv.org/abs/1505.02297}{{arXiv:1505.02297}}} {[astro-ph.SR]}

\bibitem[{{L{\"o}bling} et~al.(2020){L{\"o}bling}, {Maney}, {Rauch}, {Quinet}, {Gamrath}, {Kruk}, \& {Werner}}]{lobling2020}
{L{\"o}bling} L., {Maney} M.~A., {Rauch} T., et~al. (2020): {First discovery of trans-iron elements in a DAO-type white dwarf (BD-22$^{{\circ}}$3467)}. \mnras\, 492(1):528--548. \doi{10.1093/mnras/stz3247}, {\href{https://arxiv.org/abs/1911.09573}{{arXiv:1911.09573}}} {[astro-ph.SR]}

\bibitem[{{L{\'o}pez-Sanjuan} et~al.(2022){L{\'o}pez-Sanjuan}, {Tremblay}, {Ederoclite}, {V{\'a}zquez Rami{\'o}}, {Carrasco}, {Varela}, {Cenarro}, {Mar{\'\i}n-Franch}, {Civera}, {Daflon}, {G{\"a}nsicke}, {Gentile Fusillo}, {Jim{\'e}nez-Esteban}, {Alcaniz}, {Angulo}, {Crist{\'o}bal-Hornillos}, {Dupke}, {Hern{\'a}ndez-Monteagudo}, {Moles}, \& {Sodr{\'e}}}]{lopez-sanjuan2022}
{L{\'o}pez-Sanjuan} C., {Tremblay} P.~E., {Ederoclite} A., et~al. (2022): {J-PLUS: Spectral evolution of white dwarfs by PDF analysis}. \aap\, 658:A79. \doi{10.1051/0004-6361/202141746}, {\href{https://arxiv.org/abs/2110.14421}{{arXiv:2110.14421}}} {[astro-ph.SR]}

\bibitem[{{MacDonald} \& {Vennes}(1991)}]{macdonald1991}
{MacDonald} J., {Vennes} S. (1991): {How Much Hydrogen Is There in a White Dwarf?} \apj\, 371:719. \doi{10.1086/169937}

\bibitem[{{MacDonald} et~al.(1998){MacDonald}, {Hernanz}, \& {Jose}}]{macdonald1998}
{MacDonald} J., {Hernanz} M., {Jose} J. (1998): {Evolutionary calculations of carbon dredge-up in helium envelope white dwarfs}. \mnras\, 296(3):523--530. \doi{10.1046/j.1365-8711.1998.01392.x}, {\href{https://arxiv.org/abs/astro-ph/9803121}{{arXiv:astro-ph/9803121}}} {[astro-ph]}

\bibitem[{{Macfarlane} et~al.(2017){Macfarlane}, {Woudt}, {Dufour}, {Ramsay}, {Groot}, {Toma}, {Warner}, {Paterson}, {Kupfer}, {van Roestel}, {Berdnikov}, {Dagne}, \& {Hardy}}]{macfarlane2017}
{Macfarlane} S.~A., {Woudt} P.~A., {Dufour} P., et~al. (2017): {The OmegaWhite Survey for short-period variable stars - IV. Discovery of the warm DQ white dwarf OW J175358.85-310728.9}. \mnras\, 470(1):732--741. \doi{10.1093/mnras/stx741}, {\href{https://arxiv.org/abs/1703.08122}{{arXiv:1703.08122}}} {[astro-ph.SR]}

\bibitem[{{Manseau} et~al.(2016){Manseau}, {Bergeron}, \& {Green}}]{manseau2016}
{Manseau} P.~M., {Bergeron} P., {Green} E.~M. (2016): {A Spectroscopic Search for Chemically Stratified White Dwarfs in the Sloan Digital Sky Survey}. \apj\, 833(2):127. \doi{10.3847/1538-4357/833/2/127}

\bibitem[{{Manser} et~al.(2024){Manser}, {G{\"a}nsicke}, {Izquierdo}, {Swan}, {Najita}, {Rockosi}, {Carrillo}, {Kim}, {Xu}, {Dey}, {Aguilar}, {Ahlen}, {Blum}, {Brooks}, {Claybaugh}, {Dawson}, {de la Macorra}, {Doel}, {Gazta{\~n}aga}, {Gontcho}, {Honscheid}, {Kehoe}, {Kremin}, {Landriau}, {Le Guillou}, {Levi}, {Li}, {Meisner}, {Miquel}, {Nie}, {Rezaie}, {Rossi}, {Sanchez}, {Schubnell}, {Tarl{\'e}}, {Weaver}, {Zhou}, \& {Zou}}]{manser2024}
{Manser} C.~J., {G{\"a}nsicke} B.~T., {Izquierdo} P., et~al. (2024): {The frequency of metal-enrichment of cool helium-atmosphere white dwarfs using the DESI Early Data Release}. \mnras\, \doi{10.1093/mnrasl/slae026}, {\href{https://arxiv.org/abs/2402.18644}{{arXiv:2402.18644}}} {[astro-ph.EP]}

\bibitem[{{Martin} et~al.(2005){Martin}, {Fanson}, {Schiminovich}, {Morrissey}, {Friedman}, {Barlow}, {Conrow}, {Grange}, {Jelinsky}, {Milliard}, {Siegmund}, {Bianchi}, {Byun}, {Donas}, {Forster}, {Heckman}, {Lee}, {Madore}, {Malina}, {Neff}, {Rich}, {Small}, {Surber}, {Szalay}, {Welsh}, \& {Wyder}}]{martin2005}
{Martin} D.~C., {Fanson} J., {Schiminovich} D., et~al. (2005): {The Galaxy Evolution Explorer: A Space Ultraviolet Survey Mission}. \apjl\, 619(1):L1--L6. \doi{10.1086/426387}, {\href{https://arxiv.org/abs/astro-ph/0411302}{{arXiv:astro-ph/0411302}}} {[astro-ph]}

\bibitem[{{McCleery} et~al.(2020){McCleery}, {Tremblay}, {Gentile Fusillo}, {Hollands}, {G{\"a}nsicke}, {Izquierdo}, {Toonen}, {Cunningham}, \& {Rebassa-Mansergas}}]{mccleery2020}
{McCleery} J., {Tremblay} P.-E., {Gentile Fusillo} N.~P., et~al. (2020): {Gaia white dwarfs within 40 pc II: the volume-limited Northern hemisphere sample}. \mnras\, 499(2):1890--1908. \doi{10.1093/mnras/staa2030}, {\href{https://arxiv.org/abs/2006.00874}{{arXiv:2006.00874}}} {[astro-ph.SR]}

\bibitem[{{Melis} et~al.(2011){Melis}, {Farihi}, {Dufour}, {Zuckerman}, {Burgasser}, {Bergeron}, {Bochanski}, \& {Simcoe}}]{melis2011}
{Melis} C., {Farihi} J., {Dufour} P., et~al. (2011): {Accretion of a Terrestrial-like Minor Planet by a White Dwarf}. \apj\, 732(2):90. \doi{10.1088/0004-637X/732/2/90}, {\href{https://arxiv.org/abs/1102.0311}{{arXiv:1102.0311}}} {[astro-ph.SR]}

\bibitem[{{Michaud} et~al.(2015){Michaud}, {Alecian}, \& {Richer}}]{michaud2015}
{Michaud} G., {Alecian} G., {Richer} J. (2015): {Atomic Diffusion in Stars}. Cham: Springer International

\bibitem[{{Miller Bertolami} \& {Althaus}(2006)}]{miller-bertolami2006b}
{Miller Bertolami} M.~M., {Althaus} L.~G. (2006): {Full evolutionary models for PG 1159 stars. Implications for the helium-rich O(He) stars}. \aap\, 454(3):845--854. \doi{10.1051/0004-6361:20054723}, {\href{https://arxiv.org/abs/astro-ph/0603846}{{arXiv:astro-ph/0603846}}} {[astro-ph]}

\bibitem[{{Miller Bertolami} et~al.(2006){Miller Bertolami}, {Althaus}, {Serenelli}, \& {Panei}}]{miller-bertolami2006a}
{Miller Bertolami} M.~M., {Althaus} L.~G., {Serenelli} A.~M., et~al. (2006): {New evolutionary calculations for the born again scenario}. \aap\, 449(1):313--326. \doi{10.1051/0004-6361:20053804}, {\href{https://arxiv.org/abs/astro-ph/0511406}{{arXiv:astro-ph/0511406}}} {[astro-ph]}

\bibitem[{{Miller Bertolami} et~al.(2017){Miller Bertolami}, {Althaus}, \& {C{\'o}rsico}}]{miller-bertolami2017}
{Miller Bertolami} M.~M., {Althaus} L.~G., {C{\'o}rsico} A.~H. (2017): {On the Formation of DA White Dwarfs with low Hydrogen Contents: Preliminary Results}. In: {Tremblay} P.~E., {G{\"a}nsicke} B., {Marsh} T. (eds.) ASP Conf. Ser. 509: 20th European Workshop on White Dwarfs. San Francisco: Astronomical Society of the Pacific, p. 435, \eprint{1609.08683}

\bibitem[{{Moss} et~al.(2024){Moss}, {Bergeron}, {Kilic}, {Jewett}, {Brown}, {Kosakowski}, \& {Vincent}}]{moss2024}
{Moss} A., {Bergeron} P., {Kilic} M., et~al. (2024): {Discovery of a magnetic double-faced DBA white dwarf}. \mnras\, 527(4):10111--10122. \doi{10.1093/mnras/stad3825}

\bibitem[{{Napiwotzki}(1992)}]{napiwotzki1992}
{Napiwotzki} R. (1992): {Analysis of central stars of old planetary nebulae: Problems with the Balmer lines}. In: {Heber} U., {Jeffery} C.~S. (eds.) The Atmospheres of Early-Type Stars, vol. 401. Berlin: Springer, p. 310, \doi{10.1007/3-540-55256-1\_328}

\bibitem[{{Napiwotzki}(1999)}]{napiwotzki1999}
{Napiwotzki} R. (1999): {Spectroscopic investigation of old planetaries. IV. Model atmosphere analysis}. \aap\, 350:101--119. {\href{https://arxiv.org/abs/astro-ph/9908181}{{arXiv:astro-ph/9908181}}} {[astro-ph]}

\bibitem[{{Napiwotzki} \& {Rauch}(1994)}]{napiwotzki1994}
{Napiwotzki} R., {Rauch} T. (1994): {The Balmer line problem of hot stars and the impact of ion-dynamical effects on the Stark broadening of HI and HeII lines}. \aap\, 285:603--608

\bibitem[{{O'Brien} et~al.(2023){O'Brien}, {Tremblay}, {Gentile Fusillo}, {Hollands}, {G{\"a}nsicke}, {Koester}, {Pelisoli}, {Cukanovaite}, {Cunningham}, {Doyle}, {Elms}, {Farihi}, {Hermes}, {Holberg}, {Jordan}, {Klein}, {Kleinman}, {Manser}, {De Martino}, {Marsh}, {McCleery}, {Melis}, {Nitta}, {Parsons}, {Raddi}, {Rebassa-Mansergas}, {Schreiber}, {Silvotti}, {Steeghs}, {Toloza}, {Toonen}, {Torres}, {Weinberger}, \& {Zuckerman}}]{obrien2023}
{O'Brien} M.~W., {Tremblay} P.~E., {Gentile Fusillo} N.~P., et~al. (2023): {Gaia white dwarfs within 40 pc - III. Spectroscopic observations of new candidates in the Southern hemisphere}. \mnras\, 518(2):3055--3073. \doi{10.1093/mnras/stac3303}, {\href{https://arxiv.org/abs/2210.01608}{{arXiv:2210.01608}}} {[astro-ph.SR]}

\bibitem[{{O'Brien} et~al.(2024){O'Brien}, {Tremblay}, {Klein}, {Koester}, {Melis}, {B{\'e}dard}, {Cukanovaite}, {Cunningham}, {Doyle}, {G{\"a}nsicke}, {Gentile Fusillo}, {Hollands}, {McCleery}, {Pelisoli}, {Toonen}, {Weinberger}, \& {Zuckerman}}]{obrien2024}
{O'Brien} M.~W., {Tremblay} P.~E., {Klein} B.~L., et~al. (2024): {The 40 pc sample of white dwarfs from Gaia}. \mnras\, 527(3):8687--8705. \doi{10.1093/mnras/stad3773}, {\href{https://arxiv.org/abs/2312.02735}{{arXiv:2312.02735}}} {[astro-ph.SR]}

\bibitem[{{Oswalt} et~al.(1996){Oswalt}, {Smith}, {Wood}, \& {Hintzen}}]{oswalt1996}
{Oswalt} T.~D., {Smith} J.~A., {Wood} M.~A., et~al. (1996): {A lower limit of 9.5 Gyr on the age of the Galactic disk from the oldest white dwarf stars}. \nat\, 382(6593):692--694. \doi{10.1038/382692a0}

\bibitem[{{Ourique} et~al.(2019){Ourique}, {Romero}, {Kepler}, {Koester}, \& {Amaral}}]{ourique2019}
{Ourique} G., {Romero} A.~D., {Kepler} S.~O., et~al. (2019): {A study of cool white dwarfs in the Sloan Digital Sky Survey Data Release 12}. \mnras\, 482(1):649--657. \doi{10.1093/mnras/sty2751}, {\href{https://arxiv.org/abs/1810.03554}{{arXiv:1810.03554}}} {[astro-ph.SR]}

\bibitem[{{Ourique} et~al.(2020){Ourique}, {Kepler}, {Romero}, {Klippel}, \& {Koester}}]{ourique2020}
{Ourique} G., {Kepler} S.~O., {Romero} A.~D., et~al. (2020): {Evidence of spectral evolution on the white dwarf sample from the Gaia mission}. \mnras\, 492(4):5003--5010. \doi{10.1093/mnras/staa120}, {\href{https://arxiv.org/abs/2001.04378}{{arXiv:2001.04378}}} {[astro-ph.SR]}

\bibitem[{{Paquette} et~al.(1986){Paquette}, {Pelletier}, {Fontaine}, \& {Michaud}}]{paquette1986b}
{Paquette} C., {Pelletier} C., {Fontaine} G., et~al. (1986): {Diffusion in White Dwarfs: New Results and Comparative Study}. \apjs\, 61:197. \doi{10.1086/191112}

\bibitem[{{Paxton} et~al.(2011){Paxton}, {Bildsten}, {Dotter}, {Herwig}, {Lesaffre}, \& {Timmes}}]{paxton2011}
{Paxton} B., {Bildsten} L., {Dotter} A., et~al. (2011): {Modules for Experiments in Stellar Astrophysics (MESA)}. \apjs\, 192(1):3. \doi{10.1088/0067-0049/192/1/3}, {\href{https://arxiv.org/abs/1009.1622}{{arXiv:1009.1622}}} {[astro-ph.SR]}

\bibitem[{{Pelletier} et~al.(1986){Pelletier}, {Fontaine}, {Wesemael}, {Michaud}, \& {Wegner}}]{pelletier1986}
{Pelletier} C., {Fontaine} G., {Wesemael} F., et~al. (1986): {Carbon Pollution in Helium-rich White Dwarf Atmospheres: Time-dependent Calculations of the Dredge-up Process}. \apj\, 307:242. \doi{10.1086/164410}

\bibitem[{{Pereira} et~al.(2005){Pereira}, {Bergeron}, \& {Wesemael}}]{pereira2005}
{Pereira} C., {Bergeron} P., {Wesemael} F. (2005): {Discovery of Spectroscopic Variations in the DAB White Dwarf GD 323}. \apj\, 623(2):1076--1082. \doi{10.1086/429219}, {\href{https://arxiv.org/abs/astro-ph/0501620}{{arXiv:astro-ph/0501620}}} {[astro-ph]}

\bibitem[{{Petitclerc} et~al.(2005){Petitclerc}, {Wesemael}, {Kruk}, {Chayer}, \& {Bill{\`e}res}}]{petitclerc2005}
{Petitclerc} N., {Wesemael} F., {Kruk} J.~W., et~al. (2005): {FUSE Observations of DB White Dwarfs}. \apj\, 624(1):317--330. \doi{10.1086/428750}

\bibitem[{{Preval} et~al.(2013){Preval}, {Barstow}, {Holberg}, \& {Dickinson}}]{preval2013}
{Preval} S.~P., {Barstow} M.~A., {Holberg} J.~B., et~al. (2013): {A comprehensive near- and far-ultraviolet spectroscopic study of the hot DA white dwarf G191-B2B}. \mnras\, 436(1):659--674. \doi{10.1093/mnras/stt1604}, {\href{https://arxiv.org/abs/1308.4825}{{arXiv:1308.4825}}} {[astro-ph.SR]}

\bibitem[{{Preval} et~al.(2019){Preval}, {Barstow}, {Bainbridge}, {Reindl}, {Ayres}, {Holberg}, {Barrow}, {Lee}, {Webb}, \& {Hu}}]{preval2019}
{Preval} S.~P., {Barstow} M.~A., {Bainbridge} M., et~al. (2019): {A far-UV survey of three hot, metal-polluted white dwarf stars: WD0455-282, WD0621-376, and WD2211-495}. \mnras\, 487(3):3470--3487. \doi{10.1093/mnras/stz1506}, {\href{https://arxiv.org/abs/1905.12350}{{arXiv:1905.12350}}} {[astro-ph.SR]}

\bibitem[{{Provencal} et~al.(2000){Provencal}, {Shipman}, {Thejll}, \& {Vennes}}]{provencal2000}
{Provencal} J.~L., {Shipman} H.~L., {Thejll} P., et~al. (2000): {Carbon and Hydrogen in Hot DB White Dwarfs}. \apj\, 542(2):1041--1056. \doi{10.1086/317030}

\bibitem[{{Quirion} et~al.(2012){Quirion}, {Fontaine}, \& {Brassard}}]{quirion2012}
{Quirion} P.~O., {Fontaine} G., {Brassard} P. (2012): {Wind Competing Against Settling: A Coherent Model of the GW Virginis Instability Domain}. \apj\, 755(2):128. \doi{10.1088/0004-637X/755/2/128}

\bibitem[{{Raddi} et~al.(2015){Raddi}, {G{\"a}nsicke}, {Koester}, {Farihi}, {Hermes}, {Scaringi}, {Breedt}, \& {Girven}}]{raddi2015}
{Raddi} R., {G{\"a}nsicke} B.~T., {Koester} D., et~al. (2015): {Likely detection of water-rich asteroid debris in a metal-polluted white dwarf}. \mnras\, 450(2):2083--2093. \doi{10.1093/mnras/stv701}, {\href{https://arxiv.org/abs/1503.07864}{{arXiv:1503.07864}}} {[astro-ph.SR]}

\bibitem[{{Rauch} et~al.(1998){Rauch}, {Dreizler}, \& {Wolff}}]{rauch1998}
{Rauch} T., {Dreizler} S., {Wolff} B. (1998): {Spectral analysis of O(He)-type post-AGB stars}. \aap\, 338:651--660

\bibitem[{{Rauch} et~al.(2013){Rauch}, {Werner}, {Bohlin}, \& {Kruk}}]{rauch2013}
{Rauch} T., {Werner} K., {Bohlin} R., et~al. (2013): {The virtual observatory service TheoSSA: Establishing a database of synthetic stellar flux standards. I. NLTE spectral analysis of the DA-type white dwarf G191-B2B}. \aap\, 560:A106. \doi{10.1051/0004-6361/201322336}, {\href{https://arxiv.org/abs/1308.6450}{{arXiv:1308.6450}}} {[astro-ph.SR]}

\bibitem[{{Rauch} et~al.(2014{\natexlab{a}}){Rauch}, {Werner}, {Quinet}, \& {Kruk}}]{rauch2014a}
{Rauch} T., {Werner} K., {Quinet} P., et~al. (2014{\natexlab{a}}): {Stellar laboratories. II. New Zn iv and Zn v oscillator strengths and their validation in the hot white dwarfs G191-B2B and RE 0503-289}. \aap\, 564:A41. \doi{10.1051/0004-6361/201423491}, {\href{https://arxiv.org/abs/1403.2183}{{arXiv:1403.2183}}} {[astro-ph.SR]}

\bibitem[{{Rauch} et~al.(2014{\natexlab{b}}){Rauch}, {Werner}, {Quinet}, \& {Kruk}}]{rauch2014b}
{Rauch} T., {Werner} K., {Quinet} P., et~al. (2014{\natexlab{b}}): {Stellar laboratories. III. New Ba v, Ba vi, and Ba vii oscillator strengths and the barium abundance in the hot white dwarfs G191-B2B and RE 0503-289}. \aap\, 566:A10. \doi{10.1051/0004-6361/201423878}, {\href{https://arxiv.org/abs/1404.6094}{{arXiv:1404.6094}}} {[astro-ph.SR]}

\bibitem[{{Rauch} et~al.(2015{\natexlab{a}}){Rauch}, {Hoyer}, {Quinet}, {Gallardo}, \& {Raineri}}]{rauch2015b}
{Rauch} T., {Hoyer} D., {Quinet} P., et~al. (2015{\natexlab{a}}): {Stellar laboratories. V. The Xe vi ultraviolet spectrum and the xenon abundance in the hot DO-type white dwarf RE 0503-289}. \aap\, 577:A88. \doi{10.1051/0004-6361/201526078}, {\href{https://arxiv.org/abs/1504.01991}{{arXiv:1504.01991}}} {[astro-ph.SR]}

\bibitem[{{Rauch} et~al.(2015{\natexlab{b}}){Rauch}, {Werner}, {Quinet}, \& {Kruk}}]{rauch2015a}
{Rauch} T., {Werner} K., {Quinet} P., et~al. (2015{\natexlab{b}}): {Stellar laboratories. IV. New Ga iv, Ga v, and Ga vi oscillator strengths and the gallium abundance in the hot white dwarfs G191-B2B and RE 0503-289}. \aap\, 577:A6. \doi{10.1051/0004-6361/201425326}, {\href{https://arxiv.org/abs/1501.07751}{{arXiv:1501.07751}}} {[astro-ph.SR]}

\bibitem[{{Rauch} et~al.(2016{\natexlab{a}}){Rauch}, {Quinet}, {Hoyer}, {Werner}, {Demleitner}, \& {Kruk}}]{rauch2016a}
{Rauch} T., {Quinet} P., {Hoyer} D., et~al. (2016{\natexlab{a}}): {Stellar laboratories. VI. New Mo iv-vii oscillator strengths and the molybdenum abundance in the hot white dwarfs G191-B2B and RE 0503-289}. \aap\, 587:A39. \doi{10.1051/0004-6361/201527324}, {\href{https://arxiv.org/abs/1512.07525}{{arXiv:1512.07525}}} {[astro-ph.SR]}

\bibitem[{{Rauch} et~al.(2016{\natexlab{b}}){Rauch}, {Quinet}, {Hoyer}, {Werner}, {Richter}, {Kruk}, \& {Demleitner}}]{rauch2016b}
{Rauch} T., {Quinet} P., {Hoyer} D., et~al. (2016{\natexlab{b}}): {Stellar laboratories. VII. New Kr iv - vii oscillator strengths and an improved spectral analysis of the hot, hydrogen-deficient DO-type white dwarf RE 0503-289}. \aap\, 590:A128. \doi{10.1051/0004-6361/201628131}, {\href{https://arxiv.org/abs/1603.00701}{{arXiv:1603.00701}}} {[astro-ph.SR]}

\bibitem[{{Rauch} et~al.(2017{\natexlab{a}}){Rauch}, {Gamrath}, {Quinet}, {L{\"o}bling}, {Hoyer}, {Werner}, {Kruk}, \& {Demleitner}}]{rauch2017a}
{Rauch} T., {Gamrath} S., {Quinet} P., et~al. (2017{\natexlab{a}}): {Stellar laboratories . VIII. New Zr iv-vii, Xe iv-v, and Xe vii oscillator strengths and the Al, Zr, and Xe abundances in the hot white dwarfs G191-B2B and RE 0503-289}. \aap\, 599:A142. \doi{10.1051/0004-6361/201629794}, {\href{https://arxiv.org/abs/1611.07364}{{arXiv:1611.07364}}} {[physics.atom-ph]}

\bibitem[{{Rauch} et~al.(2017{\natexlab{b}}){Rauch}, {Quinet}, {Kn{\"o}rzer}, {Hoyer}, {Werner}, {Kruk}, \& {Demleitner}}]{rauch2017b}
{Rauch} T., {Quinet} P., {Kn{\"o}rzer} M., et~al. (2017{\natexlab{b}}): {Stellar laboratories . IX. New Se v, Sr iv-vii, Te vi, and I vi oscillator strengths and the Se, Sr, Te, and I abundances in the hot white dwarfs G191-B2B and RE 0503-289}. \aap\, 606:A105. \doi{10.1051/0004-6361/201730383}, {\href{https://arxiv.org/abs/1706.09215}{{arXiv:1706.09215}}} {[astro-ph.SR]}

\bibitem[{{Rauch} et~al.(2020){Rauch}, {Gamrath}, {Quinet}, {Demleitner}, {Kn{\"o}rzer}, {Werner}, \& {Kruk}}]{rauch2020}
{Rauch} T., {Gamrath} S., {Quinet} P., et~al. (2020): {Stellar laboratories. X. New Cu IV-VII oscillator strengths and the first detection of copper and indium in hot white dwarfs}. \aap\, 637:A4. \doi{10.1051/0004-6361/201936620}, {\href{https://arxiv.org/abs/2004.01633}{{arXiv:2004.01633}}} {[astro-ph.SR]}

\bibitem[{{Reindl} et~al.(2014{\natexlab{a}}){Reindl}, {Rauch}, {Werner}, {Kepler}, {G{\"a}nsicke}, \& {Gentile Fusillo}}]{reindl2014b}
{Reindl} N., {Rauch} T., {Werner} K., et~al. (2014{\natexlab{a}}): {Analysis of cool DO-type white dwarfs from the Sloan Digital Sky Survey data release 10}. \aap\, 572:A117. \doi{10.1051/0004-6361/201424861}, {\href{https://arxiv.org/abs/1410.7666}{{arXiv:1410.7666}}} {[astro-ph.SR]}

\bibitem[{{Reindl} et~al.(2014{\natexlab{b}}){Reindl}, {Rauch}, {Werner}, {Kruk}, \& {Todt}}]{reindl2014a}
{Reindl} N., {Rauch} T., {Werner} K., et~al. (2014{\natexlab{b}}): {On helium-dominated stellar evolution: the mysterious role of the O(He)-type stars}. \aap\, 566:A116. \doi{10.1051/0004-6361/201423498}, {\href{https://arxiv.org/abs/1405.1589}{{arXiv:1405.1589}}} {[astro-ph.SR]}

\bibitem[{{Reindl} et~al.(2023){Reindl}, {Islami}, {Werner}, {Kepler}, {Pritzkuleit}, {Dawson}, {Dorsch}, {Istrate}, {Pelisoli}, {Geier}, {Uzundag}, {Provencal}, \& {Justham}}]{reindl2023}
{Reindl} N., {Islami} R., {Werner} K., et~al. (2023): {The bright blue side of the night sky: Spectroscopic survey of bright and hot (pre-) white dwarfs}. \aap\, 677:A29. \doi{10.1051/0004-6361/202346865}, {\href{https://arxiv.org/abs/2307.03721}{{arXiv:2307.03721}}} {[astro-ph.SR]}

\bibitem[{{Renedo} et~al.(2010){Renedo}, {Althaus}, {Miller Bertolami}, {Romero}, {C{\'o}rsico}, {Rohrmann}, \& {Garc{\'\i}a-Berro}}]{renedo2010}
{Renedo} I., {Althaus} L.~G., {Miller Bertolami} M.~M., et~al. (2010): {New Cooling Sequences for Old White Dwarfs}. \apj\, 717(1):183--195. \doi{10.1088/0004-637X/717/1/183}, {\href{https://arxiv.org/abs/1005.2170}{{arXiv:1005.2170}}} {[astro-ph.SR]}

\bibitem[{{Rogers} et~al.(2024){Rogers}, {Bonsor}, {Xu}, {Dufour}, {Klein}, {Buchan}, {Hodgkin}, {Hardy}, {Kissler-Patig}, {Melis}, {Weinberger}, \& {Zuckerman}}]{rogers2024}
{Rogers} L.~K., {Bonsor} A., {Xu} S., et~al. (2024): {Seven white dwarfs with circumstellar gas discs I: white dwarf parameters and accreted planetary abundances}. \mnras\, 527(3):6038--6054. \doi{10.1093/mnras/stad3557}, {\href{https://arxiv.org/abs/2311.14048}{{arXiv:2311.14048}}} {[astro-ph.EP]}

\bibitem[{{Rolland} et~al.(2018){Rolland}, {Bergeron}, \& {Fontaine}}]{rolland2018}
{Rolland} B., {Bergeron} P., {Fontaine} G. (2018): {On the Spectral Evolution of Helium-atmosphere White Dwarfs Showing Traces of Hydrogen}. \apj\, 857(1):56. \doi{10.3847/1538-4357/aab713}, {\href{https://arxiv.org/abs/1803.05965}{{arXiv:1803.05965}}} {[astro-ph.SR]}

\bibitem[{{Rolland} et~al.(2020){Rolland}, {Bergeron}, \& {Fontaine}}]{rolland2020}
{Rolland} B., {Bergeron} P., {Fontaine} G. (2020): {A Convective Dredge-up Model as the Origin of Hydrogen in DBA White Dwarfs}. \apj\, 889(2):87. \doi{10.3847/1538-4357/ab6602}, {\href{https://arxiv.org/abs/2001.01085}{{arXiv:2001.01085}}} {[astro-ph.SR]}

\bibitem[{{Romero} et~al.(2012){Romero}, {C{\'o}rsico}, {Althaus}, {Kepler}, {Castanheira}, \& {Miller Bertolami}}]{romero2012}
{Romero} A.~D., {C{\'o}rsico} A.~H., {Althaus} L.~G., et~al. (2012): {Toward ensemble asteroseismology of ZZ Ceti stars with fully evolutionary models}. \mnras\, 420(2):1462--1480. \doi{10.1111/j.1365-2966.2011.20134.x}, {\href{https://arxiv.org/abs/1109.6682}{{arXiv:1109.6682}}} {[astro-ph.SR]}

\bibitem[{{Salaris} \& {Cassisi}(2017)}]{salaris2017}
{Salaris} M., {Cassisi} S. (2017): {Chemical element transport in stellar evolution models}. RSOS\, 4(8):170192. \doi{10.1098/rsos.170192}, {\href{https://arxiv.org/abs/1707.07454}{{arXiv:1707.07454}}} {[astro-ph.SR]}

\bibitem[{{Salaris} et~al.(2022){Salaris}, {Cassisi}, {Pietrinferni}, \& {Hidalgo}}]{salaris2022}
{Salaris} M., {Cassisi} S., {Pietrinferni} A., et~al. (2022): {The updated BASTI stellar evolution models and isochrones - III. White dwarfs}. \mnras\, 509(4):5197--5208. \doi{10.1093/mnras/stab3359}, {\href{https://arxiv.org/abs/2111.09285}{{arXiv:2111.09285}}} {[astro-ph.SR]}

\bibitem[{{Saumon} et~al.(2022){Saumon}, {Blouin}, \& {Tremblay}}]{saumon2022}
{Saumon} D., {Blouin} S., {Tremblay} P.-E. (2022): {Current challenges in the physics of white dwarf stars}. \physrep\, 988:1--63. \doi{10.1016/j.physrep.2022.09.001}, {\href{https://arxiv.org/abs/2209.02846}{{arXiv:2209.02846}}} {[astro-ph.SR]}

\bibitem[{{Schreiber} et~al.(2019){Schreiber}, {G{\"a}nsicke}, {Toloza}, {Hernandez}, \& {Lagos}}]{schreiber2019}
{Schreiber} M.~R., {G{\"a}nsicke} B.~T., {Toloza} O., et~al. (2019): {Cold Giant Planets Evaporated by Hot White Dwarfs}. \apjl\, 887(1):L4. \doi{10.3847/2041-8213/ab42e2}, {\href{https://arxiv.org/abs/1912.02345}{{arXiv:1912.02345}}} {[astro-ph.SR]}

\bibitem[{{Schuh} et~al.(2002){Schuh}, {Dreizler}, \& {Wolff}}]{schuh2002}
{Schuh} S.~L., {Dreizler} S., {Wolff} B. (2002): {Equilibrium abundances in hot DA white dwarfs as derived from self-consistent diffusion models. I. Analysis of spectroscopic EUVE data}. \aap\, 382:164--173. \doi{10.1051/0004-6361:20011588}, {\href{https://arxiv.org/abs/astro-ph/0111245}{{arXiv:astro-ph/0111245}}} {[astro-ph]}

\bibitem[{{Sc{\'o}ccola} et~al.(2006){Sc{\'o}ccola}, {Althaus}, {Serenelli}, {Rohrmann}, \& {C{\'o}rsico}}]{scoccola2006}
{Sc{\'o}ccola} C.~G., {Althaus} L.~G., {Serenelli} A.~M., et~al. (2006): {DQ white-dwarf stars with low C abundance: possible progenitors}. \aap\, 451(1):147--155. \doi{10.1051/0004-6361:20053769}, {\href{https://arxiv.org/abs/astro-ph/0602196}{{arXiv:astro-ph/0602196}}} {[astro-ph]}

\bibitem[{{Sion}(1984)}]{sion1984}
{Sion} E.~M. (1984): {Implications of the absolute magnitude distribution functions of DA and non-DA white dwarfs.} \apj\, 282:612--614. \doi{10.1086/162240}

\bibitem[{{Sion}(1999)}]{sion1999}
{Sion} E.~M. (1999): {White Dwarfs in Cataclysmic Variables}. \pasp\, 111(759):532--555. \doi{10.1086/316361}

\bibitem[{{Sion} et~al.(1983){Sion}, {Greenstein}, {Landstreet}, {Liebert}, {Shipman}, \& {Wegner}}]{sion1983}
{Sion} E.~M., {Greenstein} J.~L., {Landstreet} J.~D., et~al. (1983): {A proposed new white dwarf spectral classification system.} \apj\, 269:253--257. \doi{10.1086/161036}

\bibitem[{{Subasavage} et~al.(2017){Subasavage}, {Jao}, {Henry}, {Harris}, {Dahn}, {Bergeron}, {Dufour}, {Dunlap}, {Barlow}, {Ianna}, {L{\'e}pine}, \& {Margheim}}]{subasavage2017}
{Subasavage} J.~P., {Jao} W.-C., {Henry} T.~J., et~al. (2017): {The Solar Neighborhood. XXXIX. Parallax Results from the CTIOPI and NOFS Programs: 50 New Members of the 25 parsec White Dwarf Sample}. \aj\, 154(1):32. \doi{10.3847/1538-3881/aa76e0}, {\href{https://arxiv.org/abs/1706.00709}{{arXiv:1706.00709}}} {[astro-ph.SR]}

\bibitem[{{Swan} et~al.(2019){Swan}, {Farihi}, {Koester}, {Hollands}, {Parsons}, {Cauley}, {Redfield}, \& {G{\"a}nsicke}}]{swan2019}
{Swan} A., {Farihi} J., {Koester} D., et~al. (2019): {Interpretation and diversity of exoplanetary material orbiting white dwarfs}. \mnras\, 490(1):202--218. \doi{10.1093/mnras/stz2337}, {\href{https://arxiv.org/abs/1908.08047}{{arXiv:1908.08047}}} {[astro-ph.EP]}

\bibitem[{{Swan} et~al.(2023){Swan}, {Farihi}, {Melis}, {Dufour}, {Desch}, {Koester}, \& {Guo}}]{swan2023}
{Swan} A., {Farihi} J., {Melis} C., et~al. (2023): {Planetesimals at DZ stars - I. Chondritic compositions and a massive accretion event}. \mnras\, 526(3):3815--3831. \doi{10.1093/mnras/stad2867}, {\href{https://arxiv.org/abs/2309.06467}{{arXiv:2309.06467}}} {[astro-ph.EP]}

\bibitem[{{Tassoul} et~al.(1990){Tassoul}, {Fontaine}, \& {Winget}}]{tassoul1990}
{Tassoul} M., {Fontaine} G., {Winget} D.~E. (1990): {Evolutionary Models for Pulsation Studies of White Dwarfs}. \apjs\, 72:335. \doi{10.1086/191420}

\bibitem[{{Torres} et~al.(2023){Torres}, {Cruz}, {Murillo-Ojeda}, {Jim{\'e}nez-Esteban}, {Rebassa-Mansergas}, {Solano}, {Camisassa}, {Raddi}, \& {Doliguez Le Lourec}}]{torres2023}
{Torres} S., {Cruz} P., {Murillo-Ojeda} R., et~al. (2023): {White dwarf spectral type-temperature distribution from Gaia DR3 and the Virtual Observatory}. \aap\, 677:A159. \doi{10.1051/0004-6361/202346977}, {\href{https://arxiv.org/abs/2307.13629}{{arXiv:2307.13629}}} {[astro-ph.SR]}

\bibitem[{{Tremblay} \& {Bergeron}(2008)}]{tremblay2008}
{Tremblay} P.~E., {Bergeron} P. (2008): {The Ratio of Helium- to Hydrogen-Atmosphere White Dwarfs: Direct Evidence for Convective Mixing}. \apj\, 672(2):1144--1152. \doi{10.1086/524134}, {\href{https://arxiv.org/abs/0710.1073}{{arXiv:0710.1073}}} {[astro-ph]}

\bibitem[{{Tremblay} et~al.(2011){Tremblay}, {Bergeron}, \& {Gianninas}}]{tremblay2011}
{Tremblay} P.~E., {Bergeron} P., {Gianninas} A. (2011): {An Improved Spectroscopic Analysis of DA White Dwarfs from the Sloan Digital Sky Survey Data Release 4}. \apj\, 730(2):128. \doi{10.1088/0004-637X/730/2/128}, {\href{https://arxiv.org/abs/1102.0056}{{arXiv:1102.0056}}} {[astro-ph.SR]}

\bibitem[{{Tremblay} et~al.(2013){Tremblay}, {Ludwig}, {Steffen}, \& {Freytag}}]{tremblay2013}
{Tremblay} P.~E., {Ludwig} H.~G., {Steffen} M., et~al. (2013): {Pure-hydrogen 3D model atmospheres of cool white dwarfs}. \aap\, 552:A13. \doi{10.1051/0004-6361/201220813}, {\href{https://arxiv.org/abs/1302.2013}{{arXiv:1302.2013}}} {[astro-ph.SR]}

\bibitem[{{Tremblay} et~al.(2014){Tremblay}, {Kalirai}, {Soderblom}, {Cignoni}, \& {Cummings}}]{tremblay2014}
{Tremblay} P.~E., {Kalirai} J.~S., {Soderblom} D.~R., et~al. (2014): {White Dwarf Cosmochronology in the Solar Neighborhood}. \apj\, 791(2):92. \doi{10.1088/0004-637X/791/2/92}, {\href{https://arxiv.org/abs/1406.5173}{{arXiv:1406.5173}}} {[astro-ph.SR]}

\bibitem[{{Tremblay} et~al.(2015){Tremblay}, {Ludwig}, {Freytag}, {Fontaine}, {Steffen}, \& {Brassard}}]{tremblay2015}
{Tremblay} P.~E., {Ludwig} H.~G., {Freytag} B., et~al. (2015): {Calibration of the Mixing-length Theory for Convective White Dwarf Envelopes}. \apj\, 799(2):142. \doi{10.1088/0004-637X/799/2/142}, {\href{https://arxiv.org/abs/1412.1789}{{arXiv:1412.1789}}} {[astro-ph.SR]}

\bibitem[{{Tremblay} et~al.(2019){Tremblay}, {Cukanovaite}, {Gentile Fusillo}, {Cunningham}, \& {Hollands}}]{tremblay2019b}
{Tremblay} P.~E., {Cukanovaite} E., {Gentile Fusillo} N.~P., et~al. (2019): {Fundamental parameter accuracy of DA and DB white dwarfs in Gaia Data Release 2}. \mnras\, 482(4):5222--5232. \doi{10.1093/mnras/sty3067}, {\href{https://arxiv.org/abs/1811.03084}{{arXiv:1811.03084}}} {[astro-ph.SR]}

\bibitem[{{Unglaub}(2008)}]{unglaub2008}
{Unglaub} K. (2008): {Mass-loss and diffusion in subdwarf B stars and hot white dwarfs: do weak winds exist?} \aap\, 486(3):923--940. \doi{10.1051/0004-6361:20078019}, {\href{https://arxiv.org/abs/0808.1072}{{arXiv:0808.1072}}} {[astro-ph]}

\bibitem[{{Unglaub} \& {Bues}(1998)}]{unglaub1998}
{Unglaub} K., {Bues} I. (1998): {The effect of diffusion and mass loss on the helium abundance in hot white dwarfs and subdwarfs}. \aap\, 338:75--84

\bibitem[{{Unglaub} \& {Bues}(2000)}]{unglaub2000}
{Unglaub} K., {Bues} I. (2000): {The chemical evolution of hot white dwarfs in the presence of diffusion and mass loss}. \aap\, 359:1042--1058

\bibitem[{{Vennes} \& {Fontaine}(1992)}]{vennes1992b}
{Vennes} S., {Fontaine} G. (1992): {An Interpretation of the Spectral Properties of Hot Hydrogen-rich White Dwarfs with Stratified H/He Model Atmospheres}. \apj\, 401:288. \doi{10.1086/172060}

\bibitem[{{Vennes} et~al.(1988){Vennes}, {Pelletier}, {Fontaine}, \& {Wesemael}}]{vennes1988}
{Vennes} S., {Pelletier} C., {Fontaine} G., et~al. (1988): {The Presence of Helium in Hot DA White Dwarfs: The Role of Radiative Levitation and the Case for Stratified Atmospheres}. \apj\, 331:876. \doi{10.1086/166606}

\bibitem[{{Vennes} et~al.(2005){Vennes}, {Chayer}, \& {Dupuis}}]{vennes2005}
{Vennes} S., {Chayer} P., {Dupuis} J. (2005): {Discovery of Photospheric Germanium in Hot DA White Dwarfs}. \apjl\, 622(2):L121--L124. \doi{10.1086/429667}

\bibitem[{{Vennes} et~al.(2006){Vennes}, {Chayer}, {Dupuis}, \& {Lanz}}]{vennes2006}
{Vennes} S., {Chayer} P., {Dupuis} J., et~al. (2006): {Iron in Hot DA White Dwarfs}. \apj\, 652(2):1554--1562. \doi{10.1086/508509}, {\href{https://arxiv.org/abs/astro-ph/0608416}{{arXiv:astro-ph/0608416}}} {[astro-ph]}

\bibitem[{{Vennes} et~al.(2024){Vennes}, {Kawka}, {Klein}, {Zuckerman}, {Weinberger}, \& {Melis}}]{vennes2024}
{Vennes} S., {Kawka} A., {Klein} B.~L., et~al. (2024): {A cool, magnetic white dwarf accreting planetary debris}. \mnras\, 527(2):3122--3138. \doi{10.1093/mnras/stad3370}, {\href{https://arxiv.org/abs/2311.07937}{{arXiv:2311.07937}}} {[astro-ph.SR]}

\bibitem[{{Veras}(2021)}]{veras2021}
{Veras} D. (2021): {Planetary Systems Around White Dwarfs}. In: Oxford Research Encyclopedia of Planetary Science. Oxford: Oxford University Press, p.~1, \doi{10.1093/acrefore/9780190647926.013.238}

\bibitem[{{Veras} et~al.(2014){Veras}, {Shannon}, \& {G{\"a}nsicke}}]{veras2014}
{Veras} D., {Shannon} A., {G{\"a}nsicke} B.~T. (2014): {Hydrogen delivery onto white dwarfs from remnant exo-Oort cloud comets}. \mnras\, 445(4):4175--4185. \doi{10.1093/mnras/stu2026}, {\href{https://arxiv.org/abs/1409.7691}{{arXiv:1409.7691}}} {[astro-ph.SR]}

\bibitem[{{Vincent} et~al.(2024){Vincent}, {Barstow}, {Jordan}, {Mander}, {Bergeron}, \& {Dufour}}]{vincent2024}
{Vincent} O., {Barstow} M.~A., {Jordan} S., et~al. (2024): {Classification and parameterization of a large Gaia sample of white dwarfs using XP spectra}. \aap\, 682:A5. \doi{10.1051/0004-6361/202347694}, {\href{https://arxiv.org/abs/2308.05572}{{arXiv:2308.05572}}} {[astro-ph.SR]}

\bibitem[{{Voss} et~al.(2007){Voss}, {Koester}, {Napiwotzki}, {Christlieb}, \& {Reimers}}]{voss2007}
{Voss} B., {Koester} D., {Napiwotzki} R., et~al. (2007): {High-resolution UVES/VLT spectra of white dwarfs observed for the ESO SN Ia progenitor survey. II. DB and DBA stars}. \aap\, 470(3):1079--1088. \doi{10.1051/0004-6361:20077285}

\bibitem[{{Wachlin} et~al.(2017){Wachlin}, {Vauclair}, {Vauclair}, \& {Althaus}}]{wachlin2017}
{Wachlin} F.~C., {Vauclair} G., {Vauclair} S., et~al. (2017): {Importance of fingering convection for accreting white dwarfs in the framework of full evolutionary calculations: the case of the hydrogen-rich white dwarfs GD 133 and G 29-38}. \aap\, 601:A13. \doi{10.1051/0004-6361/201630094}, {\href{https://arxiv.org/abs/1612.09320}{{arXiv:1612.09320}}} {[astro-ph.SR]}

\bibitem[{{Wachlin} et~al.(2022){Wachlin}, {Vauclair}, {Vauclair}, \& {Althaus}}]{wachlin2022}
{Wachlin} F.~C., {Vauclair} G., {Vauclair} S., et~al. (2022): {New simulations of accreting DA white dwarfs: Inferring accretion rates from the surface contamination}. \aap\, 660:A30. \doi{10.1051/0004-6361/202142289}, {\href{https://arxiv.org/abs/2109.11370}{{arXiv:2109.11370}}} {[astro-ph.SR]}

\bibitem[{{Weidemann} \& {Koester}(1995)}]{weidemann1995}
{Weidemann} V., {Koester} D. (1995): {Surface carbon abundances and compositional stratification of cool helium-rich white dwarfs.} \aap\, 297:216--222

\bibitem[{{Werner}(1996{\natexlab{a}})}]{werner1996a}
{Werner} K. (1996{\natexlab{a}}): {On the Balmer Line Problem}. \apjl\, 457:L39. \doi{10.1086/309889}

\bibitem[{{Werner}(1996{\natexlab{b}})}]{werner1996b}
{Werner} K. (1996{\natexlab{b}}): {Search for trace amounts of hydrogen in hot DO white dwarfs.} \aap\, 309:861--866

\bibitem[{{Werner} \& {Dreizler}(1994)}]{werner1994}
{Werner} K., {Dreizler} S. (1994): {On the nickel abundance in hot hydrogen-rich white dwarfs.} \aap\, 286:L31--L34

\bibitem[{{Werner} \& {Herwig}(2006)}]{werner2006}
{Werner} K., {Herwig} F. (2006): {The Elemental Abundances in Bare Planetary Nebula Central Stars and the Shell Burning in AGB Stars}. \pasp\, 118(840):183--204. \doi{10.1086/500443}, {\href{https://arxiv.org/abs/astro-ph/0512320}{{arXiv:astro-ph/0512320}}} {[astro-ph]}

\bibitem[{{Werner} \& {Rauch}(2014)}]{werner2014b}
{Werner} K., {Rauch} T. (2014): {Weak metal lines in optical high-resolution Very Large Telescope and Keck spectra of ``cool'' PG 1159 stars}. \aap\, 569:A99. \doi{10.1051/0004-6361/201424051}

\bibitem[{{Werner} et~al.(1991){Werner}, {Heber}, \& {Hunger}}]{werner1991}
{Werner} K., {Heber} U., {Hunger} K. (1991): {Non-LTE analysis of four PG1159 stars.} \aap\, 244:437

\bibitem[{{Werner} et~al.(2007){Werner}, {Rauch}, \& {Kruk}}]{werner2007}
{Werner} K., {Rauch} T., {Kruk} J.~W. (2007): {Discovery of photospheric argon in very hot central stars of planetary nebulae and white dwarfs}. \aap\, 466(1):317--322. \doi{10.1051/0004-6361:20077101}, {\href{https://arxiv.org/abs/astro-ph/0702387}{{arXiv:astro-ph/0702387}}} {[astro-ph]}

\bibitem[{{Werner} et~al.(2012){Werner}, {Rauch}, {Ringat}, \& {Kruk}}]{werner2012}
{Werner} K., {Rauch} T., {Ringat} E., et~al. (2012): {First Detection of Krypton and Xenon in a White Dwarf}. \apjl\, 753(1):L7. \doi{10.1088/2041-8205/753/1/L7}

\bibitem[{{Werner} et~al.(2014){Werner}, {Rauch}, \& {Kepler}}]{werner2014a}
{Werner} K., {Rauch} T., {Kepler} S.~O. (2014): {New hydrogen-deficient (pre-) white dwarfs in the Sloan Digital Sky Survey Data Release 10}. \aap\, 564:A53. \doi{10.1051/0004-6361/201423441}

\bibitem[{{Werner} et~al.(2017){Werner}, {Rauch}, \& {Kruk}}]{werner2017}
{Werner} K., {Rauch} T., {Kruk} J.~W. (2017): {Far-UV spectroscopy of two extremely hot, helium-rich white dwarfs}. \aap\, 601:A8. \doi{10.1051/0004-6361/201630266}

\bibitem[{{Werner} et~al.(2018{\natexlab{a}}){Werner}, {Rauch}, {Kn{\"o}rzer}, \& {Kruk}}]{werner2018b}
{Werner} K., {Rauch} T., {Kn{\"o}rzer} M., et~al. (2018{\natexlab{a}}): {First detection of bromine and antimony in hot stars}. \aap\, 614:A96. \doi{10.1051/0004-6361/201832723}, {\href{https://arxiv.org/abs/1803.04809}{{arXiv:1803.04809}}} {[astro-ph.SR]}

\bibitem[{{Werner} et~al.(2018{\natexlab{b}}){Werner}, {Rauch}, \& {Kruk}}]{werner2018a}
{Werner} K., {Rauch} T., {Kruk} J.~W. (2018{\natexlab{b}}): {Metal abundances in hot white dwarfs with signatures of a superionized wind}. \aap\, 609:A107. \doi{10.1051/0004-6361/201731740}, {\href{https://arxiv.org/abs/1711.04138}{{arXiv:1711.04138}}} {[astro-ph.SR]}

\bibitem[{{Werner} et~al.(2019){Werner}, {Rauch}, \& {Reindl}}]{werner2019}
{Werner} K., {Rauch} T., {Reindl} N. (2019): {Spectral analysis of the extremely hot DA white dwarf PG 0948+534}. \mnras\, 483(4):5291--5300. \doi{10.1093/mnras/sty3408}, {\href{https://arxiv.org/abs/1812.07486}{{arXiv:1812.07486}}} {[astro-ph.SR]}

\bibitem[{{Wesemael} et~al.(1993){Wesemael}, {Greenstein}, {Liebert}, {Lamontagne}, {Fontaine}, {Bergeron}, \& {Glaspey}}]{wesemael1993}
{Wesemael} F., {Greenstein} J.~L., {Liebert} J., et~al. (1993): {An Atlas of Optical Spectra of White-Dwarf Stars}. \pasp\, 105:761. \doi{10.1086/133228}

\bibitem[{{Wesemael} et~al.(1994){Wesemael}, {Bergeron}, {Lamontagne}, {Fontaine}, {Beauchamp}, {Demers}, {Irwin}, {Holberg}, {Kepler}, \& {Vennes}}]{wesemael1994}
{Wesemael} F., {Bergeron} P., {Lamontagne} R.~L., et~al. (1994): {Hot Degenerates in the Montreal-Cambridge-Tololo Survey. II. Two New Hybrid White Dwarfs, MCT 0128-3846 and MCT 0453-2933, and the Nature of the DAB Stars}. \apj\, 429:369. \doi{10.1086/174326}

\bibitem[{{Williams} et~al.(2013){Williams}, {Winget}, {Montgomery}, {Dufour}, {Kepler}, {Hermes}, {Falcon}, {Winget}, {Bolte}, {Rubin}, \& {Liebert}}]{williams2013}
{Williams} K.~A., {Winget} D.~E., {Montgomery} M.~H., et~al. (2013): {Photometric Variability in a Warm, Strongly Magnetic DQ White Dwarf, SDSS J103655.39+652252.2}. \apj\, 769(2):123. \doi{10.1088/0004-637X/769/2/123}, {\href{https://arxiv.org/abs/1304.3165}{{arXiv:1304.3165}}} {[astro-ph.SR]}

\bibitem[{{Williams} et~al.(2016){Williams}, {Montgomery}, {Winget}, {Falcon}, \& {Bierwagen}}]{williams2016}
{Williams} K.~A., {Montgomery} M.~H., {Winget} D.~E., et~al. (2016): {Variability in Hot Carbon-dominated Atmosphere (Hot DQ) White Dwarfs: Rapid Rotation?} \apj\, 817(1):27. \doi{10.3847/0004-637X/817/1/27}, {\href{https://arxiv.org/abs/1511.08834}{{arXiv:1511.08834}}} {[astro-ph.SR]}

\bibitem[{{Wilson} et~al.(2015){Wilson}, {G{\"a}nsicke}, {Koester}, {Toloza}, {Pala}, {Breedt}, \& {Parsons}}]{wilson2015}
{Wilson} D.~J., {G{\"a}nsicke} B.~T., {Koester} D., et~al. (2015): {The composition of a disrupted extrasolar planetesimal at SDSS J0845+2257 (Ton 345)}. \mnras\, 451(3):3237--3248. \doi{10.1093/mnras/stv1201}, {\href{https://arxiv.org/abs/1505.07466}{{arXiv:1505.07466}}} {[astro-ph.EP]}

\bibitem[{{Winget} et~al.(1987){Winget}, {Hansen}, {Liebert}, {van Horn}, {Fontaine}, {Nather}, {Kepler}, \& {Lamb}}]{winget1987}
{Winget} D.~E., {Hansen} C.~J., {Liebert} J., et~al. (1987): {An Independent Method for Determining the Age of the Universe}. \apjl\, 315:L77. \doi{10.1086/184864}

\bibitem[{{Wolff} et~al.(2000){Wolff}, {Jordan}, {Koester}, \& {Reimers}}]{wolff2000}
{Wolff} B., {Jordan} S., {Koester} D., et~al. (2000): {The nature of the DAB white dwarf HS 0209+0832}. \aap\, 361:629--640

\bibitem[{{Xu} et~al.(2013){Xu}, {Jura}, {Klein}, {Koester}, \& {Zuckerman}}]{xu2013}
{Xu} S., {Jura} M., {Klein} B., et~al. (2013): {Two Beyond-primitive Extrasolar Planetesimals}. \apj\, 766(2):132. \doi{10.1088/0004-637X/766/2/132}, {\href{https://arxiv.org/abs/1302.4799}{{arXiv:1302.4799}}} {[astro-ph.EP]}

\bibitem[{{Xu} et~al.(2014){Xu}, {Jura}, {Koester}, {Klein}, \& {Zuckerman}}]{xu2014}
{Xu} S., {Jura} M., {Koester} D., et~al. (2014): {Elemental Compositions of Two Extrasolar Rocky Planetesimals}. \apj\, 783(2):79. \doi{10.1088/0004-637X/783/2/79}, {\href{https://arxiv.org/abs/1401.4252}{{arXiv:1401.4252}}} {[astro-ph.EP]}

\bibitem[{{Xu} et~al.(2017){Xu}, {Zuckerman}, {Dufour}, {Young}, {Klein}, \& {Jura}}]{xu2017}
{Xu} S., {Zuckerman} B., {Dufour} P., et~al. (2017): {The Chemical Composition of an Extrasolar Kuiper-Belt-Object}. \apjl\, 836(1):L7. \doi{10.3847/2041-8213/836/1/L7}, {\href{https://arxiv.org/abs/1702.02868}{{arXiv:1702.02868}}} {[astro-ph.EP]}

\bibitem[{{Xu} et~al.(2019){Xu}, {Dufour}, {Klein}, {Melis}, {Monson}, {Zuckerman}, {Young}, \& {Jura}}]{xu2019}
{Xu} S., {Dufour} P., {Klein} B., et~al. (2019): {Compositions of Planetary Debris around Dusty White Dwarfs}. \aj\, 158(6):242. \doi{10.3847/1538-3881/ab4cee}, {\href{https://arxiv.org/abs/1910.07197}{{arXiv:1910.07197}}} {[astro-ph.SR]}

\bibitem[{{York} et~al.(2000){York}, {Adelman}, {Anderson}, {Anderson}, {Annis}, {Bahcall}, {Bakken}, {Barkhouser}, {Bastian}, {Berman}, {Boroski}, {Bracker}, {Briegel}, {Briggs}, {Brinkmann}, {Brunner}, {Burles}, {Carey}, {Carr}, {Castander}, {Chen}, {Colestock}, {Connolly}, {Crocker}, {Csabai}, {Czarapata}, {Davis}, {Doi}, {Dombeck}, {Eisenstein}, {Ellman}, {Elms}, {Evans}, {Fan}, {Federwitz}, {Fiscelli}, {Friedman}, {Frieman}, {Fukugita}, {Gillespie}, {Gunn}, {Gurbani}, {de Haas}, {Haldeman}, {Harris}, {Hayes}, {Heckman}, {Hennessy}, {Hindsley}, {Holm}, {Holmgren}, {Huang}, {Hull}, {Husby}, {Ichikawa}, {Ichikawa}, {Ivezi{\'c}}, {Kent}, {Kim}, {Kinney}, {Klaene}, {Kleinman}, {Kleinman}, {Knapp}, {Korienek}, {Kron}, {Kunszt}, {Lamb}, {Lee}, {Leger}, {Limmongkol}, {Lindenmeyer}, {Long}, {Loomis}, {Loveday}, {Lucinio}, {Lupton}, {MacKinnon}, {Mannery}, {Mantsch}, {Margon}, {McGehee}, {McKay}, {Meiksin}, {Merelli}, {Monet}, {Munn}, {Narayanan}, {Nash}, {Neilsen}, {Neswold}, {Newberg}, {Nichol}, {Nicinski},
  {Nonino}, {Okada}, {Okamura}, {Ostriker}, {Owen}, {Pauls}, {Peoples}, {Peterson}, {Petravick}, {Pier}, {Pope}, {Pordes}, {Prosapio}, {Rechenmacher}, {Quinn}, {Richards}, {Richmond}, {Rivetta}, {Rockosi}, {Ruthmansdorfer}, {Sandford}, {Schlegel}, {Schneider}, {Sekiguchi}, {Sergey}, {Shimasaku}, {Siegmund}, {Smee}, {Smith}, {Snedden}, {Stone}, {Stoughton}, {Strauss}, {Stubbs}, {SubbaRao}, {Szalay}, {Szapudi}, {Szokoly}, {Thakar}, {Tremonti}, {Tucker}, {Uomoto}, {Vanden Berk}, {Vogeley}, {Waddell}, {Wang}, {Watanabe}, {Weinberg}, {Yanny}, {Yasuda}, \& {SDSS Collaboration}}]{york2000}
{York} D.~G., {Adelman} J., {Anderson} J.John~E., et~al. (2000): {The Sloan Digital Sky Survey: Technical Summary}. \aj\, 120(3):1579--1587. \doi{10.1086/301513}, {\href{https://arxiv.org/abs/astro-ph/0006396}{{arXiv:astro-ph/0006396}}} {[astro-ph]}

\bibitem[{{Zuckerman} \& {Reid}(1998)}]{zuckerman1998}
{Zuckerman} B., {Reid} I.~N. (1998): {Metals in Cool DA White Dwarfs}. \apjl\, 505(2):L143--L146. \doi{10.1086/311608}

\bibitem[{{Zuckerman} et~al.(2003){Zuckerman}, {Koester}, {Reid}, \& {H{\"u}nsch}}]{zuckerman2003}
{Zuckerman} B., {Koester} D., {Reid} I.~N., et~al. (2003): {Metal Lines in DA White Dwarfs}. \apj\, 596(1):477--495. \doi{10.1086/377492}

\bibitem[{{Zuckerman} et~al.(2007){Zuckerman}, {Koester}, {Melis}, {Hansen}, \& {Jura}}]{zuckerman2007}
{Zuckerman} B., {Koester} D., {Melis} C., et~al. (2007): {The Chemical Composition of an Extrasolar Minor Planet}. \apj\, 671(1):872--877. \doi{10.1086/522223}, {\href{https://arxiv.org/abs/0708.0198}{{arXiv:0708.0198}}} {[astro-ph]}

\bibitem[{{Zuckerman} et~al.(2010){Zuckerman}, {Melis}, {Klein}, {Koester}, \& {Jura}}]{zuckerman2010}
{Zuckerman} B., {Melis} C., {Klein} B., et~al. (2010): {Ancient Planetary Systems are Orbiting a Large Fraction of White Dwarf Stars}. \apj\, 722(1):725--736. \doi{10.1088/0004-637X/722/1/725}, {\href{https://arxiv.org/abs/1007.2252}{{arXiv:1007.2252}}} {[astro-ph.SR]}

\bibitem[{{Zuckerman} et~al.(2011){Zuckerman}, {Koester}, {Dufour}, {Melis}, {Klein}, \& {Jura}}]{zuckerman2011}
{Zuckerman} B., {Koester} D., {Dufour} P., et~al. (2011): {An Aluminum/Calcium-rich, Iron-poor, White Dwarf Star: Evidence for an Extrasolar Planetary Lithosphere?} \apj\, 739(2):101. \doi{10.1088/0004-637X/739/2/101}, {\href{https://arxiv.org/abs/1107.2167}{{arXiv:1107.2167}}} {[astro-ph.SR]}

\end{thebibliography}

\end{document}